\documentclass[a4paper,titlepage,11pt]{article}
\pdfoutput=1
\usepackage[utf8]{inputenc}
\usepackage{geometry}
\usepackage[english]{babel}
\usepackage{amsmath, amsfonts, graphicx}
\usepackage{mathtools}
\usepackage{graphicx}
\usepackage{float}
\usepackage{pstricks}
\usepackage[labelfont=bf,font={small}]{caption}
\usepackage{pgfplots}
\usepackage{tikz}
\usepackage{jheppubedit}
\usetikzlibrary{positioning, arrows}
\usetikzlibrary{external}
\usepackage{footmisc}
\usepackage{multirow}
\usepackage{url}
\usepackage[breakable, theorems, skins]{tcolorbox}
\usepackage{enumerate}
\usepackage{physics}
\usepackage{bm}
\usepackage{epstopdf}
\usepackage{tensor}
\usepackage{tikz}
\usepackage{simplewick}

\newcommand{\mj}{\mathcal{J}}

\newcommand{\mo}{\mathcal{O}}
\newcommand{\slr}{\text{SL}(2,\mathbb{R})}
\newcommand{\slrp}{\text{SL}^+(2,\mathbb{R})}

\newcommand{\tj}[6]{ \begin{pmatrix}
   #1 & #2 & #3 \\
   #4 & #5 & #6 
  \end{pmatrix}}
  \newcommand{\sj}[6]{ \begin{Bmatrix}
   #1 & #2 & #3 \\
   #4 & #5 & #6 
  \end{Bmatrix}}

\begin{document}
\title{Particle on Group and Schwarzian: A Wilson Line Perspective}
\def\nspc{{\hspace{-2pt}}}
\newcommand{\threej}[6]{\left(\mbox{\small$\!\begin{array}{ccc} #1 \! & \!\! #3 \! & \!\! #5 \nspc \\[-1mm]  #2 \!  & \!\! #4 \!  & \!\! #6 \nspc \end{array}\!$}\!\right)}

\begin{titlepage}

\setcounter{page}{1} \baselineskip=15.5pt \thispagestyle{empty}

\vfil

${}$
\vspace{1cm}

\begin{center}

\def\thefootnote{\fnsymbol{footnote}}
\begin{changemargin}{0.05cm}{0.05cm} 
\begin{center}
{\Large \bf The Schwarzian Theory - A Wilson Line Perspective} \\
\end{center} 
\end{changemargin}

~\\[1cm]
{Andreas Blommaert\footnote{\href{mailto:andreas.blommaert@ugent.be}{\protect\path{andreas.blommaert@ugent.be}}}, Thomas G. Mertens\footnote{\href{mailto:thomas.mertens@ugent.be}{\protect\path{thomas.mertens@ugent.be}}} and Henri Verschelde\footnote{\href{mailto:henri.verschelde@ugent.be}{\protect\path{henri.verschelde@ugent.be}}}
}
\\[0.3cm]

{\normalsize { \sl Department of Physics and Astronomy
\\[1.0mm]
Ghent University, Krijgslaan, 281-S9, 9000 Gent, Belgium}}\\[3mm]

\end{center}

 \vspace{0.2cm}
\begin{changemargin}{01cm}{1cm} 
{\small  \noindent 
\begin{center} 
\textbf{Abstract}
\end{center} }
We provide a holographic perspective on correlation functions in Schwarzian quantum mechanics, as boundary-anchored Wilson line correlators in Jackiw-Teitelboim gravity. We first study compact groups and identify the diagrammatic representation of bilocal correlators of the particle-on-a-group model as Wilson line correlators in its 2d holographic BF description. We generalize to the Hamiltonian reduction of $\slr$ and derive the Schwarzian correlation functions. Out-of-time ordered correlators are determined by crossing Wilson lines, giving a $6j$-symbol, in agreement with 2d CFT results.
\end{changemargin}
 \vspace{0.3cm}
\vfil
\begin{flushleft}
\today
\end{flushleft}

\end{titlepage}

\newpage
\tableofcontents

\setcounter{footnote}{0}

\section{Introduction}

\begin{figure}[h]
\centering
\includegraphics[width=0.55\textwidth]{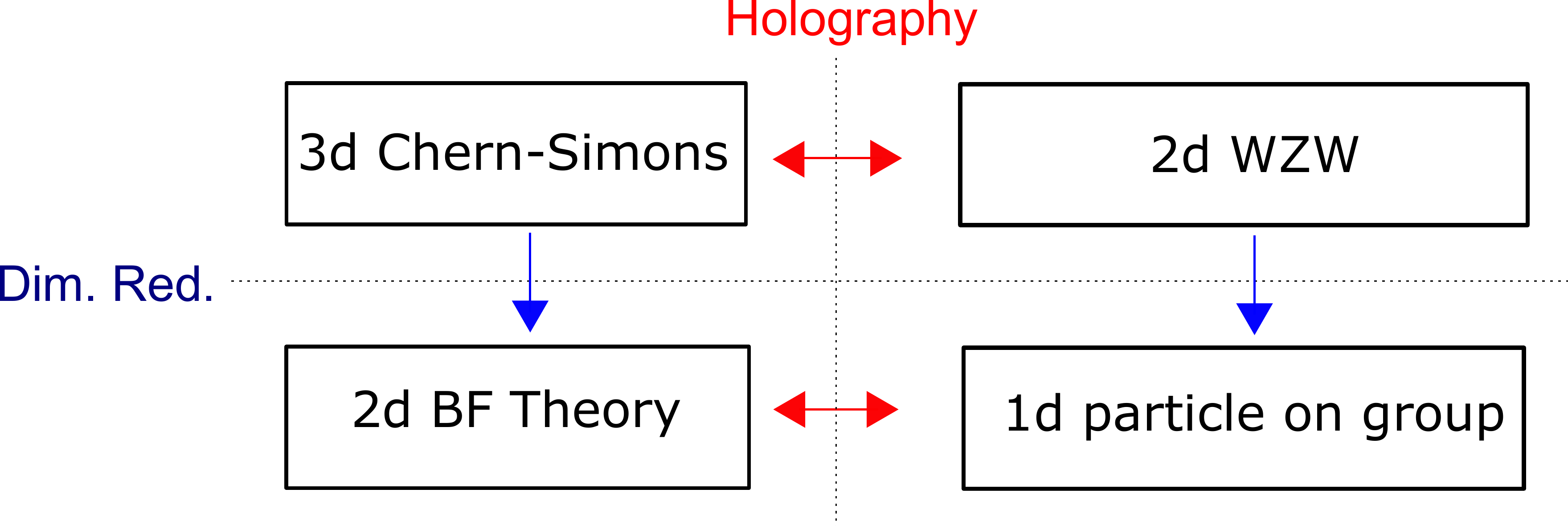}
\caption{The 3d Chern-Simons / 2d WZW holographic relation, which dimensionally reduces to the 2d BF / 1d particle-on-a-group holography.}
\label{schemedimholin}
\end{figure}

Exactly solvable quantum systems that exhibit holographic features are very scarce. Therefore every example one finds should be thoroughly understood. The past few years, it became clear that one such example is the Sachdev-Ye-Kitaev (SYK) model (and its relatives) \cite{KitaevTalks, Sachdev:1992fk, Polchinski:2016xgd, Jevicki:2016bwu, Maldacena:2016hyu,Jevicki:2016ito,randommatrix,Turiaci:2017zwd,Gross:2017hcz,Gross:2017aos,Das:2017pif,Das:2017wae,Kitaev:2017awl,Berkooz:2018qkz}. These models describe $N$ Majorana fermions interacting through a random all-to-all 4-point interaction. The model is diagrammatically tractable at large $N$, and many features are understood by now. 
A very interesting regime is that of strong coupling, where the model is effectively described by Schwarzian quantum mechanics:
\begin{equation}
\label{SSch}
S = -C\int dt \, \left\{f,t\right\},
\end{equation}
where $\left\{f,t\right\} = \frac{f'''}{f'} - \frac{3}{2}\frac{f''^2}{f'^2}$ denotes the Schwarzian derivative of $f$. It was immediately realized that a 2d dilaton gravity theory in AdS$_2$ - the Jackiw-Teitelboim (JT) model - has dynamics described by precisely this same Schwarzian action \cite{Jackiw:1984je, Teitelboim:1983ux,Jackiw:1992bw,Almheiri:2014cka, Jensen:2016pah, Maldacena:2016upp, Engelsoy:2016xyb, Cvetic:2016eiv,Mandal:2017thl,Nayak:2018qej}, such that the Schwarzian theory can be used to study a 2d quantum gravity.
Motivated by this, using the structural link of Schwarzian quantum mechanics with 2d Virasoro CFT, the quantum dynamics of the Schwarzian theory was discussed in \cite{Mertens:2017mtv,Mertens:2018fds}, with the link with the semi-classical regime in JT gravity explicitly uncovered in \cite{Lam:2018pvp}. Several avenues towards Schwarzian correlation functions of bilocal operator insertions exist by now: the 2d Liouville perspective was explored in \cite{Mertens:2017mtv,Mertens:2018fds}, a 1d Liouville approach was analyzed in \cite{altland,Bagrets:2017pwq}, and a particle on $H_2^+$ in an infinite (imaginary) magnetic field is explored in \cite{Kitaev:2018wpr, Zhenbin}. Whereas each approach has its own value, this does leave open the question of whether these results can be found directly from Jackiw-Teitelboim gravity. In particular, we would like to know what is the bulk JT gravity interpretation of the bilocal Schwarzian operators. We will answer this question using the topological BF formulation of JT gravity. 
\begin{figure}[h]
\centering
\includegraphics[width=0.55\textwidth]{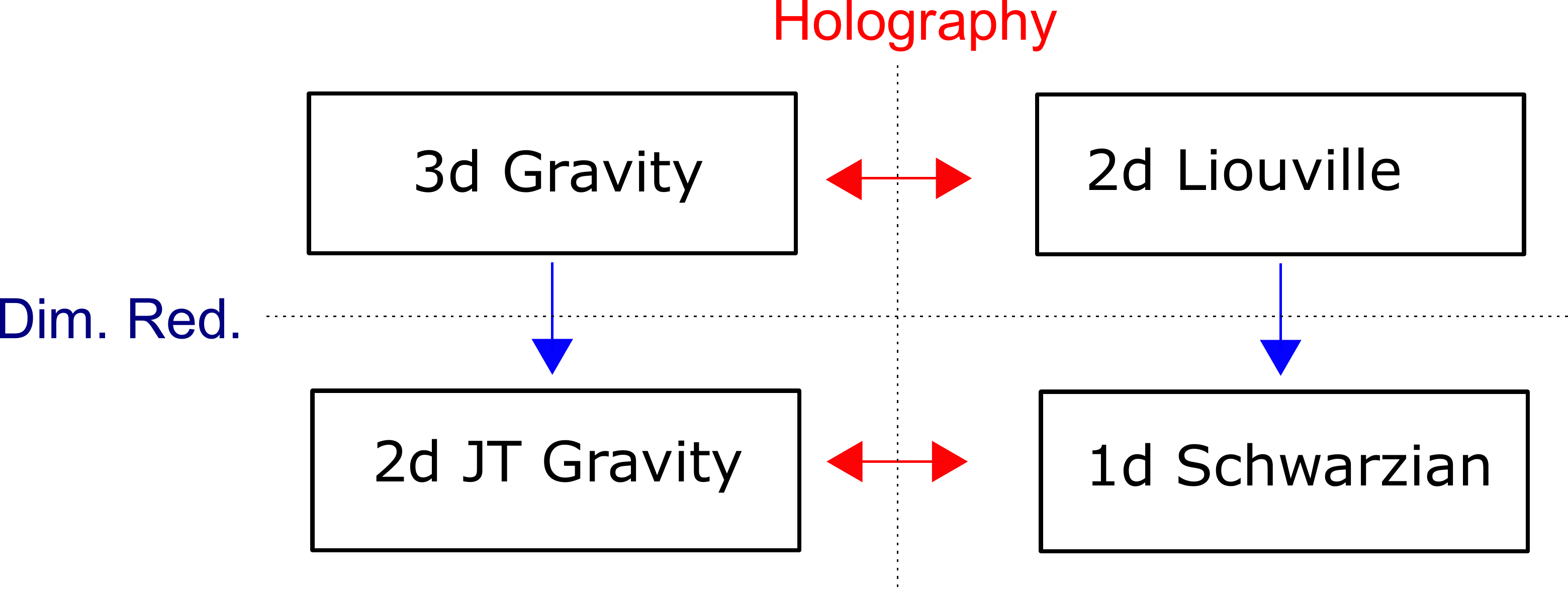}
\caption{Relation through holography and dimensional reduction of 3d $\Lambda<0$ gravity, 2d JT gravity, 2d Liouville CFT and 1d Schwarzian QM.}
\label{schwarzianembed}
\end{figure}
The relation of the Schwarzian with Liouville and JT gravity is depicted in Figure \ref{schwarzianembed}.
\\~\\
It was realized soon after in \cite{Stanford:2017thb,Mertens:2018fds}, that the above gravitational story is only the irrational case (in a 2d CFT sense) of a more generic story that includes Wess-Zumino-Witten (WZW) CFT and Chern-Simons (CS) / BF bulk theories. The analogue of the Schwarzian model is played by the 1d particle-on-a-group theory (Figure \ref{schemedimholin}). This model also appears in SYK models with internal symmetries \cite{Davison:2016ngz,Yoon:2017nig,Choudhury:2017tax,Narayan:2017hvh,Gaikwad:2018dfc}, as a building block of the supersymmetric theories \cite{Fu:2016vas}, and in higher spin generalizations \cite{Gonzalez:2018enk}. Correlation functions of the 1d particle-on-a-group theory were determined using the identification of particle-on-a-group as the dimensional reduction at $k\to +\infty$ of WZW \cite{Mertens:2018fds}. This is the rightmost track of Figure \ref{schemedimholin}.
In the first part of this work, we will arrive at these same correlation functions using holography of the 2d BF theory (using the bottom track in Figure \ref{schemedimholin}). This will provide a complementary perspective on the previous derivations, and in particular on the out-of-time ordered (OTO) correlators. The results presented here for BF theory were already anticipated in \cite{Blommaert:2018oue}, where we investigated the particle-on-a-group model as describing the edge dynamics of 2d Yang-Mills on a disk. 

It is well-known that AdS$_3$ gravity / Liouville are constrained versions of respectively $\slr$ CS / WZW \cite{Witten:1988hc, Bershadsky:1989mf, Verlinde:1989ua, Forgacs:1989ac, Balog:1990mu}, generalizing the well-known compact story \cite{Witten:1988hf,Elitzur:1989nr,Moore:1989yh}. In this work, we will exploit the dimensional reduction of this gauge theory perspective on AdS$_3$ gravity and discuss JT gravity / Schwarzian as a constrained version of $\slr$ BF / particle-on-a-group. The result is a bulk JT derivation of Schwarzian correlators. 
\\~\\
In section \ref{s:compact} we discuss the bulk BF interpretation of correlators in particle-on-a-group. After a brief discussion on non-compact generalizations in section \ref{sect:interlude}, we generalize to the Schwarzian / Liouville theories in section \ref{s:gravity}. We resort to the conventional Hamiltonian reduction of $\slr$ to find out how Schwarzian correlators are related to a constrained $\slrp$ BF theory. We end in section \ref{sect:concl} with some possibilities to pursue in the future, in particular we make a short detour to higher spin Schwarzian systems, focusing on the SL$(n,\mathbb{R})$ Hamiltonian reduction. In order to make the presentation self-contained, the appendices contain some review material on what we need of $\slr$ representation theory and the higher spin generalization.
\subsubsection*{Note Added}
A related approach is being developed in \cite{IPW}.

\section{Particle-on-a-group Theory}
\label{s:compact}
We consider in this section correlators of boundary-anchored Wilson lines in BF theory with specific boundary conditions and prove a 1:1 relation with bilocal correlators in the 1d particle on a group theory.

\begin{figure}[h]
\centering
\includegraphics[width=0.75\textwidth]{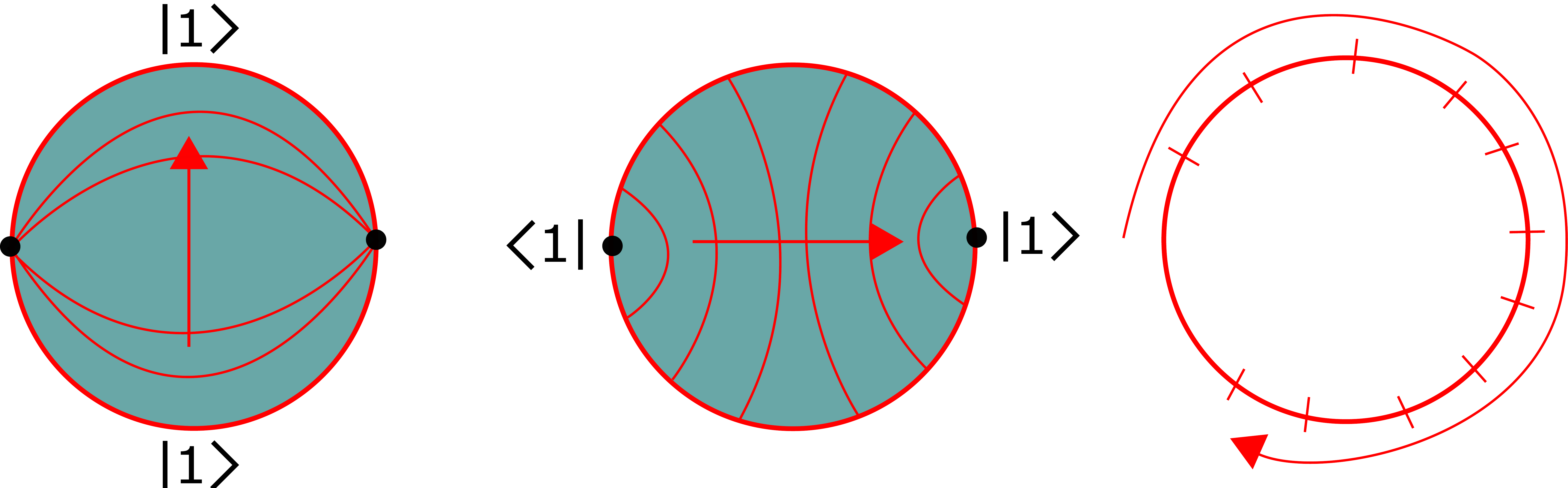}
\caption{Left: disk amplitude in the open channel. Middle: disk amplitude in the crossed channel. Right: Boundary circle with evolution along the circle.}
\label{fig:Channels}
\end{figure}

Dimensional reduction of 3d Chern-Simons with a boundary results in 2d BF theory \cite{Mertens:2018fds} (using the left track in figure \ref{schemedimholin}):
\begin{equation}
S[\chi,A]=\int \Tr (\chi F)+\frac{1}{2}\int d\tau \Tr(\chi A_\tau). \label{BF}
\end{equation}
Introducing a 1d metric $\gamma_{\tau\tau}$ along the boundary curve, we impose the boundary condition $\chi\rvert_\text{bdy} = C \sqrt{\gamma_{\tau\tau}} A_\tau\rvert_\text{bdy}$ for some fixed constant $C$. The bulk is topological, but this is broken by the boundary condition, and one finds the energy-momentum tensor:
\begin{equation}
T_{\tau\tau}(\tau) \equiv \frac{2}{\sqrt{\gamma}}\frac{\delta S}{\delta \gamma^{\tau\tau}} = \frac{1}{2C} \text{Tr}(\chi^2) = L_{\text{bdy}},
\end{equation}
setting $\gamma_{\tau\tau}=1$ in the end. Hence the choice of boundary term is crucial. This is manifest when choosing alternatively e.g. $A_\tau\rvert_\text{bdy}=0$: this puts $H(\tau)=0$ and destroys the boundary dynamics. In that case, we are left with the completely topological theory 
\begin{equation}
S[\chi,A]=\int \Tr (\chi F).\label{bf}
\end{equation}
Notice that this is just the $e\to 0$ limit of 2d Yang-Mills theory with boundary conditions $A_\tau\rvert_\text{bdy}=0$ \cite{Witten:1991we,Cordes:1994fc}.\footnote{The equivalence is proven in the path integral by integrating out $\chi$ in the action \begin{equation}
S[\chi,A]=\frac{e}{2}\int d^2 x \sqrt{g}\Tr( \chi^2)+\int \Tr(\chi F),
\end{equation}
which becomes \eqref{bf} in the limit $e\to 0$.} Structurally, the theory \eqref{BF} is then very similar to 2d Yang-Mills (YM) theory, a feature which will become more apparent below. We explore this completely topological BF-theory and its observables in Appendix \ref{app:knots} applying traditional techniques. The observables in Chern-Simons are Wilson lines and this continues to hold after dimensional reduction to BF. As in CS, observables of BF split into topological bulk observables (knots discussed in Appendix \ref{app:knots}), and dynamical observables associated with boundary dynamics: boundary-anchored Wilson lines. Here we will only be interested in the latter. In particular we will restrict ourselves to Wilson lines with both endpoints on the boundary: networks of Wilson lines are left for future work.\footnote{Such networks of Wilson lines have natural holographic duals in the particle-on-group and Schwarzian models, though they are not the bilocal boundary operators we are interested in here.}
\\~\\
We will mainly be interested in the thermal theory \eqref{BF} defined on a disk. In order to compute the partition function and the correlation functions, we will employ a slicing of the theory in intervals (instead of the usual circles employed in e.g. \cite{Migdal:1975zg,Witten:1991we,Witten:1992xu,Cordes:1994fc}). We will utilize several different descriptions of the same amplitude, as in Figure \ref{fig:Channels}.
The left Figure describes an open channel slicing with identity boundary states that we will explore in \ref{sect:ocd}. The middle Figure is, due to the topological invariance of the model manifestly the same, and is an instructive slicing to link with the brane description of \cite{Mertens:2017mtv,Mertens:2018fds}. The right Figure describes a boundary-intrinsic particle-on-a-group perspective that we will discuss in section \ref{sect:holobf}.

\subsection{Correlation Functions in BF}
\label{s:wilsonbf}
Let us now calculate boundary anchored Wilson line amplitudes in 2d BF theory, and 2d YM, with a boundary. This is an extension of the known correlators \cite{Migdal:1975zg,Witten:1991we,Witten:1992xu,Cordes:1994fc}.

\subsubsection{Open Channel Description}
\label{sect:ocd}
\begin{figure}[h]
\centering
\includegraphics[width=0.6\textwidth]{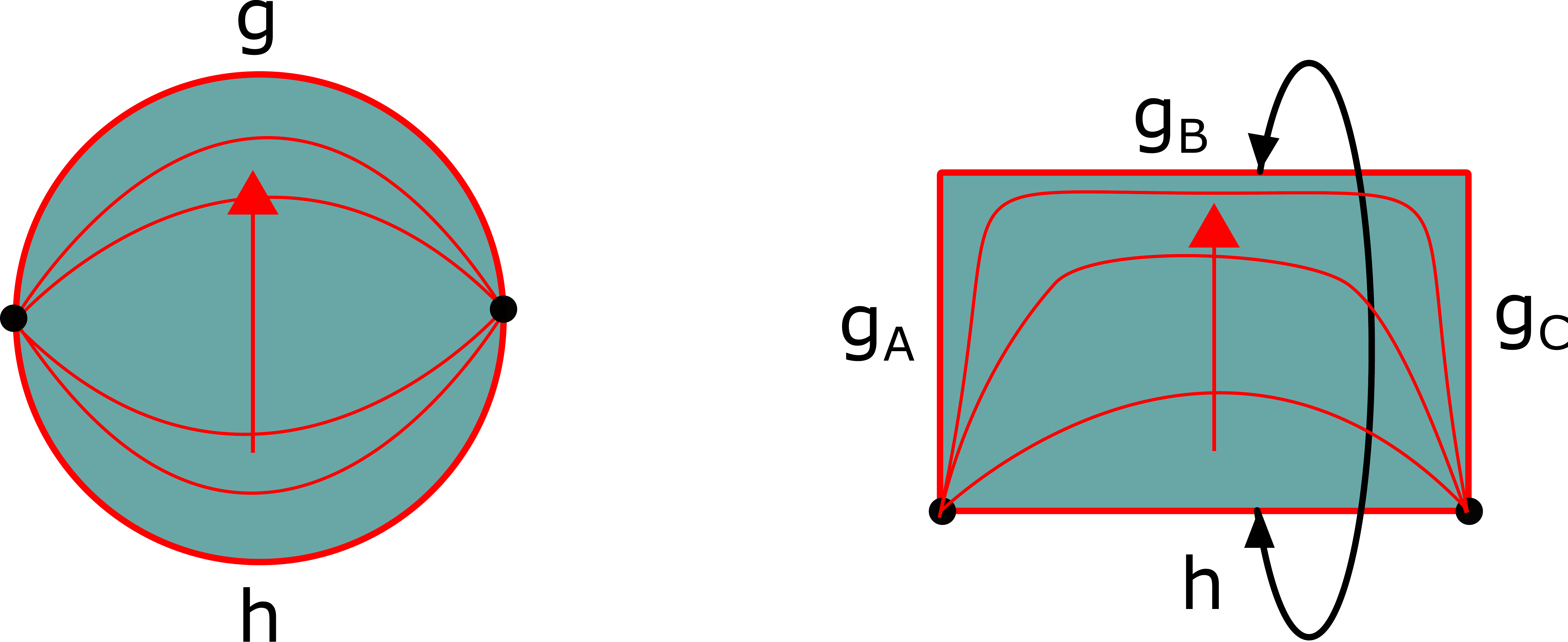}
\caption{The disk amplitude in the open channel (left). Hilbert space slicings are denoted in red. Deforming the amplitude (either preserving the area (YM), or the boundary length (BF)) into a rectangle (right), and identifying the upper and lower boundaries, one arrives at the fundamental cylinder amplitude.}
\label{fig:diskrect}
\end{figure}

\begin{figure}[h]
\centering
\includegraphics[width=0.55\textwidth]{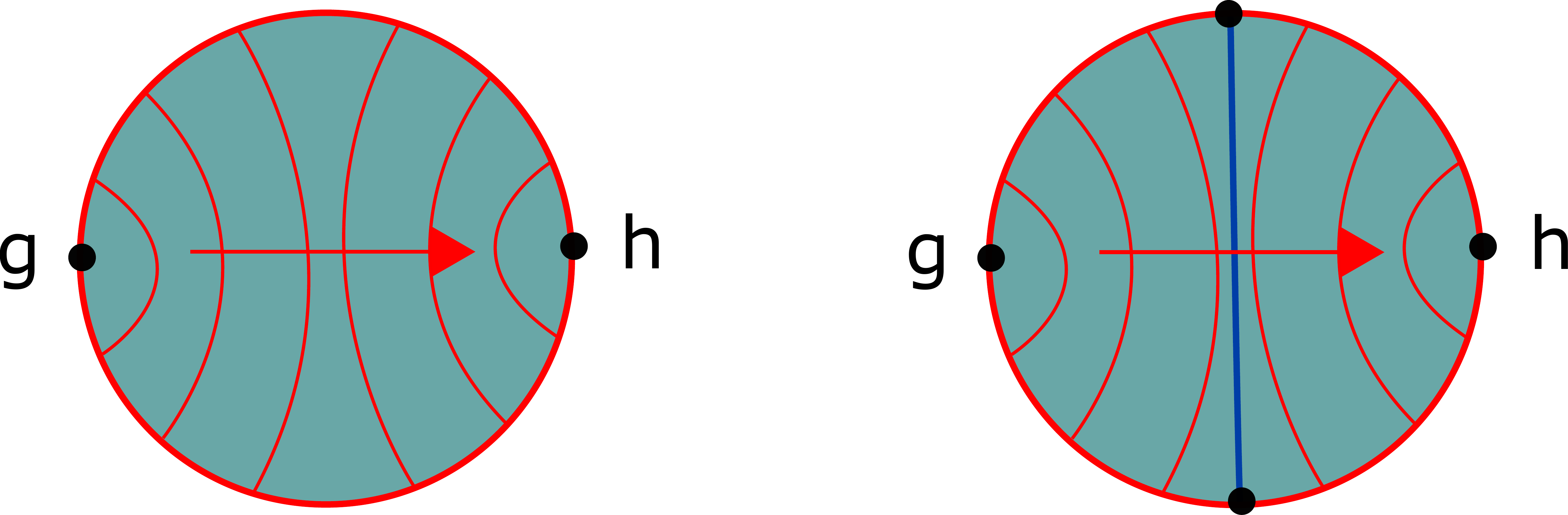}
\caption{Left: Hilbert space slicing in the point-defect channel, representing propagation between pointlike defects. Right: Amplitude with a Wilson line inserted.}
\label{WDI}
\end{figure}

Though 2d YM and BF are distinct theories (the first is quasi-topological and the second fully topological), quantizing them on a manifold with boundaries requires the same techniques. This is due to the lack of propagating degrees of freedom in two dimensions and the fact that both theories have an identical Hilbert space structure, determined by the Peter-Weyl theorem; when quantizing the theory on a circle, the Hilbert space consists of the irreps whose normalized wavefunctions are the characters $\chi_R(g)=\bra{g}\ket{R}=\Tr R(g)$. 
\\~\\
For the purpose of calculating boundary-anchored Wilson lines, we are interested instead in quantizing the theory on an interval: the Peter-Weyl theorem now states that the Hilbert space is spanned by all matrix elements of the unitary irreducible representations of the group. The same logic was used to write 2d Yang-Mills on a sphere as an open string theory in \cite{Donnelly:2016jet}. The normalized wavefunctions are given by:
\begin{equation}
\boxed{\bra{g}\ket{R,ab}= \sqrt{\text{dim R}}\, R_{ab}(g)},\label{wavefctrational}
\end{equation}
where $R_{ab}(g) = \left\langle a\right| g \left|b\right\rangle$ is the matrix element in the representation $R$. Two conjugate bases for the Hilbert space of BF theory (or particle-on-a-group) are hence $\left|R,ab\right\rangle$ and the group elements $\left|g\right\rangle$. For example, for $U(1)$ these would be the discretized charges $\left|q\right\rangle$ or the angular coordinate $\left|\phi\right\rangle$ respectively. The appropriate normalization of the wavefunctions is deduced from the grand orthogonality theorem:\footnote{We gather some relevant equations in Appendix \ref{app:groupth}.}
\begin{equation}
\int dg R_{ab}(g)R'_{cd}(g^{-1}) = \frac{\delta_{RR'}}{\text{dim R}}\delta_{ad}\delta_{bc},\label{Rotho}
\end{equation}
and orthonormality of the wavefunctions. Note that \eqref{Rotho} is fixed by the normalization $R_{a b}(\mathbf{1})=\delta_{ab}$ which in turn is enforced by the defining property of representation matrices $R_{ab}(\mathbf{1})R_{bc}(\mathbf{1})=R_{ac}(\mathbf{1})$. 
\\~\\
Both for YM and BF, the Hamiltonian reduces to the Casimir of the algebra, hence the basis states $\ket{R,ab}$ are also eigenstates of the Hamiltonian. Specifically, for BF the boundary-supported Hamiltonian propagation factor evaluated in a state $\ket{R,ab}$ is:
\begin{equation}
T H(R) = L \frac{\mathcal{C}_R}{C},\label{bfham}
\end{equation}
with $L$ the length of the boundary segment propagated over, and $\mathcal{C}_R$ the Casimir in the irrep $R$. Note that only the physical boundaries of the 2d manifold come with a boundary Hamiltonian contribution,\footnote{Gluing different patches together results in a net cancellation of the Hamiltonian density on the gluing curve, such that gluing is consistent with the concept of a boundary Hamiltonian.} and that the Hamiltonian is independent of the labels $a$ and $b$.  Likewise, for 2d YM with action 
\begin{equation}
S[A]=\frac{1}{2e}\int d^2 x \sqrt{g}\Tr F^2,
\end{equation}
the Hamiltonian propagation factor scales with the area $A$ and the coupling $e$:
\begin{equation}
T H(R)=e A \mathcal{C}_R.
\end{equation}
Consider now a disk-shaped surface and divide the circular boundary into two intervals. Associate a state $\ket{g}$ with the first interval and a state $\ket{h}$ with the second. The situation is shown on the left of Figure \ref{fig:diskrect}. The partition function of this disk is:
\begin{equation}
Z_\text{disk}(g,h)=\bra{g}e^{-TH}\ket{h}.\label{zdisk}
\end{equation}
The states $\ket{g}$ and $\ket{h}$ in \eqref{zdisk} can be viewed as \emph{boundary states}, that can be expanded in the representation basis using \eqref{wavefctrational}:
\begin{equation}
\ket{g} = \sum_{R,m,n}\sqrt{\text{dim R}}\, R_{mn}(g)\ket{R,mn}.
\end{equation}
Let us now demonstrate the correctness of this open channel perspective provided in \eqref{zdisk} by rederiving the known 2d YM results \cite{Cordes:1994fc,Witten:1991we} from it. Inserting a completeness relation in \eqref{zdisk}, we obtain:
\begin{equation}
\label{diska}
Z_\text{disk}(g,h)= \sum_{R,a,b} \bra{g}\ket{R,a,b}\bra{R,a,b}\ket{h}e^{-TH(R)} = \sum_{R,a,b} \text{dim R} \, R_{ab}(g)R_{ba}(h^{-1}) \, e^{-TH(R)}.
\end{equation}
Denoting the holonomy $U=g\cdot h^{-1}$ along the entire boundary of the disk and using $\chi_R(U)=\Tr R(U)$ this matches with the calculation of the disk partition function from the closed channel \cite{Cordes:1994fc,Witten:1991we}:
\begin{equation}
Z_\text{disk}(U)=\sum_{R} \text{dim R} \, \chi_R(U) e^{-TH}.
\end{equation}
Alternatively, using the definition of irrep matrices and splitting $g$ as $g=g_A \cdot g_B\cdot g_C$ we can rewrite the disk amplitude as a rectangle amplitude as in the right of Figure \ref{fig:diskrect}: 
\begin{equation}
Z_\text{rect}(h,g_A,g_B,g_C)=\sum_{R,a,b,x,y} \text{dim R} \, R_{ax}(g_A)R_{xy}(g_B)R_{yb}(g_C)R_{ba}(h^{-1}) e^{-TH(R)}.
\end{equation}
This can be glued into a cylinder by gluing along the boundaries and using \eqref{Rotho} as
\begin{equation}
Z_\text{cyl}(g_A,g_C)=\int d h d g_B \, \delta(h-g_B) \, Z_\text{rect}(h,g_A,g_B,g_C)=\sum_{R,a,b}\chi_R(g_A) \, \chi_R(g_C) \, e^{-TH(R)},
\end{equation}
Again this matches with the closed channel calculation of the annulus. Reproducing the disk and the annulus proves by induction (gluing) that the open-channel calculations result in the correct partition function on a generic 2d manifold \cite{Migdal:1975zg,Witten:1991we,Witten:1992xu,Cordes:1994fc}, and thus confirms the validity of our open-channel approach. Physical boundary segments of 2d YM or BF are characterized by $g=h=\mathbf{1}$ \cite{Blommaert:2018oue}. The disk partition function becomes: 
\begin{equation}
Z_\text{disk} = \sum_{R,m} \text{dim R}\, e^{-\beta \frac{\mathcal{C}_R}{C}} = \sum_{R} (\text{dim R})^2\, e^{-\beta \frac{\mathcal{C}_R}{C}},
\end{equation}
indeed equal to the partition function of a particle on a group manifold \cite{Marinov:1979gm,Chu:1994hm}.
\\~\\ 
It is useful to look at these same calculations from the perspective of yet another Hilbert space slicing, this to make contact with the computations performed in \cite{Mertens:2017mtv,Mertens:2018fds}, where particle-on-a-group correlators were obtained by dimensionally reducing WZW (see Appendix \ref{app:wzwcorr}). Consider the ``point-defect'' slicing of Figure \ref{WDI} left. The calculation of the disk amplitude is precisely the same as the above. Now however, the boundary states are associated with point-like defects. Setting again $g=h=\mathbf{1}$ identifies these defects as the dimensional reduction of the vacuum brane inserted in WZW, which is also characterized by $g=h=\mathbf{1}$.\footnote{The boundary state $\left|\mathbf{1}\right\rangle$ is also closely related to the so-called $\Omega$-state in 2d YM, introduced by Gross and Taylor in \cite{Gross:1993yt}, and interpreted in \cite{Donnelly:2016jet} as an entanglement brane. Whereas the latter is a state in the closed (chiral) Hilbert space (i.e. a class function), and is expanded in a basis of irreps as
\begin{equation}
\left|\Omega\right\rangle = \sum_R \text{dim R}\, \left|R\right\rangle,
\end{equation}
the former resides in the open Hilbert space, and is expanded as
\begin{equation}
\left|\mathbf{1}\right\rangle = \sum_{R,m} \sqrt{\text{dim R}}\, \left|R,mm\right\rangle.
\end{equation}
} The thermal boundary circle of Figure \ref{WDI} denotes the doubled space obtained after performing the doubling trick; before dimensional reduction this was the torus on which the chiral (effect of the branes) WZW lives. This same observation will turn out to hold for the JT gravity / Schwarzian: the boundary defects are the dimensional reduction of ZZ-branes $\phi\to \infty$. 

\subsubsection{Wilson Lines}
\begin{figure}[h]
\centering
\includegraphics[width=0.55\textwidth]{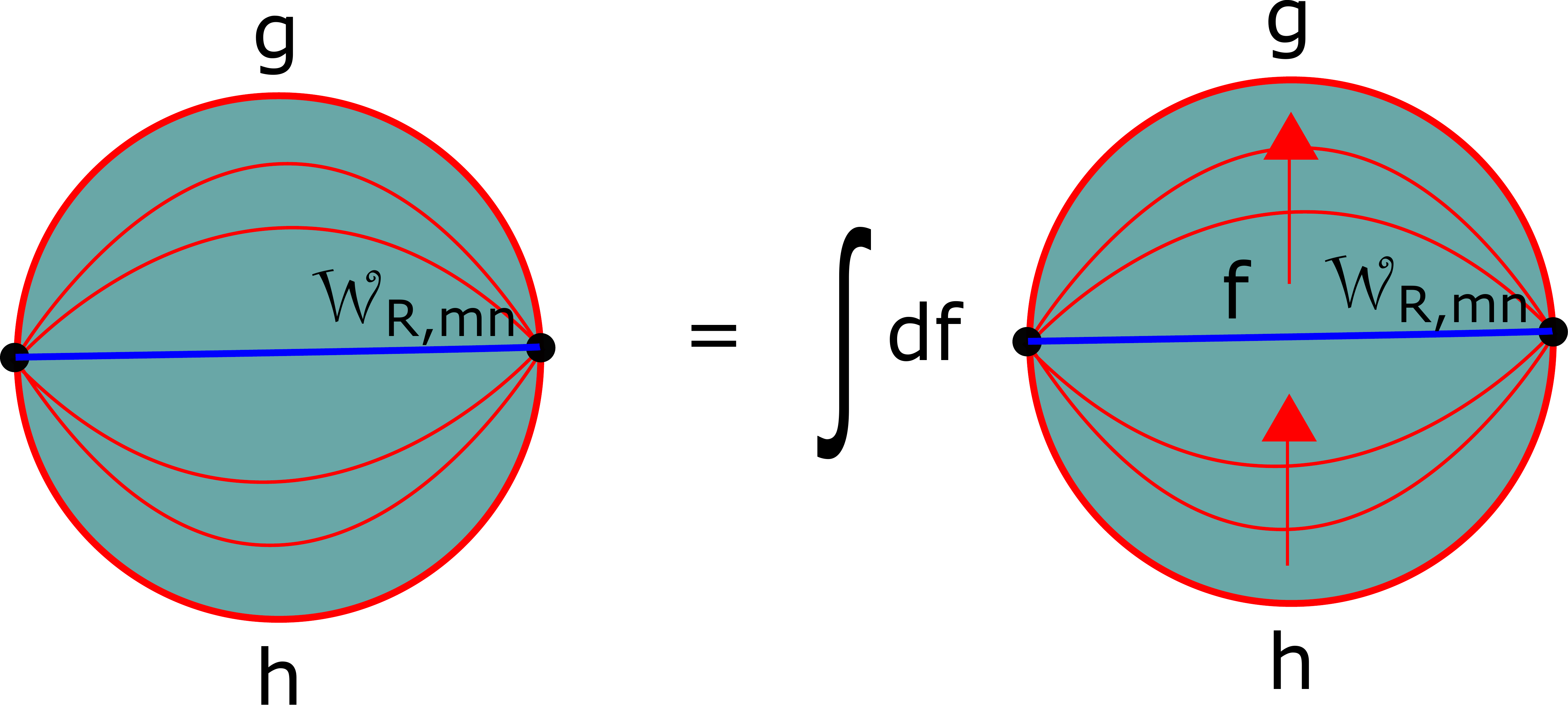}
\caption{Left: Disk with a single Wilson line inserted. Red lines are Hilbert space slicings. Right: Computation done by splitting the disk into two semidisks with the Wilson line in the middle.}
\label{WLI}
\end{figure}

Armed with this open-channel formalism we can now directly calculate generic correlators of boundary-anchored Wilson lines in 2d YM and BF, in the same manner as was done in the past for closed Wilson loops in 2d YM. Basically the only tools we need are the disk amplitudes \eqref{diska}, the identification of a Wilson line segment in irrep $R$ with starting point and endpoint labeled by $m$ respectively $n$ evaluated on group element $g$ as $R_{mn}(g)$, and the integral over three representation matrices:
\begin{equation}
\int d g R_{1,m_1n_1}(g) R_{2,m_2n_2}(g) R_{3,m_3n_3}(g) = \tj{R_1}{R_2}{R_3}{m_1}{m_2}{m_3}\tj{R_1}{R_2}{R_3}{n_1}{n_2}{n_3}.\label{3R}
\end{equation}
As an instructive example, consider the diagram with a single Wilson line in the bulk (Figure \ref{WLI} left).
To compute the amplitude, we separate the disk in two halves, and apply our previous computation to both sides:
\begin{equation}
\bra{g} e^{-TH} \mathcal{W}_{R,mn} \ket{h} = \int df \bra{g} e^{-T_1 H} \ket{f} \mathcal{W}_{R,mn}(f) \bra{f} e^{-T_2 H} \ket{h}.
\end{equation}
The gluing integral over the intermediate group variable $f$ is then computable by \eqref{3R}. In formulas:
\begin{align}
&\sum_{R_i,m_i,n_i} \!\!\! \bra{g}\left.\!\! R_1,m_1n_1\right\rangle  e^{-T_1 \frac{\mathcal{C}_{R_1}}{C}} \!\int df \left\langle R_1,m_1n_1\right.\!\ket{f} R_{mn}(f) \bra{f}\left. \! R_2,m_2n_2\right\rangle e^{-T_2 \frac{\mathcal{C}_{R_2}}{C}} \left\langle R_2,m_2n_2\right.\!\ket{h} \nonumber \\
&= \sum_{R_i,m_i,n_i} \text{dim R$_1$}\text{dim R$_2$} \, R_{1,m_1n_1}(g) R_{2,m_2n_2}(h^{-1})e^{-T_1 \frac{\mathcal{C}_{R_1}}{C}} e^{-T_2 \frac{\mathcal{C}_{R_2}}{C}}\nonumber \\
&\qquad \times  \int df R_{1,m_1n_1}(f^{-1}) R_{mn}(f) R_{2,m_2n_2}(f).
\end{align}
Setting again $g=h=\mathbf{1}$, one recovers the particle-on-a-group result \cite{Mertens:2018fds}:
\begin{equation}
\Big\langle \mathcal{O}_{R,m n}(\tau_1,\tau_2)\Big\rangle= \delta_{mn}\sum_{R_i,m_1,m_2} \text{dim R$_1$}\text{dim R$_2$} \, e^{-T_1 \mathcal{C}_{R_1}} e^{-T_2 \mathcal{C}_{R_2}} \tj{R_1}{R}{R_2}{m_1}{m}{m_2}^2. 
\end{equation}
Again, it is instructive to rephrase this amplitude in the point-defect channel (Figure \ref{WDI} right). The Wilson line operator has one end on each side of the boundary, in line with the general idea that local non-chiral CFT operators map to bilocal operators in particle-on-a-group on account of the doubling that takes the non-chiral CFT between branes to the chiral CFT. 
\\~\\
A bulk crossing of Wilson lines can be dealt with using methods already developed in the 2d YM context \cite{Witten:1991we}. 
\begin{align}
\label{6jpict}
\begin{tikzpicture}[scale=1.2, baseline={([yshift=0cm]current bounding box.center)}]
\draw[thick] (-1.05,1.05) -- (1.05,-1.05);
\draw[thick] (-1.05,-1.05) -- (1.05,1.05);
\draw[thick] (0,0) circle (1.5);
\draw[thick,red] (1.05,-1.05) arc (60:120:2.08);
\draw[thick,red] (1.05,-1.05) arc (45:135:1.5);
\draw[thick,red] (1.05,-1.05) arc (75:105:4.0);
\draw[fill,black] (-1.05,-1.05) circle (0.1);
\draw[fill,black] (1.05,-1.05) circle (0.1);
\draw[fill,black] (-1.05,1.05) circle (0.1);
\draw[fill,black] (1.05,1.05) circle (0.1);
\draw (-.8,.35) node {\footnotesize  $R_A$};
\draw (.88,.35) node {\footnotesize  $R_B$};
\end{tikzpicture} \qquad\qquad\qquad
\begin{tikzpicture}[scale=1.2, baseline={([yshift=0cm]current bounding box.center)}]
\draw[thick] (-1.0,1.15) -- (0,0.15);
\draw[thick] (0.15,0) -- (1.15,-1.0);
\draw[thick] (-1.05,1.05) -- (1.05,-1.05);
\draw[thick] (-1.15,1.00) -- (-0.15,0);
\draw[thick] (0,-0.15) -- (1.00,-1.15);
\draw[thick] (-1.05,-1.05) -- (1.05,1.05);
\draw[thick] (-1.15,-1.00) -- (-0.15,0);
\draw[thick] (0,0.15) -- (1.00,1.15);
\draw[thick] (-1.00,-1.15) -- (0,-0.15);
\draw[thick] (0.15,0) -- (1.15,1.00);
\draw[thick] (0,0) circle (1.5);
\draw[fill,black] (-1.05,-1.05) circle (0.1);
\draw[fill,black] (1.05,-1.05) circle (0.1);
\draw[fill,black] (-1.05,1.05) circle (0.1);
\draw[fill,black] (1.05,1.05) circle (0.1);
\draw (-0.9,-0.4) node {\footnotesize  $g_1$};
\draw (0.9,-0.4) node {\footnotesize  $g_2$};
\draw (0.9,0.4) node {\footnotesize  $g_3$};
\draw (-0.9,0.4) node {\footnotesize  $g_4$};
\end{tikzpicture}
\end{align}
On the left, we have drawn some slices of the open channel Hilbert space in the bottom wedge. 
\begin{figure}[h]
\centering
\includegraphics[width=0.35\textwidth]{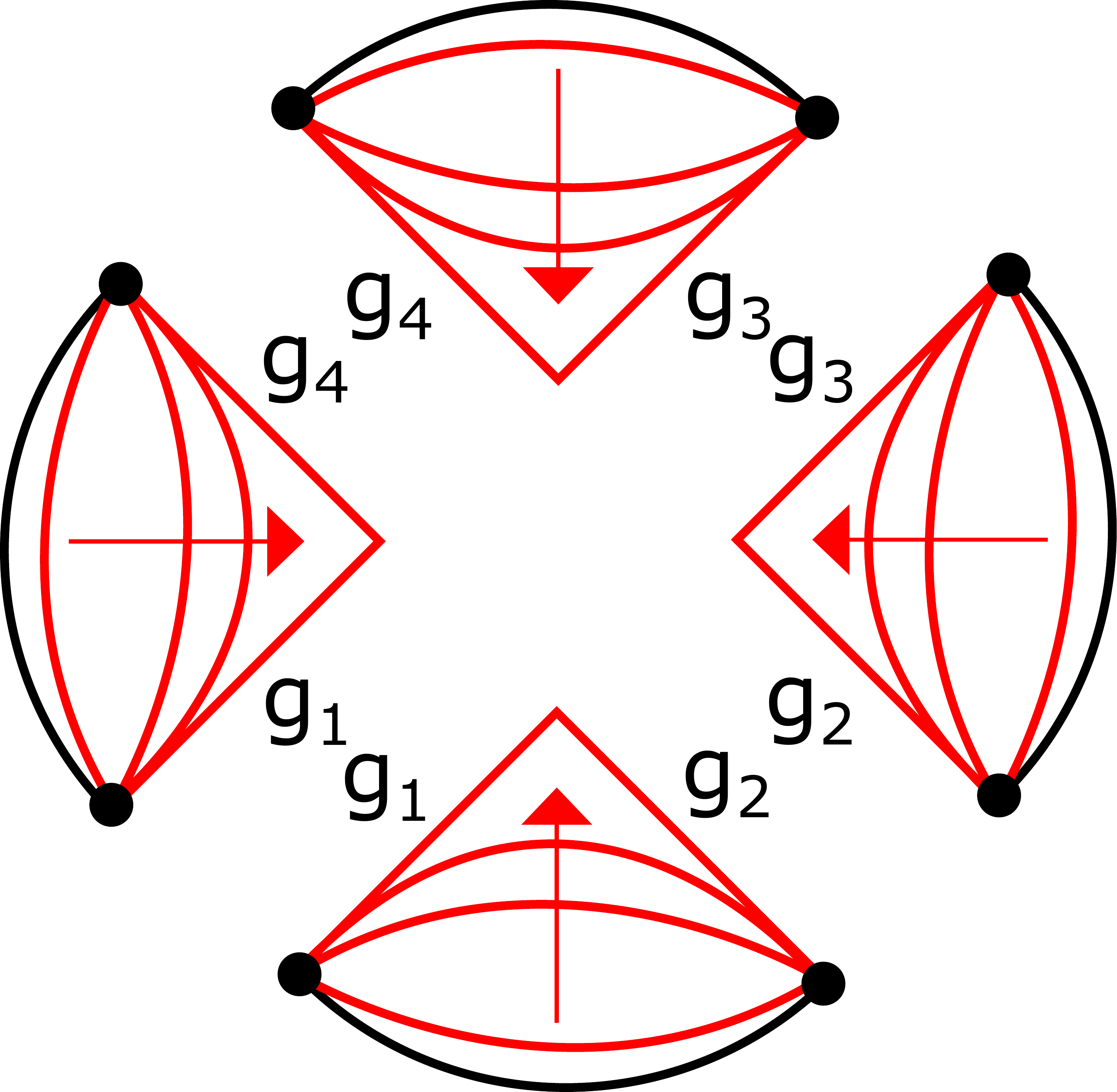}
\caption{Hilbert space slicing within each of the wedges. We are interested in the Hilbert space description on the final slice, consisting of two connected straight lines.}
\label{fig:wedgeslice}
\end{figure}
Zooming on this bottom wedge only (Figure \ref{fig:wedgeslice}), we can write the wavefunction on the top edges of this wedge by decomposing the representation matrices as
\begin{equation}
\label{grouplaw}
R_{m\bar{m}}(g_1 \cdot g_2) = \sum_n \, R_{m n}(g_1)R_{n\bar{m}}(g_2).
\end{equation}
This is repeated for each wedge. Combining the resulting representation matrices with the $R_{m n}(g)$ associated to the Wilson line itself, one needs to integrate products of three representation matrices on a single group element $g_i, i=1 \hdots 4$, as in \eqref{3R}. This procedure can be suggestively written graphically by explicitly drawing lines associated to the $R_{mn}$ within each wedge, and artificially separating the wedge contributions from the Wilson line contributions, as in the right of \eqref{6jpict}. These group integrals lead to four $3j$-symbols associated to the four outer vertices (the four big black dots in \eqref{6jpict}), and four $3j$-symbols associated to the internal vertex. The latter still contains a sum over $m$-labels, resulting in a $6j$-symbol:
\begin{align}
\sj{R_B}{R_1}{R_4}{R_A}{R_3}{R_2} = \sum_{m_i,m_A,m_B} \tj{R_1}{R_2}{R_A}{m_1}{m_2}{m_A} \tj{R_2}{R_3}{R_B}{m_2}{m_3}{m_B} \tj{R_3}{R_4}{R_A}{m_3}{m_4}{m_A} \tj{R_4}{R_1}{R_B}{m_1}{m_2}{m_B},\label{sixj}
\end{align}
for each crossing of Wilson lines.
\\~\\
At the beginning of this section \ref{s:compact}, we commented that Wilson lines can be freely deformed: moving Wilson lines through each other, leaves the amplitudes unchanged. Armed with the specific expressions, we check this explicitly in Appendix \ref{app:A3}.

\subsection{Holography in BF}
\label{sect:holobf}
In this section, we will make the link with the particle-on-a-group model more explicit, and provide another derivation of the correlators directly in this language. Starting with \eqref{BF}, path-integrating over the bulk $\chi$ forces $F=0$ and hence renders $A$ flat: $A=-dg g^{-1}$. This holds even when Wilson line operators are inserted into the path integral. The Hamiltonian of the theory lives on the boundary and is completely determined by the \emph{choice} of boundary conditions. Choosing $\chi\rvert_\text{bdy} = C A_\tau\rvert_\text{bdy}$, the BF theory reduces to a boundary theory with particle on group action \cite{Mertens:2018fds}:
\begin{equation}
S[g]=\frac{C}{2}\int d\tau \Tr (g^{-1}\partial_\tau g g^{-1}\partial_\tau g),\label{PG}
\end{equation}
with coupling $C$ determined by the choice of boundary conditions and the path integral over $g\in LG/G$, the right coset of the loop group of $G$.\footnote{An alternative way to obtain the particle on group action from BF theory is to add a term to the BF action that enforces a specific boundary condition:
\begin{equation}
S[\chi,A]=\int \Tr (\chi F)+\frac{1}{2}\int d\tau \Tr(\chi A_\tau)+\frac{1}{4C}\int d\tau \sqrt{h}\Tr(\chi^2).
\end{equation}
Path integrating over $\chi$ entirely results again in \eqref{PG}.}
As the theory fully reduces to the boundary particle-on-a-group, this provides an alternative and direct explanation of the Hilbert space structure as the Peter-Weyl decomposition of an arbitrary function in $L^2(G)$, the group element $g$ being just the dynamical variable of the boundary quantum mechanical model. 
\\~\\
The matrix elements of a Wilson line from boundary point $\tau_i$ to $\tau_f$ in representation $R$ are:
\begin{equation}
\label{poexp}
\mathcal{W}_{R,mn}(\tau_i,\tau_f)=\mathcal{P}\exp{-\int_{\tau_i}^{\tau_f} R(A) }_{mn}.
\end{equation}
The identification of correlators of boundary-anchored Wilson lines in BF as correlators of bilocals in particle on group can be proven by observing that upon integrating out the bulk $\chi$ of BF theory which renders A flat, the path-ordered exponential \eqref{poexp} reduces to a bilocal operator:
\begin{equation}
\boxed{\mathcal{W}_{R,mn}(\tau_i,\tau_f)=R_{mn}(g(\tau_f)g^{-1}(\tau_i))=\mathcal{O}_{R,m n}(\tau_i,\tau_f)}.\label{wilsonbilocal}
\end{equation}
These bilocals are precisely the operators whose particle-on-a-group correlators were calculated from WZW correlators in \cite{Mertens:2018fds}. 
\\~\\
We can use this dictionary to directly compute correlation functions of the particle-on-a-group theory, without resorting to the holographic bulk and Wilson line description. Writing two bilocal operators as:
\begin{align}
\mathcal{W}_{R_A,m_A\bar{m}_A} &= R_A(g_2)_{m_A \alpha} R_A(g_1^{-1})_{\alpha \bar{m}_A}, \nonumber \\
\mathcal{W}_{R_B,m_B\bar{m}_B} &= R_B(g_4)_{m_B \beta} R_B(g_3^{-1})_{\beta \bar{m}_B},
\end{align}
the time-ordered four-point correlator is computed in the particle-on-a-group model as the thermal trace:
\begin{equation}
\text{Tr}\left[e^{-\beta H} R_B(\hat{g}_4)_{m_B \beta } R_B(\hat{g}_3^{-1})_{\beta \bar{m}_B} R_A(\hat{g}_2)_{m_A \alpha} R_A(\hat{g}_1^{-1})_{\alpha \bar{m}_A} \right],
\end{equation}
with $H$ the Hamiltonian associated to \eqref{PG}. The computation itself involves inserting complete sets of states and is defered to Appendix \ref{app:pog}. Instead taking crossing bilocals as
\begin{align}
\mathcal{W}_{R_A,m_A\bar{m}_A} &= R_A(g_3)_{m_A \alpha} R_A(g_1^{-1})_{\alpha \bar{m}_A}, \nonumber \\
\mathcal{W}_{R_B,m_B\bar{m}_B} &= R_B(g_4)_{m_B \beta} R_B(g_2^{-1})_{\beta \bar{m}_B}.
\end{align}
the crossed correlator is computed as
\begin{equation}
\text{Tr}\left[e^{-\beta H} R_B(\hat{g}_4)_{m_B \beta } R_A(\hat{g}_3)_{m_A \alpha} R_B(\hat{g}_2^{-1})_{\beta \bar{m}_B} R_A(\hat{g}_1^{-1})_{\alpha \bar{m}_A} \right].
\end{equation}
Also this computation is defered to the Appendix \ref{app:pog}. The results of these computations match the previous expressions.

\subsection{Diagrammatic Expansion}
\label{s:pgdiag}
The calculations of section \ref{s:wilsonbf} can be summarized in a set of diagrammatic rules which can be used to write down any amplitude of an arbitrary number of boundary-anchored Wilson line insertions in BF.

\begin{itemize}
\item The starting point is just the disk on which the Wilson lines are drawn. 
\item Each region in the disk is assigned an irrep $R_i$, and contributes a weight $\dim R_i$. Assign a $m_i$ label to each boundary segment. Eventually these labels $R_i$ and $m_i$ are to be summed over.
\item Each boundary segment carries a Hamiltonian propagation factor \eqref{bfham} proportional to the length $L_i$ of the relevant segment $i$. Each intersection of an endpoint of a Wilson line with the boundary has associated with it 3 irreps and 3 labels and is weighted with the $3j$-symbol associated with these labels.
\begin{align}
\label{frules}
\begin{tikzpicture}[scale=0.8, baseline={([yshift=-0.1cm]current bounding box.center)}]
\draw[thick] (-0.2,0) arc (170:10:1.53);
\draw[fill,black] (-0.2,0.0375) circle (0.1);
\draw[fill,black] (2.8,0.0375) circle (0.1);
\draw (3.4, 0) node {\footnotesize $\tau_1$};
\draw (-0.7,0) node {\footnotesize $\tau_2$};
\draw (1.25, 1.6) node {\footnotesize $\textcolor{red}{m}$};
\draw (1.25, 0.5) node {\footnotesize $R$};
\draw (5, 0) node {$\raisebox{10mm}{$\qquad\qquad= \exp{- (\tau_2-\tau_1) \frac{\mathcal{C}_R}{C}},$}$};
\end{tikzpicture}
\end{align}
\begin{align}
\begin{tikzpicture}[scale=1, baseline={([yshift=-0.1cm]current bounding box.center)}]
\draw[thick] (-.2,.9) arc (25:-25:2.2);
\draw[fill,black] (0,0) circle (0.08);
 \draw[thick,blue](-1.5,0) -- (0,0);
\draw (.5,-1) node {\footnotesize $\textcolor{red}{m_2}$};
\draw (.5,1) node {\footnotesize$\textcolor{red}{m_1}$};
\draw (-0.3,.3) node {\footnotesize$\textcolor{red}{m}$};
\draw (-1,.3) node {\footnotesize$ \color{blue}R$};
\draw (-1,.8) node {\footnotesize$ R_1$};
\draw (-1,-.5) node {\footnotesize$ R_2$};
\draw (3,0.1) node {$\raisebox{-10mm}{$\ =\  \, \tj{R_1}{R_2}{R}{m_1}{m_2}{m}.$}$}; \end{tikzpicture}\ \label{pg3j}
\end{align}
\item Each crossing of two Wilson lines is associated with 6 irreps and is weighed with the appropriate $6j$-symbol \eqref{sixj}
\begin{align}
\label{crossing}
\ \begin{tikzpicture}[scale=1, baseline={([yshift=0cm]current bounding box.center)}]
\draw[thick,blue] (-0.85,0.85) -- (0.85,-0.85);
\draw[thick,blue] (-0.85,-0.85) -- (0.85,0.85);
\draw[dotted,thick] (-0.85,-0.85) -- (-1.25,-1.25);
\draw[dotted,thick] (0.85,0.85) -- (1.25,1.25);
\draw[dotted,thick] (-0.85,0.85) -- (-1.25,1.25);
\draw[dotted,thick] (0.85,-0.85) -- (1.25,-1.25);
\draw (1.5,0) node {\scriptsize $R_4$};
\draw (-1.5,0) node {\scriptsize $R_2$};
\draw (-.75,.33) node {\scriptsize \color{blue}$R_A$};
\draw (.78,.33) node {\scriptsize \color{blue}$R_B$};
\draw (0,1.5) node {\scriptsize  $R_1$};
\draw (0,-1.5) node {\scriptsize $R_3$};
\end{tikzpicture}~~\raisebox{-3pt}{$\ \ \  = \ \ \sj{R_B}{R_1}{R_4}{R_A}{R_3}{R_2}$}~~~
\end{align}

\end{itemize}
These rules agree with the diagrammatic rules established for the correlators of bilocals in particle-on-a-group \eqref{wilsonbilocal} in \cite{Mertens:2018fds}, which confirms the full equivalence of BF correlators and particle-on-a-group correlators through the dictionary \eqref{wilsonbilocal}.
\\~\\
Some examples are instructive to clarify the above rules. Using \eqref{wilsonbilocal}, the two-point function and four-point function are diagrammatically:
\begin{align}
&\begin{tikzpicture}[scale=1, baseline={([yshift=0cm]current bounding box.center)}]
\draw[thick] (0,0) circle (1.5);
\draw[thick,blue] (-1.5,0) -- (1.5,0);
\draw[fill,black] (-1.5,0) circle (0.1);
\draw[fill,black] (1.5,0) circle (0.1);
\draw (0,1.7) node {\small \color{red}$m_1$};
\draw (1.2,0.2) node {\small \color{red}$m$};
\draw (-1.2,0.2) node {\small \color{red}$\bar{m}$};
\draw (0,-1.7) node {\small \color{red}$m_2$};
\draw (-2,0) node {\small $\tau_2$};
\draw (2,0) node {\small $\tau_1$};
\draw (0,.25) node {\small \color{blue}$R$};
\draw (0,1) node {\small $R_1$};
\draw (0,-1) node {\small $R_2$};
\end{tikzpicture} = \Big\langle \mathcal{O}_{R,m\bar{m}}(\tau_1,\tau_2)\Big\rangle \nonumber \\ 
&= \sum_{R_1,R_2}\dim R_1\dim R_2 e^{-\tau_{21}\frac{\mathcal{C}_{R_1}}{C}}e^{-(\beta-\tau_{21})\frac{\mathcal{C}_{R_2}}{C}} \cross \sum_{m_1,m_2}\tj{R_1}{R}{R_2}{m_1}{m}{m_2}\tj{R_1}{R}{R_2}{m_1}{\bar{m}}{m_2},\label{pg2pt}
\end{align}

\begin{align}
&\begin{tikzpicture}[scale=1, baseline={([yshift=0cm]current bounding box.center)}]
\draw[thick] (0,0) circle (1.5);
\draw[thick,blue] (1.3,.7) arc (300:240:2.6);
\draw[thick,blue] (-1.3,-.7) arc (120:60:2.6);
\draw[fill,black] (-1.3,-.68) circle (0.1);
\draw[fill,black] (1.3,-.68) circle (0.1);
\draw[fill,black] (-1.3,0.68) circle (0.1);
\draw[fill,black] (1.3,0.68) circle (0.1);
\draw (0,0) node {\footnotesize  $R_3$};
\draw (0,.55) node {\footnotesize \color{blue}$R_A$};
\draw (0,-.55) node {\footnotesize \color{blue}$R_B$};
\draw (0,1) node {\footnotesize  $R_1$};
\draw (0,-1) node {\footnotesize  $R_2$};
\draw (0,1.88) node {\small \color{red}$m_1$};
\draw (0,-1.85) node {\small \color{red}$m_2$};
\draw (1.85,0) node {\small \color{red}$m_3$};
\draw (-1.85,0) node {\small \color{red}$\tilde{m}_3$};
\draw (1.09,0.35) node {\small \color{red}$m_A$};
\draw (-1,0.35) node {\small \color{red}$\bar{m}_A$};
\draw (1.09,-0.35) node {\small \color{red}$\bar{m}_B$};
\draw (-1,-0.35) node {\small \color{red}$m_B$};
\draw (-1.75,-.75) node {\footnotesize $\tau_3$};
\draw (-1.75,.75) node {\footnotesize $\tau_2$};
\draw (1.75,-.75) node {\footnotesize $\tau_4$};
\draw (1.75,.75) node {\footnotesize $\tau_1$};
\end{tikzpicture} = \Big\langle \mathcal{O}_{R_A,m_A\bar{m}_A}(\tau_1,\tau_2) \mathcal{O}_{R_B,m_B\bar{m}_B}(\tau_3,\tau_4) \Big\rangle \nonumber \\
&\sum_{R_1,R_2,R_3,R_4}\prod_i \dim R_i e^{-a \tau_{21} \mathcal{C}_{R_1}} e^{-a \tau_{32} \mathcal{C}_{R_3}} e^{-a \tau_{43} \mathcal{C}_{R_2}} e^{-a (\beta - \tau_{14}) \mathcal{C}_{R_3}} \nonumber \\
&\qquad\cross\sum_{m_1,m_2,m_3,\tilde{m}_3}\tj{R_1}{R_A}{R_3}{m_1}{m_A}{m_3}\tj{R_1}{R_A}{R_3}{m_1}{\bar{m}_A}{\tilde{m}_3}\tj{R_2}{R_B}{R_3}{m_2}{m_B}{\tilde{m}_3}\tj{R_2}{R_B}{R_3}{m_2}{\bar{m}_B}{m_3}.\label{pg4pt}
\end{align}

\noindent A further example is the expectation value of two crossing Wilson lines, or equivalently the product of two bilocal operators $\mathcal{O}_{R_A,m_A\bar{m}_A}(\tau_1,\tau_3)$ and $\mathcal{O}_{R_B,m_B\bar{m}_B}(\tau_2,\tau_4)$, with time-ordering $\tau_1<\tau_2<\tau_3<\tau_4$:

\begin{align}
&\begin{tikzpicture}[scale=1, baseline={([yshift=0cm]current bounding box.center)}]
\draw[thick]  (0,0) ellipse (1.6 and 1.6);
\draw[thick,blue] (1.1,-1.2) arc (35.7955:82:4);
\draw[thick,blue] (-1.1,-1.2) arc (144.2045:98:4);
\draw (0,-0.8) node {\small $R_3$};
\draw (0,1) node {\small $R_1$};
\draw (-1,-0.2) node {\small $R_2$};
\draw (1,-0.2) node {\small $R_4$};
\draw (-0.5,0.35) node {\small \color{blue}$R_A$};
\draw (0.5,0.35) node {\small \color{blue}$R_B$};
\draw[fill,black] (-1.56,0.4) circle (0.1);
\draw[fill,black] (1.56,0.4) circle (0.1);
\draw[fill,black] (-1.11,-1.17) circle (0.1);
\draw[fill,black] (1.11,-1.17) circle (0.1);
\draw (-1.9,0.4) node {\small $\tau_1$};
\draw (-1.45,-1.4) node {\small $\tau_2$};
\draw (1.45,-1.4) node {\small $\tau_3$};
\draw (1.9,0.4) node {\small $\tau_4$};
\draw (0,1.85) node {\small \color{red}$m_1$};
\draw (0,-2) node {\small \color{red}$m_3$};
\draw (1.97,-.75) node {\small \color{red}$m_4$};
\draw (1.11,.6) node {\small \color{red}$\bar{m}_B$};
\draw (-1.11,-0.7) node {\small \color{red}$m_B$};
\draw (-1.1,.6) node {\small \color{red}$m_A$};
\draw (1.11,-0.7) node {\small \color{red}$\bar{m}_A$};
\draw (-1.97,-.75) node {\small \color{red}$m_2$};
\end{tikzpicture} = \Big\langle \mathcal{O}_{R_A,m_A\bar{m}_A}(\tau_1,\tau_3) \mathcal{O}_{R_B,m_B\bar{m}_B}(\tau_2,\tau_4) \Big\rangle \nonumber \\ \nonumber 
&= \sum_{R_1,R_2,R_3,R_4}\prod_i \dim R_i e^{-a \tau_{21} \mathcal{C}_{R_2}} e^{-a \tau_{32} \mathcal{C}_{R_3}} e^{-a \tau_{43} \mathcal{C}_{R_4}} e^{-a (\beta - \tau_{14}) \mathcal{C}_{R_1}}\times \sj{R_B}{R_1}{R_4}{R_A}{R_3}{R_2}\\
&\qquad \cross \sum_{m_1,m_2,m_3,m_4}\tj{R_1}{R_A}{R_2}{m_1}{m_A}{m_2}\tj{R_1}{R_B}{R_4}{m_1}{\bar{m}_B}{m_4}\tj{R_2}{R_B}{R_3}{m_2}{m_B}{m_3}\tj{R_3}{R_A}{R_4}{m_3}{\bar{m}_A}{m_4}
\end{align}
Schematically writing this out in terms of local operators, using \eqref{wilsonbilocal}, we would be computing something like
\begin{equation}
\contraction{}{\mathcal{O}_1}{\mathcal{O}_2}{\mathcal{O}_3}
\bcontraction{\mathcal{O}_1}{\mathcal{O}_2}{\mathcal{O}_3}{\mathcal{O}_4}
\left\langle \mathcal{O}_1\mathcal{O}_2\mathcal{O}_3\mathcal{O}_4\right\rangle.
\end{equation}
Such a configuration is closely related, but not equal to an out-of-time ordered (OTO) correlator, which in terms of local operators would be 
\begin{equation}
\contraction{}{\mathcal{O}_1}{\mathcal{O}_3}{\mathcal{O}_2}
\bcontraction{\mathcal{O}_1}{\mathcal{O}_3}{\mathcal{O}_2}{\mathcal{O}_4}
\left\langle \mathcal{O}_1\mathcal{O}_3\mathcal{O}_2\mathcal{O}_4\right\rangle.
\end{equation}
Technically, the only difference is in the exponential damping factors. For OTO correlators, the thermal circle in the diagram is to be interpreted as the \emph{unfolded} time contour. Pragmatically, when going to OTO correlators, care must be taken in the exponential factors to ensure all of them are damped in Euclidean signature, in the end leading to a specific Lorentzian OTO correlator after Wick rotation \cite{Mertens:2017mtv,Lam:2018pvp}. The diagram is
\begin{align}
&\begin{tikzpicture}[scale=1, baseline={([yshift=0cm]current bounding box.center)}]
\draw[thick]  (0,0) ellipse (1.6 and 1.6);
\draw[thick,blue] (1.1,-1.2) arc (35.7955:82:4);
\draw[thick,blue] (-1.1,-1.2) arc (144.2045:98:4);
\draw (0,-0.8) node {\small $R_3$};
\draw (0,1) node {\small $R_1$};
\draw (-1,-0.2) node {\small $R_2$};
\draw (1,-0.2) node {\small $R_4$};
\draw (-0.5,0.35) node {\small \color{blue}$R_A$};
\draw (0.5,0.35) node {\small \color{blue}$R_B$};
\draw[fill,black] (-1.56,0.4) circle (0.1);
\draw[fill,black] (1.56,0.4) circle (0.1);
\draw[fill,black] (-1.11,-1.17) circle (0.1);
\draw[fill,black] (1.11,-1.17) circle (0.1);
\draw (-1.9,0.4) node {\small $\tau_1$};
\draw (-1.45,-1.4) node {\small $\tau_3$};
\draw (1.45,-1.4) node {\small $\tau_2$};
\draw (1.9,0.4) node {\small $\tau_4$};
\draw (0,1.85) node {\small \color{red}$m_1$};
\draw (0,-2) node {\small \color{red}$m_3$};
\draw (1.97,-.75) node {\small \color{red}$m_4$};
\draw (1.11,.6) node {\small \color{red}$\bar{m}_B$};
\draw (-1.11,-0.7) node {\small \color{red}$m_B$};
\draw (-1.1,.6) node {\small \color{red}$m_A$};
\draw (1.11,-0.7) node {\small \color{red}$\bar{m}_A$};
\draw (-1.97,-.75) node {\small \color{red}$m_2$};
\end{tikzpicture} = \Big\langle \mathcal{O}_{R_A,m_A\bar{m}_A}(\tau_1,\tau_2) \mathcal{O}_{R_B,m_B\bar{m}_B}(\tau_3,\tau_4) \Big\rangle_{\text{OTO}} \nonumber \\  
&= \sum_{R_1,R_2,R_3,R_4}\prod_i \dim R_i e^{-a \tau_{31} \mathcal{C}_{R_2}} e^{-a \tau_{32} \mathcal{C}_{R_3}} e^{-a \tau_{42} \mathcal{C}_{R_4}} e^{-a (\beta - \tau_{41}) \mathcal{C}_{R_1}}\times \sj{R_B}{R_1}{R_4}{R_A}{R_3}{R_2} \nonumber \\
&\qquad \cross \sum_{m_1,m_2,m_3,m_4}\tj{R_1}{R_A}{R_2}{m_1}{m_A}{m_2}\tj{R_1}{R_B}{R_4}{m_1}{\bar{m}_B}{m_4}\tj{R_2}{R_B}{R_3}{m_2}{m_B}{m_3}\tj{R_3}{R_A}{R_4}{m_3}{\bar{m}_A}{m_4},
\end{align}
with notably $\tau_{32}=L_3$ and not $\tau_{23}$ in the second exponential ensuring proper damping.
\\~\\
All of this demonstrates that the particle-on-a-group diagrams are to be interpreted as bulk BF Wilson line diagrams, a perspective that is not clear a priori when deriving these expressions with 2d CFT techniques \cite{Mertens:2018fds}.

\section{Interlude: Non-Compact Groups}
\label{sect:interlude}

The above description is readily generalizable to non-compact groups, where one just uses the $3j$- and $6j$-symbols of the representations of the group. A new feature is the appearance of continuous representations, which require the replacement $\sum_R\to \int d R$ and $\text{dim R}\to \rho(R)$, with $\rho(R)$ the Plancherel measure of the group. The difficulty, as always, is to find out which representations actually appear in the Hilbert space. 
\\~\\
Irreducible representations of a non-compact group are not all unitary. Only the unitary representations appear in the Peter-Weyl decomposition and hence in the Hilbert spaces of the models. It is particularly interesting at this point to turn the eye towards $\slr$, which is closely linked to the gravitational systems. The unitary irreducible representations of $\slr$ have all been classified. States are labeled by irreps and eigenvalues of one generators. Depending on which $\mathfrak{sl}(2,\mathbb{R})$ generator one diagonalizes, one distinguishes the elliptic basis ($J^2$), the hyperbolic basis ($J^0$) and the parabolic basis ($J^+ \equiv J^1 + i J^2$).\footnote{Our convention of algebra generators is given in equation \eqref{alggen}.} One tends to choose the elliptic basis, as $J^3$ is the generator of a compact subgroup such that its eigenvalue indices $m$ and $n$ are discrete. The unitary irreps along with the range $m$ of $\mj^3$ eigenvalues are:
\begin{itemize}
\item Principal Continuous Representation $\mathcal{C}_{s}$, with $j=-1/2 + is$, $m \in \mathbb{Z}$.
\item Complementary Representation $\mathcal{E}_j$, with $-1 < j < 0$, $m\in \mathbb{Z}$. 
\item Principal Discrete Representation (highest or lowest weight) $\mathcal{D}_\ell^\pm$, with $\ell \in \frac{\mathbb{N}}{2}$, $m = \pm \ell, \pm (\ell+1), \hdots$.
\item Trivial Representation $j=0$.
\end{itemize}
Any $\psi \in L^2(G)$ can be decomposed by the Plancherel formula as
\begin{equation}
\label{plslr}
\psi(g) = \int_{0}^{+\infty}ds s\tanh(\pi s) \, \sum_{m,n \in \mathbb{Z}} c^{s}_{mn} R_{-\frac{1}{2}+is,mn}(g) + \sum_{\ell=0}^{+\infty} \sum_{m,n=\pm \ell}^{\pm \infty} \left(\ell+\frac{1}{2}\right) c^{\ell}_{mn} R^{\pm}_{\ell,mn}(g),
\end{equation}
where the first term contains the principal continuous representations $\mathcal{C}_{s}$, and the second term contains both the lowest and highest weight discrete representation $\mathcal{D}_\ell^\pm$. The complementary continuous representation and trivial representation, despite being unitary, are not in $L^2(G)$, and hence not in the particle-on-a-group or BF Hilbert space.
\\~\\
To obtain gravity, the crux of the matter will be to impose suitable constraints on this system (the parabolic basis will be most convenient for this), and find out which representations survive.

\section{Schwarzian Theory}
\label{s:gravity}
\begin{figure}[h]
\centering
\includegraphics[width=0.55\textwidth]{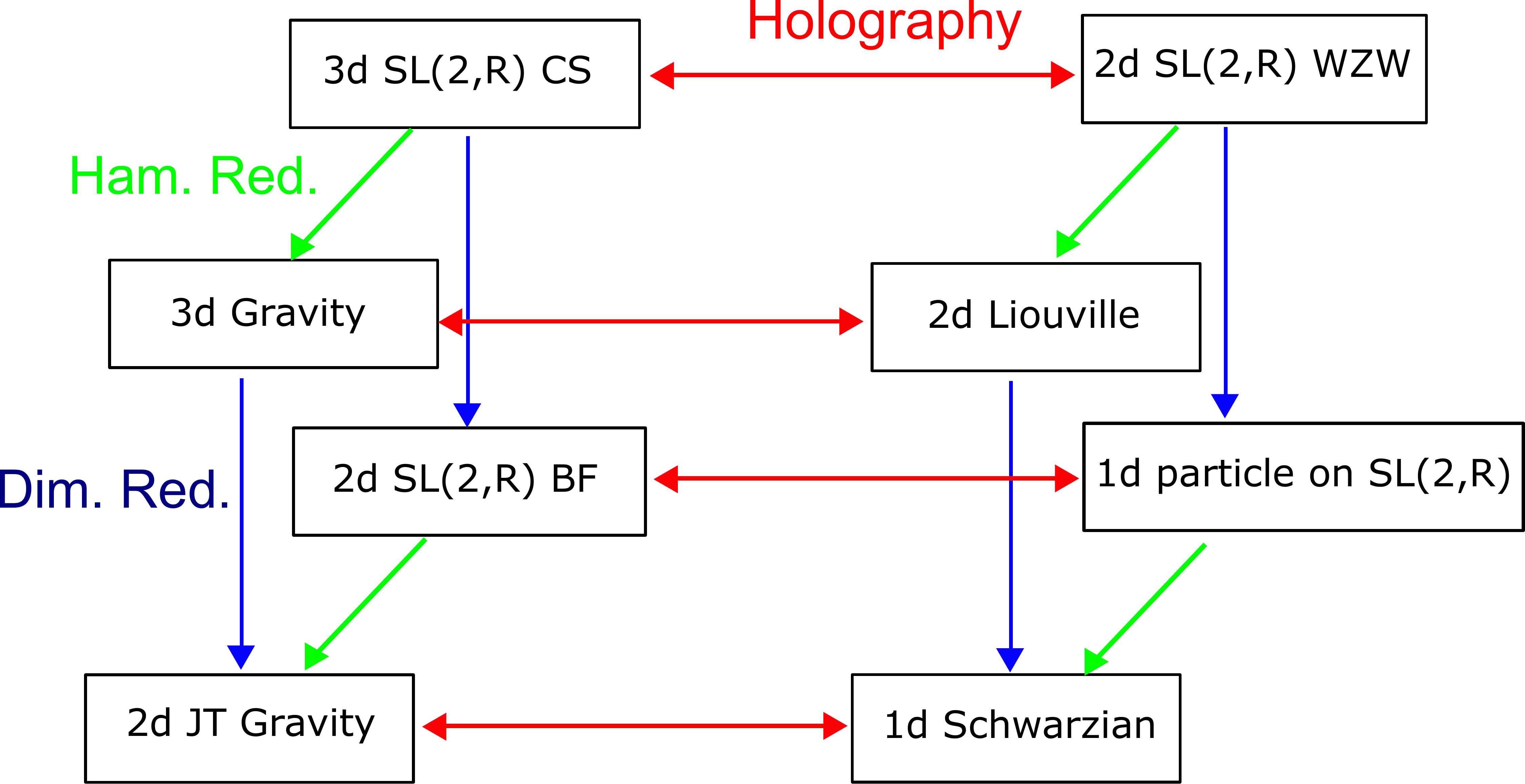}
\caption{Relation through holography, the Schwarzian limit (dimensional reduction), and Hamiltonian reduction of eight theories.}
\label{schemedimholin2}
\end{figure}

In this section we discuss the gravitational analogue of the story of section \ref{s:compact}. We start by recapitulating some of the relevant expressions from 3d / 2d holography before specializing to 2d / 1d holography. Figure \ref{schemedimholin2} summarizes the relation between the theories relevant for this section.
The discussion of this section relies heavily on how precisely the gravitational sector is encoded within gauge theory in these models. By identifying the Schwarzian as a constrained particle on $\slr$ \cite{Jackiw:1992bw,Gonzalez:2018enk},\footnote{We will further on make a distinction between $\slr$ and $\slrp$ which will be important. The gravitational model is related to the $\slrp$ model. Most arguments in fact only deduce the $\mathfrak{sl}(2,\mathbb{R})$ algebraic structure, without any need to specify in detail the exponentiated form. We will find that we need the semigroup $\slrp$, consisting of all $\slr$ matrices with only positive entries.}  we are able to generalize the conclusions of section \ref{s:compact} and identify boundary-anchored Wilson lines in JT gravity - which is just constrained $\slrp$ BF theory \cite{Jackiw:1992bw,toap} - as bilocal operators in the Schwarzian. The $6j$-symbols, associated to crossing Wilson lines in the bulk, appear in Schwarzian OTO correlators and account for gravitational shockwaves.\footnote{In particular, in \cite{Lam:2018pvp} it was proven that precisely these $\slrp$ $6j$-symbols contain the entire eikonal gravitational shockwave expression in the semiclassical $C \to \infty$ regime.} 
Since $\slr$ BF and by extension $\slr$ YM is less understood than their rational cousins, this section serves a double goal: we provide a Wilson line perspective on Schwarzian correlators by assuming the calculations of boundary-anchored Wilson lines in constrained $\slrp$ BF theory are structurally equivalent to those of their rational cousins, and \emph{prove} the resulting diagrammatic rules by explicitly matching with Schwarzian correlators.

\subsection{Liouville as Constrained WZW}
\label{sect:Liou}
It is well-known that 3d gravity with negative cosmological constant can be written as two copies of $\slr$ CS, or equivalently as a non-chiral $\slr$ WZW model. Imposing the metric to be asymptotically AdS$_3$ constrains the $\slr$ CS gauge fields. The result is that the boundary $\slr$ WZW model is also constrained / gauged, reducing the symmetry from $\slr$ Kac-Moody to Virasoro and the action to the Liouville action. 
\\~\\
Consider a 2d worldsheet $(z,\bar{z})$ on which one defines five fields $(\phi,\gamma_L,\gamma_R,\beta,\bar{\beta})$ with Lagrangian:
\begin{equation}
\label{waki}
\mathcal{L} = \partial \phi \bar{\partial} \phi + \beta \bar{\partial} \gamma_L + \bar{\beta} \partial \gamma_R + \beta \bar{\beta} e^{2\phi}.
\end{equation}
This is the Wakimoto representation of the $\slr$ WZW model \cite{Giveon:1998ns}. Integrating out $\beta$ and $\bar{\beta}$ through their equations of motion $\bar{\partial}\gamma_L + \bar{\beta}e^{2\phi} =0$, $\partial \gamma_R + \beta e^{2\phi} =0$, one retrieves the standard WZW $\slr$ Lagrangian for a metric $ds^2 = d\phi^2 + e^{-2\phi}d\gamma_L d\gamma_R$ \cite{Teschner:1997fv,Teschner:1999ug,Maldacena:2000hw}:
\begin{equation}
\mathcal{L} = \partial \phi \bar{\partial} \phi + \bar{\partial} \gamma_L\partial \gamma_R e^{-2\phi},
\end{equation}
Liouville is obtained from $\slr$ WZW by means of a Drinfeld-Sokolov Hamiltonian reduction \cite{Bershadsky:1989mf,Dijkgraaf:1991ba,Carlip:2005tz,Compere:2014cna}. Specifically, to obtain Liouville from $\slr$ WZW one constrains one component of the holomorphic Kac-Moody currents $\mj^a$ and one component of the antiholomorphic currents $\bar{\mj}^a$ as \cite{Dijkgraaf:1991ba}:
\begin{equation}
\mj^-=i\sqrt{\mu},\quad \bar{\mj}^+=-i\sqrt{\mu},\label{bc}
\end{equation}
for some constant $\mu$ that will turn out to be the Liouville cosmological constant. The constraints \eqref{bc} emerge in the bulk as the constraints that force the metric to be asymptotically AdS$_3$ \cite{Coussaert:1995zp}. Imposing these constraints in a proper path integral manner is done by gauging the parabolic subgroup of $\slr$ in the gauged WZW model.\footnote{Alternatively, one can also start with a chiral $\slr$ WZW; this is explained in detail in \cite{Alekseev:1988ce,Alekseev:1990mp} where the procedure results first in the Virasoro coadjoint orbit action, which in turn is directly related to the Liouville path integral between branes, producing a single Virasoro character.}
\\~\\
At the level of the action, the equivalence is readily observed using the Wakimoto formulation of $\slr$ WZW \eqref{waki}. The system \eqref{waki} has conserved Kac-Moody currents:
\begin{alignat}{4}
&\mj^- &&= \beta , \qquad &&\bar{\mj}^- &&= \bar{\beta}\gamma_R^2 - 2\gamma_R \bar{\partial} \phi - \bar{\partial} \gamma_R, \nonumber \\
&\mj^0 &&= \beta\gamma_L - \partial \phi , \qquad &&\bar{\mj}^0 &&=  \bar{\beta}\gamma_R - \bar{\partial} \phi, \\
&\mj^+ &&= \beta\gamma_L^2 - 2\gamma_L \partial \phi - \partial \gamma_L, \qquad &&\bar{\mj}^+ &&= \bar{\beta} \nonumber.
\end{alignat}
Imposing \eqref{bc} at the level of the action and substituting this in \eqref{waki}, one obtains the Liouville Lagrangian:
\begin{equation}
\mathcal{L} = \partial \phi \bar{\partial} \phi + \mu e^{2\phi}.
\end{equation}
Correlation functions can also be linked to each other, though this requires a bit more effort \cite{Hikida:2007tq}. The procedure identifies the parameters of Liouville and WZW as:
\begin{equation}
b^2=\frac{1}{k-2}.\label{bk}
\end{equation}
In what follows we will be interested solely in a particular dimensional reduction of this story, in which we also take $b\to 0$ or equivalently $k\to \infty$.\footnote{As a check on \eqref{bk}, note that at large $k$ the central charge $c=6k$ of the constrained $\slr$ WZW model matches the Liouville central charge, and that the conformal weight $h=P^2+\frac{1}{4b^2}$ of the Liouville states at small $b$ matches the weights $h=\frac{j(j+1)}{k-2}+\frac{k}{4}$ of the $\slr$ principal continuous irrep primaries $j=-\frac{1}{2}+is$, upon identification of the Liouville momentum as $P=b s$.}

\subsection{Schwarzian as Constrained Particle-on-a-group}
\label{ss:liouville}
The dimensional reduction of Liouville while taking simultaneously $b\to 0$ results in the Schwarzian theory \cite{Mertens:2017mtv,Mertens:2018fds}, dual to JT gravity. Similarly, the dimensional reduction while taking simultaneously $k\to\infty$ of $\slr$ WZW results in a particle on $\slr$, dual to $\slr$ BF. We are interested here in the embedding of the Schwarzian in $\slr$ BF. At the level of the action, this was already examined in \cite{Mertens:2017mtv}, where it was demonstrated that the Schwarzian
\begin{equation}
L = \left\{f,\tau\right\},\label{sch}
\end{equation}
can be rewritten through integrating in and out Lagrange multipliers both as the action of a particle on the $\slr$ group manifold:\footnote{Here the phase space is larger compared to the Schwarzian or 1d Liouville.}
\begin{equation}
L = \frac{1}{2} \dot{\phi}^2 + \dot{\gamma_L}\dot{\gamma_R}e^{-\phi},
\end{equation}
and as the 1d Liouville model:
\begin{equation}
L = \frac{1}{2} \dot{\phi}^2 - \mu e^{\phi}.
\end{equation}
Here we go deeper, and demonstrate that the perspective of the Schwarzian as constrained particle on $\slr$ is powerful enough to allow a full solution of the quantum model. In the bulk this means we solve quantum JT gravity by writing it as a BF theory.
\\~\\
The first goal is to identify the embedding of the Schwarzian states in the $\slr$ BF spectrum. Let us first return thereto to the Liouville / $\slr$ WZW story. The Hilbert space of $\slr$ WZW consists of states $\ket{j,m\bar{m},n\bar{n}}$, with $j$ labeling either a principal continuous or a discrete irrep of $\slr$. The labels $m$ and $\bar{m}$ specify a certain state in the zero-grade Hilbert space of $\slr$, generated by $\mj^a_0$. Applying the raising operators $\mj^a_{-n}$ constructs the entire module, associated with a specific irrep $j$. 

After Hamiltonian reduction, only the principal continuous spectrum survives as normalizable solutions to the constrained $\mathfrak{sl}(2,\mathbb{R})$ Casimir differential equation.\footnote{This is discussed in more detail in section \ref{s:openchansch} and Appendix \ref{app:eigenvalue}.} Consider now the module $\ket{k,m\bar{m},n\bar{n}}$ for fixed $k$ referring to a principal continuous series irrep of $\slr$ $j=-\frac{1}{2}+ik$. As shown in \cite{Bershadsky:1989mf}, the effect of the Drinfeld-Sokolov Hamiltonian reduction is to map such a module to a single Virasoro Verma module associated with the same state $k$: these are states $\ket{k,n\bar{n}}$ where now $k$ is a Virasoro primary and $n$ and $\bar{n}$ label Virasoro descendants. 

Further taking the double-scaling limit to descend to the 1d theory, requires $n=\bar{n}=0$ as descendants are suppressed. Explicitly, to obtain the spectrum of Schwarzian states, we impose the Hamiltonian reduction constraints \eqref{bc} on the $\slr$ quantum mechanics states $\ket{j,m\bar{m}}$. As explained in Appendix \ref{app:rep} and \cite{Dijkgraaf:1991ba} this forces us to consider states of mixed parabolic type, where the labels $m,\bar{m}$ are continuous eigenvalues of $\mj^-,\bar{\mj}^+$ respectively which will henceforth be labeled by Greek letters $\nu, \lambda \hdots$. If we set $\mj^- = i\nu$ and $\bar{\mj}^+ = -i\lambda$, the constraints show that the Schwarzian states $\ket{k}$ are diagonal $\slr$ states with $\nu = \lambda = \sqrt{\mu}$:
\begin{equation}
\ket{k} = \ket{k,\sqrt{\mu}\sqrt{\mu}}.\label{states}
\end{equation}

\subsection{Correlation Functions in Constrained BF}
\label{corrjt}
We proceed by again calculating generic correlators of boundary-anchored Wilson lines in JT gravity or constrained $\slrp$ BF. Later on in section \ref{sect:holosch} we identify boundary-anchored Wilson lines through holography as bilocal operators in the Schwarzian theory. The match of all amplitudes discussed here with the Schwarzian amplitudes completes the proof of the dictionary. 
A subtlety that appears here is that, whereas the algebra is indeed $\mathfrak{sl}(2,\mathbb{R})$, the relevant group structure is in fact the subsemigroup $\slrp$, where all entries of the matrix are positive. This difference does not matter for most applications, but here it is important. Its main effect is the overall normalization of the matrix elements, which leads to the correct Plancherel measure and density of states. The relevance of the subsemigroup $\slrp$ to the gravitational problem first appeared in the work of Ponsot and Teschner in \cite{Ponsot:1999uf,Ponsot:2000mt} in the context of the $q$-deformed $6j$-symbols of SL${}_q^+(2,\mathbb{R})$ as braiding and fusion matrices of Virasoro CFT. This apparent subtlety of restricting the $\slr$-matrices to only positive entries is important when looking at more ``fine-grained" properties of gravity, as appears elsewhere \cite{toap}.

\subsubsection{Open Channel Description}
\label{s:openchansch}
We present again an open-channel basis description. Representation matrices are matrix elements $R_{k,\nu\lambda}(g)=\bra{k,\nu}g\ket{k,\lambda}$ and can be found in the available literature \cite{V, VK}. In Appendix \ref{app:rep} we elaborate on the representation theory required to describe Schwarzian quantum mechanics, and on the parameterization relevant for this work.\footnote{There appears to be about as many ways to choose and parameterize the basis and the group element of $\slr$ as there are ``ways to die in the west''.} This Appendix also provides with some of the technical calculations underlying the main story presented here.
\\~\\
From \eqref{states}, we find the relevant representation matrix elements for constrained $\slrp$ as $R_{k,\sqrt{\mu}\sqrt{\mu}}(g)$. If we coordinatize the $\slr$ group manifold as\footnote{This is only the Poincar\'e patch of the full $\slr$ manifold. The $\slrp$ manifold is however fully contained within the Poincar\'e patch, making this decription complete. This should be contrasted with global effects in the Hamiltonian reduction that appear when considering $\slr$ instead, see e.g. \cite{Tsutsui:1994pp}.}
\begin{equation}
\label{slrmetr}
ds^2 = d\phi^2 + e^{-2\phi}d\gamma_L d\gamma_R,
\end{equation}
the $\slrp$ submanifold is obtained by restricting to the region $\gamma_L, \gamma_R >0$. The matrix elements of the continuous series $\slrp$ irreps, evaluated between mixed parabolic eigenstates, are explicitly given by \cite{Dijkgraaf:1991ba,VK}:
\begin{equation}
R_{k,\nu\lambda}(g) \,\, \equiv \,\, \left\langle \nu_L\right|g(\phi,\gamma_L,\gamma_R) \left|\lambda_R\right\rangle = \frac{1}{\pi}\left(\frac{\lambda}{\nu}\right)^{ik} e^{\phi} K_{2i k}(\sqrt{\nu\lambda}e^{\phi})e^{-\nu\gamma_L -\lambda \gamma_R}.\label{repcontgeneral}
\end{equation}
Specifying to the constrained case of interest, we set $\lambda = \nu = \sqrt{\mu}$, and, after stripping off the dependence on $\gamma_L$ and $\gamma_R$, we obtain the Schwarzian constrained matrix element:
\begin{equation}
R_{k}(\phi) \equiv e^{\phi} K_{2ik}(\sqrt{\mu}e^{\phi}).\label{repcont}
\end{equation}
As the metric \eqref{slrmetr} comes with the integration measure $d\phi d\gamma_L d\gamma_R e^{-2\phi}$, we obtain the orthogonality relation
\begin{equation}
\label{MacDo}
\int_{-\infty}^{+\infty} d\phi \, e^{-2\phi} R_{k}(\phi)R_{k'}(\phi) =  \frac{1}{\pi^2} \int_{-\infty}^{+\infty} d\phi \, K_{2i k}(\sqrt{\mu}e^\phi)K_{2i k'}(\sqrt{\mu}e^\phi) = \frac{1}{8k\sinh(2\pi k)} \delta(k-k'),
\end{equation}
This relation is the analogue of \eqref{Rotho}, and we identify the analogue of $\dim k$ for this noncompact example as:  
\begin{equation}
\boxed{\dim k = k\sinh 2\pi k}.\label{dimk}
\end{equation}
This is the Plancherel measure on $\slrp$, as explained in Appendix \ref{app:rep}. The open channel principal continuous series $\slrp$ wavefunctions of mixed parabolic type, are the analogues of \eqref{wavefctrational}:
\begin{equation}
\label{normwf}
\bra{\phi}\ket{k}=\sqrt{k\sinh 2\pi k}\, R_{k}(\phi),
\end{equation}
and are orthogonal with respect to the measure $d g=d\phi e^{-2\phi}$. Note that the density of states \eqref{dimk} differs from the $\slr$ Plancherel measure $\rho(k) = k \tanh(\pi k)$, arising from the fact that the structure is the subsemigroup $\slrp$ and not the $\slr$ group manifold itself (see Appendix \ref{app:rep}). The observation that this is indeed the analogue of $\dim k$ solidifies by looking back to 2d Liouville, where this role is played by the $S_{0k}$ elements of the modular S-matrix. For irreps $k$ of compact Lie groups, $S_{0k}$ reduces to the dimension $\dim k$ of the irrep in the minisuperspace limit, which matches with the $k\sinh 2\pi k$ measure in the non-compact case at hand \cite{Mertens:2017mtv}.
\\~\\
The normalized wavefunctions \eqref{normwf} can be identified with the Liouville wavefunctions \cite{Braaten:1983pz,Braaten:1983np,Thorn:2002am}:
\begin{equation}
\label{wvfliou}
\psi_k(\phi) = \sqrt{k\sinh 2\pi k}\, K_{2i k}(\sqrt{\mu}e^\phi),
\end{equation}
now with the flat integration measure $d\phi$. In fact, imposing the constraints directly on the $\mathfrak{sl}(2,\mathbb{R})$ Casimir differential equation, one lands on the 1d Liouville eigenvalue problem, i.e. the Schr\"odinger equation for a particle in an exponential potential $V(\phi) =\mu e^{2\phi}$. We give some more formulas accompanying this discussion in Appendix \ref{app:eigenvalue}.
\\~\\
Knowledge of the wavefunctions and matrix elements is sufficient to calculate the disk amplitude of JT gravity. Consider the disk amplitude from Figure \ref{fig:diskrect}, where now the boundary states are characterized by some constrained $\slrp$ group elements, $\phi_i$ and $\phi_f$:
\begin{equation}
Z_\text{disk}(\phi_i,\phi_f)=\bra{\phi_f}e^{-TH}\ket{\phi_i}.
\end{equation}
The Hamiltonian is boundary-localized and is the Casimir in the principal continuous representation of $\slr$:
\begin{equation}
H(k)=\frac{\mathcal{C}_k}{C}=\frac{k^2}{C},
\end{equation}
where the constant shift $1/4$ of the Casimir was dropped.\footnote{It drops out of expectation values when normalized by the partition function $Z_{\text{disk}}$.} Inserting a complete set of states $\ket{k}$ of the constrained theory, we obtain:
\begin{equation}
Z_\text{disk}(g,h)=\int d k \bra{\phi_f}\ket{k}\bra{k}\ket{\phi_i}e^{-TH(k)}=\int d \mu (k) R_{k}(\phi_f) R_{k}(\phi_i) e^{-TH(k)},\label{disksch}
\end{equation}
where we introduced $d\mu (k) = k\sinh 2\pi k \, d k$. For compact groups, we noted that the physical boundary of BF was characterized by $g=h=\mathbf{1}$. For JT gravity, we are using mixed parabolic matrix elements. We explain in Appendix \ref{app:rep} in equation \eqref{locus} that the analogous statement, in the Gauss decomposition of the $\slr$ group element \eqref{Gauss}, is that $\phi \to \infty$, $\gamma_L = e^{\phi} = - \gamma_R$ to find the identity element. Using the asymptotics of the Bessel function, we indeed obtain up to some $k$-independent irrelevant prefactors, $R_{k}(\phi)\to 1$.
\\
As for the compact case in section \ref{sect:ocd}, there is an analogous description of the disk amplitude in the point defect channel of Figure \ref{WDI}. The boundary defect is characterized by $\phi \to \infty$, which is precisely the ZZ-brane boundary conditions, upon again identifying $\phi$ as the 1d Liouville field. 
\\~\\
From \eqref{disksch}, the amplitude of the empty disk is:
\begin{equation}
Z_{\text{disk}} = \int d \mu (k) e^{-\frac{\beta k^2}{C}},
\end{equation}
indeed matching the known Schwarzian partition function \cite{Stanford:2017thb,Mertens:2018fds,randommatrix}.

\subsubsection{Wilson Lines}
As for compact groups, a Wilson line segment in a rep $j$ of $\slrp$ with starting point and endpoint labeled by $m$ respectively $n$ is a matrix element $R_{j,m n}$. Here we will be concerned with boundary anchored Wilson lines in a discrete series rep $j=\ell$ of $\slrp$ since, as we will discuss below, these are the only sensible ones in this theory.
\\~\\
Let us start this discussion with a derivation of the $\slrp$ mixed parabolic $3j$-symbols when one of the reps is in the principal discrete series and the two other ones are in the principal continuous series. These appear when we are calculating expectation values of boundary-anchored Wilson lines in JT gravity, as for example in Figure \ref{WLI} earlier. They are obtained using the analogue of equation \eqref{3R}:
\begin{equation}
\int  d g R_{k_1,\nu_1\nu_1}(g)R_{\ell,m m}(g)R_{k_2,\nu_2\nu_2}(g) = \tj{k_1}{\ell}{k_2}{\nu_1}{m}{\nu_2}^2,\label{3Rslr}
\end{equation}
where $d g=d\phi d\gamma_L d\gamma_R e^{-2\phi}$. For the Schwarzian / JT applications, we consider the case $\nu_i=\sqrt{\mu}$. We are then lead to restrict the Wilson line endpoints on the boundary to $J^+=0$ ($J^+$ is conserved). The $\slrp$ discrete series matrix element of mixed parabolic type is a modified Bessel function of the first kind $\sim I_{2\ell-1}$ \cite{VK}. These are eigenfunctions of the same differential equation as the Macdonald functions \eqref{repcontgeneral} and thus also eigenfunctions of the $\mathfrak{sl}(2,\mathbb{R})$ Casimir. In particular the $J^+=\bar{J}^- = 0$ matrix elements are independent of $\gamma_L$ and $\gamma_R$ and read, up to an irrelevant prefactor:
\begin{equation}
R_{\ell,00}(\phi)=\lim_{\epsilon\to 0}R_{\ell,\epsilon\epsilon}(\phi) = \lim_{\epsilon\to 0}e^\phi I_{2\ell-1}(\epsilon e^{\phi})=e^{2\ell\phi}.\label{lim}
\end{equation}
Hence the  $3j$-symbols that appear in the JT calculation for a Wilson line reaching the boundary, are obtained from
\begin{equation}
\int_{-\infty}^{+\infty} d \phi \, e^{-2\phi} \, R_{k_1}(\phi) R_{\ell,00}(\phi)R_{k_2}(\phi)=\tj{k_1}{\ell}{k_2}{\sqrt{\mu}}{0}{\sqrt{\mu}}^2.\label{3Rsch}
\end{equation}
The integral over $\phi$ can be evaluated using formula (6.576) of \cite{gradstein1963tables}:\footnote{
More general mixed parabolic matrix elements of $\slrp$ can be calculated using the integral \cite{Gervois:1985fe,Gervois:1985ff}:
\begin{align}
\int_{0}^{+\infty} d x \, K_{2ik_1}(\nu_1 x) I_{2\ell-1}(m x) K_{2ik_2}(\nu_2 x)&=\nonumber\\ \frac{m^{2\ell}{\nu_1}^{2ik_1}}{{\nu_2}^{2ik_1+2\ell}}\frac{\Gamma(\ell\pm ik_1\pm ik_2)}{\Gamma(2\ell)^2} &{}_2F_1\left(\ell+ik_1+ik_2,\ell+ik_1-ik_2,2\ell;-\frac{m}{\nu_1}\right)^2.
\end{align}
The resulting $3j$-symbols can be compared to the results of \cite{Basu:1981ju}.
}
\begin{equation}
\int_{-\infty}^{+\infty} d\phi \, e^{2\ell\phi}K_{2ik_1}(\sqrt{\mu}e^\phi)K_{2ik_2}(\sqrt{\mu}e^\phi)=\frac{\Gamma(\ell\pm ik_1 \pm ik_2)}{\Gamma(2\ell)}.\label{gradstein}
\end{equation}
Therefore the $3j$-symbols relevant for JT gravity and the Schwarzian are:
\begin{equation}
\boxed{\tj{k_1}{\ell}{k_2}{\sqrt{\mu}}{0}{\sqrt{\mu}}=\left(\frac{\Gamma(\ell\pm ik_1 \pm ik_2)}{\Gamma(2\ell)}\right)^\frac{1}{2}},\label{cgsch}
\end{equation}
which can immediately be identified with the Schwarzian diagrammatic rule \cite{Mertens:2017mtv} and is to be compared with \eqref{pg3j}. The same $3j$-symbols for $\slr$ in the parabolic basis were obtained in the Mathematics literature in \cite{Basu:1981ju}.
\\
Notice that this construction fails for continuous irrep Wilson lines: the limit \eqref{lim} is ill-defined for $j=-\frac{1}{2}+ik$, because the Macdonald function oscillates erratically near the origin. As such, the integral \eqref{3Rsch} only converges for $j=\ell$, and we are restricted to considering discrete rep Wilson lines as operators in JT gravity. Note that in the $j=\ell$ case, the Wilson line operator reduces to the familiar Liouville operators. This separation of states (continuous irreps) and operators (discrete reps) happens naturally for the Schwarzian as for its Liouville ancestor \cite{Seiberg:1990eb} by the specific boundary conditions singled out in the Hamiltonian reduction of $\mathfrak{sl}(2,\mathbb{R})$. It is instructive to compare at this point to the minisuperspace computation of the Liouville CFT three-point function. This results in precisely the same integral \eqref{gradstein} to be performed. Indeed, the bulk minisuperspace regime of Liouville yields the wavefunctions $$\bra{\phi}\ket{k}=\frac{K_{2ik}(\sqrt{\mu}e^\phi)}{\Gamma(2ik)}.$$ 
Acknowledging that $\frac{1}{\abs{\Gamma(2ik)}^2}=k\sinh 2\pi k =\dim k$, we see that these only differ by \eqref{wvfliou} by a phase factor, which is irrelevant. Matrix elements of constrained $\slrp$ can directly be identified with Liouville wavefunctions. We obtain the precise analogue of the relation \eqref{3ptminisuper}:
\begin{equation}
\bra{k_2,\sqrt{\mu},\sqrt{\mu}}\mo_{\ell,00}\ket{k_1,\sqrt{\mu},\sqrt{\mu}}=\sqrt{\dim k_1}\sqrt{\dim k_2}\tj{k_1}{\ell}{k_2}{\sqrt{\mu}}{0}{\sqrt{\mu}}^2.
\end{equation}
We have now acquired the necessary tools to calculate from the ground up generic configurations of Wilson lines in JT gravity i.e. constrained $\slrp$ BF theory. A single Wilson line in discrete series irrep $\ell$ from boundary to boundary is just $R_{\ell,00}(g)$. Following formula \eqref{3Rsch} and the expression for the disk amplitude with a generic boundary \eqref{disksch}, each endpoint of a Wilson lines on the boundary will contribute the $\slrp$ $3j$-symbol \eqref{cgsch}. 
\\~\\
What happens on a bulk intersection of two Wilson lines? We would need an analogue of equation \eqref{grouplaw} to decompose the Hilbert space into two segments. This requires a more thorough discussion of $\slrp$ representation theory, and we defer the treatment to Appendix \ref{app:rep}. Barring the details, the computation is essentially the same as that of the compact case, and leads to the appearance of a $6j$-symbol at the intersection of crossing Wilson lines, where the $6j$-symbol of $\slr$ is obtained in the usual manner as an integral over all labels $m_i$ of products of four generic $\slrp$ $3j$-symbols. The main novelty is that the latter are written down in the hyperbolic basis of $\slrp$. The $6j$-symbol of $\slr$ with 2 discrete and 4 continuous representation labels can be deduced from the mathematical literature \cite{groenevelt}. It also agrees with formula (5.13) in \cite{Mertens:2017mtv}, determined from the 2d Liouville braiding $R$-matrix:
\begin{align}
\sj{\ell_1}{k_2}{k_1}{\ell_2}{k_4}{k_3} =& \left(\Gamma(\ell_1\pm i k_2 \pm ik_1)\Gamma(\ell_2 \pm ik_2\pm ik_3)\Gamma(\ell_1\pm i k_4 \pm i k_3)\Gamma(\ell_2\pm i k_4 \pm i k_1)\right)^\frac{1}{2}\nonumber \\
&\qquad\cross\mathbb{W}(k_1, k_3 ; \ell_1 + i k_4,\ell_1 - i k_4, \ell_2 - i k_2,\ell_2 + i k_2),
\end{align}
where $\mathbb{W}(\alpha,\beta;a,b,c,d)$ is the Wilson function \cite{groenevelt,groenevelt2}.\footnote{Explicitly this is:
\begin{equation}
\mathbb{W}(\alpha,\beta;a,b,c,d) \equiv \frac{\Gamma(d-a)~_4F_3\Big[\mbox{\small$\begin{array}{cccc}\! a+i\beta \, & \, a-i\beta \, &\, \tilde{a}+i \alpha &\, \tilde{a}-i\alpha \!\nspc \\[-1mm] \!\!\!\! a+b \!\! &\!\!\!\!\! a+c \!\! &\!\! 1+a-d \!\nspc \end{array}$} \,\,; 1 \Big]}{\Gamma(a+b)\Gamma(a+c)\Gamma(d\pm i \beta) \Gamma(\tilde{d}\pm i\alpha)}  + (a\leftrightarrow d).
\end{equation}}

\subsection{Holography in Constrained BF}
\label{sect:holosch}
We proceed as in the compact case of section \ref{sect:holobf} by deriving the boundary Schwarzian action and bilocal operator vertex, directly from the bulk constrained chiral $\mathfrak{sl}(2,\mathbb{R})$ BF action and Wilson lines. The relation between the bulk $\mathfrak{sl}(2,\mathbb{R})$ BF theory and the boundary Schwarzian was given essentially in \cite{Gonzalez:2018enk}, which we phrase here in our language. As before, we start with the BF-action \eqref{BF}, and integrating out the bulk $\chi$ forces the connection $A$ to be flat. 
\\~\\
The gravitational boundary conditions gravely influence the resulting boundary dynamics, restricting the degrees of freedom. Going up one dimension, in chiral $\mathfrak{sl}(2,\mathbb{R})$ Chern-Simons in coordinates $(\tau,\phi,r)$, the gravitational boundary conditions are:\footnote{For matrix representations of the generators see appendix \ref{app:rep}.}
\begin{equation}
\left.A_z\right|_{r\to \infty} = i J^- + T(\tau) i J^+ = \left(\begin{array}{cc}
0 & T(\tau) \\
1 & 0 \\
\end{array}\right), \qquad \left.A_{\bar{z}}\right|_{r\to \infty} = 0,
\end{equation}
or 
\begin{equation}
\left.A_\tau\right|_{r\to \infty} = \left.A_{\phi}\right|_{r\to \infty} = \left(\begin{array}{cc}
0 & T(\tau) \\
1 & 0 \\
\end{array}\right).
\end{equation}
Upon dimensional reduction to BF theory, as $A_\phi = \chi$ this constrains $\chi\rvert_\text{bdy}
$. Integrating out the bulk $\chi$, one writes down the resulting BF boundary action:
\begin{equation}
\label{LSch}
L = H = \frac{1}{2}\text{Tr}\chi A_\tau = T(\tau),
\end{equation}
identifying the quantity $T(\tau)$ with both the total energy $H$ in the spacetime, and the Lagrangian $L$. The remaining degrees of freedom in the theory are $T(\tau)$ and are path-integrated over. This is the analogue of $g(\tau)$ for the unconstrained group models.
\\~\\
To rewrite the gravitational Wilson line in terms of this variable $T(\tau)$, we follow the recent approach of \cite{Fitzpatrick:2016mtp}, which was done in the 3d / 2d case. A Wilson line in a discrete representation can be conveniently written in the Borel-Weil representation of the algebra \eqref{BW}. A discrete lowest weight state in representation $j$, labeled $0$, has wavefunction
\begin{equation}
\bra{x}\left. j,0 \right\rangle = \frac{1}{x^{2j}}, \quad \bra{j,0}\left. x \right\rangle = \delta(x).
\end{equation}
The lowest weight diagonal matrix element of the Wilson line operator in this representation $j$, can then be written as
\begin{align}
\label{mexp}
\mo_j(\tau_1,\tau_2)=\mathcal{W}_{j,00}(\tau_1,\tau_2) &= \int dx \, \delta(x) \, \mathcal{P} e^{\int_{\tau_1}^{\tau_2}d \tau \left[\partial_x - T(\tau) \left(-x^2\partial_x - 2jx\right)\right]}\frac{1}{x^{2j}} \nonumber \\
&= \left.\mathcal{P} e^{\int_{\tau_1}^{\tau_2}d\tau \left[\partial_x - T(\tau) \left(-x^2\partial_x - 2jx\right)\right]}\frac{1}{x^{2j}}\right|_{x=0},
\end{align}
where the explicit constrained boundary connection was used. The remaining technical computation was done in \cite{Fitzpatrick:2016mtp}, by interpreting this path-ordered exponential as a Hamiltonian evolution operator, and then rewriting the latter in a path integral fashion. A field redefinition allows to write out \eqref{mexp} in a more suggestive manner. Upon introducing a new function $f(t)$, determined as the solution of 
\begin{equation}
\label{fT}
\left\{f,\tau\right\} = T(\tau),
\end{equation}
the result can be written as
\begin{equation}
\label{Scorr}
\mo_\ell(\tau_1,\tau_2) = \left(\frac{f'_1f'_2}{(f_1-f_2)^2}\right)^{\ell}.
\end{equation}
The field redefinition from the boundary condition $T(\tau)$ to $f(\tau)$ using equation \eqref{fT} identifies $f$ as the time reparametrization from the Poincar\'e patch into a generic frame of reference, due to a time-dependent bulk total energy $T(\tau)$ in spacetime at time $\tau$. This field redefinition is only determined up to a $\slr$ redundancy. Moreover, demanding no infinite energy configuration $T(\tau)<\infty$ implies $f'$ cannot have sign switches, meaning either $f'>0$ or $f'<0$ everywhere. As $\left\{f,\tau\right\}$ does not change upon $f' \to -f'$, we choose to take $f'>0$ everywhere. This identifies the integration space as $\text{diff }\mathbb{R} / \slr$. The Lagrangian \eqref{LSch} becomes a Schwarzian derivative of $f$ \eqref{sch}, and the Wilson line \eqref{Scorr} becomes the Schwarzian bilocal operator.\footnote{The correct path integration measure over $f$ also follows from the particle-on-a-group path integral, though we will not explicitly check this here.}
The nonzero temperature parametrization can be found by imposing the additional boundary condition $T(\tau+\beta)=T(\tau)$, allowing a further reparametrization $f \to F=\tan \frac{\pi f}{\beta}$, and resulting in the integration space $\text{diff } S^1 / \slr$.
\\~\\
As a check on this computation, for constant $T(\tau) = E$, representing black hole geometries in the bulk, the time-ordering $\mathcal{P}$ becomes irrelevant, and the matrix exponential \eqref{mexp} can be directly computed using \eqref{repgroup}, with the result:
\begin{equation}
\mo_{\ell}(\tau_1,\tau_2) = \left(\frac{\frac{E}{2}}{\sin^2\left(\tau_{12}\sqrt{\frac{2}{E}}\right)}\right)^{\ell},
\end{equation}
which is precisely the two-point function on a thermal background with temperature $\frac{1}{\beta} = \frac{1}{\pi} \sqrt{\frac{E}{2}}$, matching the Hawking temperature of a mass $E$ black hole in JT \cite{Almheiri:2014cka, Jensen:2016pah, Maldacena:2016upp, Engelsoy:2016xyb}.
\\~\\
Summarizing we have re-established that the holographic dual of the constrained $\slrp$ BF is the Schwarzian, but have also proven that Wilson lines in JT gravity are dual to the bilocal operators in the Schwarzian:
\begin{equation}
\boxed{\mathcal{W}_{\ell,00}(\tau_1,\tau_2)=\mo_\ell(\tau_1,\tau_2) = \left(\frac{f'_1f'_2}{(f_1-f_2)^2}\right)^{\ell}}.\label{dictsch}
\end{equation}

\subsection{Diagrammatic Expansion}
The calculation of a generic correlators of Wilson lines in JT gravity discussed in section \ref{corrjt} can be summarized in a set of diagrammatic rules very similar to those of section \ref{s:pgdiag}.

\begin{itemize}
\item The starting point is just the disk on which the Wilson lines are drawn.
\item Each region in the disk is assigned a state $k_i$ associated with a principal continuous series irrep of $\slrp$, and contributes a weight $\dim k_i=k_i \sinh 2\pi k_i$, which is the Plancherel measure of $\slrp$. These labels $k_i$ are integrated over.
\item Each boundary segment carries a Schwarzian Hamiltonian propagation factor $e^{-L_i\frac{\mathcal{C}_{k_i}}{C}}$. Each intersection of a Wilson line with the boundary has associated with it 3 $\slrp$ irreps of which one is discrete and the two other are continuous. Such a crossing is weighted with the Schwarzian $3j$-symbol \eqref{cgsch}.
\begin{align}
\label{frules}
\begin{tikzpicture}[scale=0.8, baseline={([yshift=-0.1cm]current bounding box.center)}]
\draw[thick] (-0.2,0) arc (170:10:1.53);
\draw[fill,black] (-0.2,0.0375) circle (0.1);
\draw[fill,black] (2.8,0.0375) circle (0.1);
\draw (3.4, 0) node {\footnotesize $\tau_1$};
\draw (-0.7,0) node {\footnotesize $\tau_2$};
\draw (1.25, 0.5) node {\footnotesize $k$};
\draw (5, 0) node {$\raisebox{10mm}{$\qquad\qquad= \exp{- (\tau_2-\tau_1) \frac{\mathcal{C}_k}{C}}$}$};
\end{tikzpicture}
\end{align}
\begin{align}
\begin{tikzpicture}[scale=1, baseline={([yshift=-0.1cm]current bounding box.center)}]
\draw[thick] (-.2,.9) arc (25:-25:2.2);
\draw[fill,black] (0,0) circle (0.08);
 \draw[thick,blue](-1.5,0) -- (0,0);
\draw (-1,.3) node {\footnotesize$ \color{blue}\ell$};
\draw (-1,.8) node {\footnotesize$ k_1$};
\draw (-1,-.5) node {\footnotesize$ k_2$};
\draw (3,0.1) node {$\mbox{$\ =\  \, \left(\frac{\Gamma(\ell\pm ik_1 \pm ik_2)}{\Gamma(2\ell)}\right)^{\frac{1}{2}} .$}$}; \end{tikzpicture}\ \label{sch3j}
\end{align}
\item Each crossing of two Wilson lines is associated with 6 $\slr$ irreps and is weighed with the appropriate $\slr$ $6j$-symbol:
\begin{align}
\label{crossing}
\ \begin{tikzpicture}[scale=1, baseline={([yshift=0cm]current bounding box.center)}]
\draw[thick,blue] (-0.85,0.85) -- (0.85,-0.85);
\draw[thick,blue] (-0.85,-0.85) -- (0.85,0.85);
\draw[dotted,thick] (-0.85,-0.85) -- (-1.25,-1.25);
\draw[dotted,thick] (0.85,0.85) -- (1.25,1.25);
\draw[dotted,thick] (-0.85,0.85) -- (-1.25,1.25);
\draw[dotted,thick] (0.85,-0.85) -- (1.25,-1.25);
\draw (1.5,0) node {\scriptsize $k_4$};
\draw (-1.5,0) node {\scriptsize $k_2$};
\draw (-.75,.33) node {\scriptsize \color{blue}$\ell_1$};
\draw (.78,.33) node {\scriptsize \color{blue}$\ell_2$};
\draw (0,1.5) node {\scriptsize  $k_1$};
\draw (0,-1.5) node {\scriptsize $k_3$};
\end{tikzpicture}~~\raisebox{-3pt}{$\ \ \  = \ \ \sj{k_1}{\ell_1}{k_2}{k_3}{\ell_2}{k_4}.$}~~~
\end{align}

\end{itemize}
These rules agree with the diagrammatic rules for the correlators of bilocals \eqref{dictsch} in the Schwarzian theory, which were established from the 2d Liouville CFT perspective in \cite{Mertens:2017mtv}, and extensively analyzed in that work. One interesting example to mention here is the OTO four-point function The expectation value of the product of two bilocal operators $\mathcal{O}_{\ell_1}(\tau_1,\tau_2)$ and $\mathcal{O}_{\ell_2}(\tau_3,\tau_4)$, with time-ordering $\tau_1<\tau_2<\tau_3<\tau_4$ but applied out-of-time order, is then:
\begin{align}
\begin{tikzpicture}[scale=1, baseline={([yshift=0cm]current bounding box.center)}]
\draw[thick]  (0,0) ellipse (1.6 and 1.6);
\draw[thick,blue] (1.1,-1.2) arc (35.7955:82:4);
\draw[thick,blue] (-1.1,-1.2) arc (144.2045:98:4);
\draw (0,-0.8) node {\footnotesize $k_3$};
\draw (0,1) node {\footnotesize $k_1$};
\draw (-1,-0.2) node {\footnotesize $k_2$};
\draw (1,-0.2) node {\footnotesize $k_4$};
\draw (-0.5,0.35) node {\footnotesize \color{blue}$\ell_1$};
\draw (0.5,0.35) node {\footnotesize \color{blue}$\ell_2$};
\draw[fill,black] (-1.56,0.4) circle (0.1);
\draw[fill,black] (1.56,0.4) circle (0.1);
\draw[fill,black] (-1.11,-1.17) circle (0.1);
\draw[fill,black] (1.11,-1.17) circle (0.1);
\draw (-1.9,0.4) node {\footnotesize $\tau_1$};
\draw (-1.45,-1.4) node {\footnotesize $\tau_3$};
\draw (1.45,-1.4) node {\footnotesize $\tau_2$};
\draw (1.9,0.4) node {\footnotesize $\tau_4$};
\end{tikzpicture} 
\end{align}
Using the diagrammatic rules, this becomes:
\begin{align}
\nonumber \Big\langle \mathcal{O}_{\ell_1}(\tau_1,\tau_3) \mathcal{O}_{\ell_2}(\tau_2,\tau_4) \Big\rangle_{\text{OTO}} =&\prod_i \left(d\mu (k_i)e^{-L_i \frac{\mathcal{C}_{k_i}}{C}}\right)\sj{k_1}{\ell_1}{k_2}{k_3}{\ell_2}{k_4}\\
\nonumber &\cross \left(\frac{\Gamma(\ell_1\pm ik_1 \pm ik_2)}{\Gamma(2\ell_1)}\right)^{\frac{1}{2}}\left(\frac{\Gamma(\ell_1\pm ik_3 \pm ik_4)}{\Gamma(2\ell_1)}\right)^{\frac{1}{2}}\\
&\cross \left(\frac{\Gamma(\ell_2\pm ik_1 \pm ik_4)}{\Gamma(2\ell_2)}\right)^{\frac{1}{2}}\left(\frac{\Gamma(\ell_2\pm ik_2 \pm ik_3)}{\Gamma(2\ell_2)}\right)^{\frac{1}{2}}.\label{schOTO4pt}
\end{align}
Summarizing, by matching correlators of bilocals in the Schwarzian with correlators of Wilson lines in the gauge theory formulation of JT gravity, we have confirmed that these operators are dual.

\section{Concluding Remarks}
\label{sect:concl}
Boundary-anchored Wilson lines in 2d BF theory were shown to map to bilocal operators in the particle-on-a-group model. By making explicit the embedding of the Schwarzian in constrained particle on the $\slrp$ manifold or equivalently the embedding of JT gravity in $\slrp$ BF, we were able to provide a bulk derivation of correlators in the Schwarzian, relying heavily on the Hamiltonian reduction formulas. The holographic duals of bilocals in the Schwarzian are boundary-anchored Wilson lines in JT gravity. OTO-correlation functions of the particle-on-a-group and Schwarzian were determined in terms of crossing Wilson lines in the bulk, giving $6j$-symbols in the process. For the case of the Schwarzian / JT gravity, it was shown in \cite{Lam:2018pvp} that these contain the entire gravitational shockwave expansion, including Lyapunov behavior.
\\~\\
Explicitly, we provided a constructive derivation of correlators of boundary-anchored Wilson lines in 2d BF and 2d Yang-Mills, first for compact groups and afterwards for $\slr$. Since expectation values of boundary-anchored Wilson lines in $\slr$ YM and BF did not appear before, matching them with the Schwarzian correlators \cite{Mertens:2017mtv} served a double purpose: next to establishing the holographic dual of Schwarzian correlators, it also verifies the bulk calculation. \\
An important aspect of our story is the appearance of the Plancherel measure $\rho(k) = k \sinh(2\pi k)$ of $\slrp$. The seeming subtlety of $\slr$ versus $\slrp$ is crucial in obtaining this specific measure. We elaborate on this in \cite{toap}.
\\~\\
In the process, we explained the structural similarity of 2d YM and BF with boundary conditions: the only difference is in the Hamiltonian propagation factor, but the Hilbert space of both theories is identical. Both Hamiltonians are proportional to the Casimir, but in Yang-Mills the propagation scales with the bulk area $A$, whereas the BF Hamiltonian lives on the boundary and propagation hence scales with the boundary length $L$. BF theory is obtained by taking $e\to 0$ in YM, and choosing specific boundary conditions that introduce a boundary action. 

In \cite{Blommaert:2018oue}, it was established that certain classes of boundary anchored Wilson lines in YM have an interpretation in terms of particle-on-a-group correlators: the mapping is simply $e A_i = L_i$ where $A_i$ is the area of the bulk region touching the boundary segment $i$. This is made explicit by considering wedge-shaped Wilson line configurations in YM \cite{Blommaert:2018oue}. Likewise, the Schwarzian correlators have an interpretation as wedge-diagrams in constrained $\slr$ Yang-Mills.
\\~\\
We remarked at various points in section \ref{s:gravity} that the matrix elements of the constrained $\mathfrak{sl}(2,\mathbb{R})$ system are precisely the Liouville wavefunctions, and this means that the resulting integrations are technically identical. This is a generic feature of the Hamiltonian reduction where the Casimir operator explicitly turns into the Liouville minisuperspace Hamiltonian, as reviewed in appendix \ref{app:eigenvalue}. This extends to close cousins of Liouville CFT. E.g. the $\mathcal{N}=1$ OSp$(1|2)$ Casimir operator was diagonalized in \cite{Hikida:2007sz}, leading to the wavefunctions
\begin{equation}
\psi_{k,\pm}(\phi) \sim e^{\phi/2}\left(K_{ik+\frac{1}{2}}(e^{\phi}) \pm K_{ik-\frac{1}{2}}(e^{\phi}) \right),
\end{equation}
which can be directly identified with the $\mathcal{N}=1$ Liouville minisuperspace wavefunctions as determined in \cite{Douglas:2003up}. These wavefunctions were used in \cite{Mertens:2017mtv} to determine $\mathcal{N}=1$ super-Schwarzian correlators. Hence, the $\mathcal{N}=1$ Schwarzian correlators can be calculated from Wilson lines in constrained OSp$(1|2)$ BF.

This holds for 2d Toda CFT as well, for which the $W_N$-symmetry is the pivotal structure. This CFT is expected to contain information on the higher spin Schwarzian systems. E.g. the (Euclidean) higher spin Schwarzian action for the $W_3$ case is given by \cite{Gonzalez:2018enk}:
\begin{equation}
S = -C \int d\tau \left( \frac{f'''}{f'} - \frac{4}{3} \left(\frac{f''}{f'}\right)^2 + \frac{e'''}{e'} - \frac{4}{3} \left(\frac{e''}{e'}\right)^2 - \frac{1}{3} \frac{f''e''}{f'e'}\right),
\end{equation}
in terms of two time reparametrizations $f$ and $e$. This would be an interesting system to understand further. In light of our treatment of Hamiltonian reduction, we collect the relevant equations in appendix \ref{app:higher} to obtain Toda / higher spin Schwarzian from the SL$(n,\mathbb{R})$ WZW / particle-on-a-group model, with in particular the so-called Whittaker functions playing role both as wavefunctions of the Toda minisuperspace model, and as SL$(n,\mathbb{R})$ Casimir eigenfunctions \cite{Gerasimov:1996zk}.
\\~\\
As mentioned in section \ref{sect:holobf}, it is possible to consider more general configurations of boundary-anchored Wilson than the ones discussed here, by allowing three or more Wilson lines to end in a bulk vertex. Small gauge invariance requires that these couple to the identity, such that a three-vertex comes with an additional $3j$-symbol, a four-vertex comes with a sum over two $3j$-symbols etcetera. The net effect is that a three-vertex in the bulk becomes a $6j$-symbol in diagrammatic language, a four-vertex becomes a $9j$-symbol summed over one of the irreps appearing in it i.e. by summing over all intermediate states, etcetera. This opens up the possibility to investigate networks of Wilson lines in the bulk. In 3d CS certain such networks consisting of three-vertices were shown to calculate conformal blocks in 2d WZW \cite{Fitzpatrick:2016mtp,Bhatta:2016hpz,Besken:2016ooo,Bhatta:2018gjb}, and to be related with the OPE as well as with bulk reconstruction using HKLL, where one of the Wilson lines connects the vertex to a local bulk operator \cite{Guica:2016pid}. It could be interesting to make explicit the dimensional reduction of this story.

\section*{Acknowledgements}
The authors thank Nick Bultinck, Nele Callebaut, Luca Iliesiu, Jared Kaplan, Ho Tat Lam, Gustavo Turiaci and Herman Verlinde for several valuable discussions. AB and TM gratefully acknowledge financial support from Research Foundation Flanders (FWO Vlaanderen).

\appendix

\section{Knots in BF}
\label{app:knots}
The Hamiltonian of BF theory has support only on the boundary and hence all interesting dynamics takes place on the boundary. As discussed in the main text, the interesting observables are boundary-anchored Wilson lines or boundary bilocals. Aside from this dynamical sector, there is an interesting set of purely topological knot observables that are analogous to the knot expectation values in 3d Chern-Simons \cite{Witten:1988hf}. 
\\~\\
In arbitrary dimensions, BF theory is defined as:
\begin{equation}
S = \frac{k}{4\pi}\int_{\mathcal{M}} \text{Tr} B \wedge F, \quad F = dA + A \wedge A,
\end{equation}
where we omitted the boundary term irrelevant for the purely bulk objects considered here. The interesting bulk operators in this theory are Wilson loops / surfaces associated to both fields $A$ and $B$, with Wilson lines of the same type commuting and hence uninteresting.

\subsection{Punctures}
\begin{figure}[h]
\centering
\includegraphics[width=0.8\textwidth]{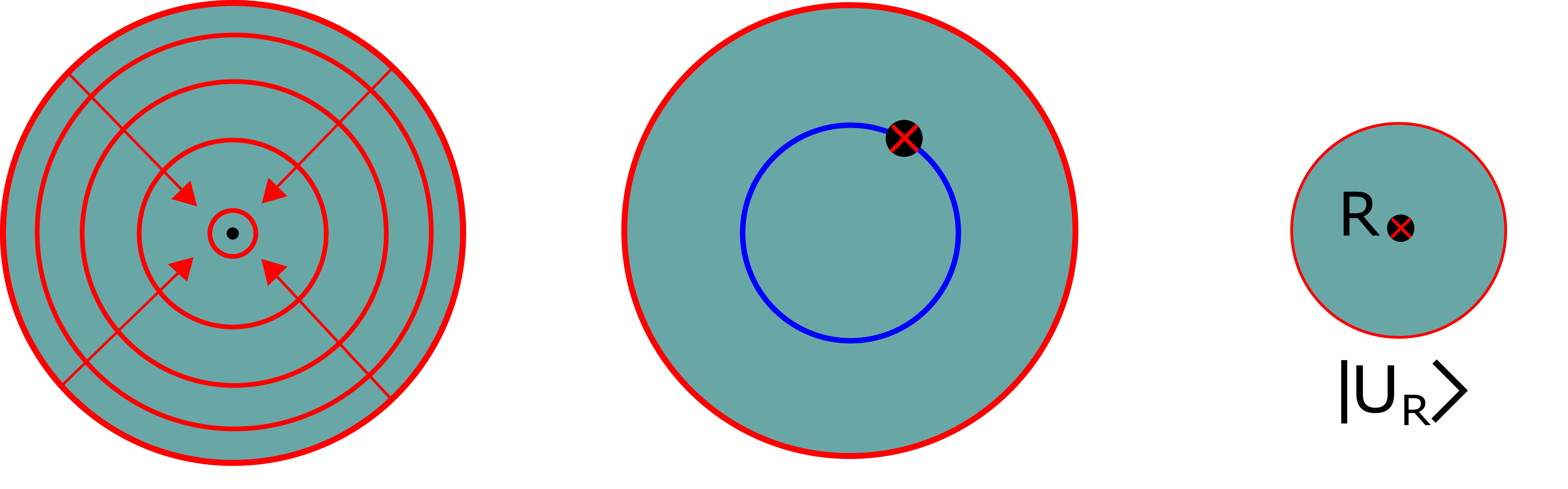}
\caption{Left: Disk slicing as concentric circles in the character Hilbert basis. Middle: intersecting Wilson line operators. Right: State-operator correspondence associating a fixed holonomy state $\left|U_\lambda\right\rangle$ to a local puncture in the representation $R$.}
\label{ClosedChannel}
\end{figure}

\begin{figure}[h]
\centering
\includegraphics[width=0.2\textwidth]{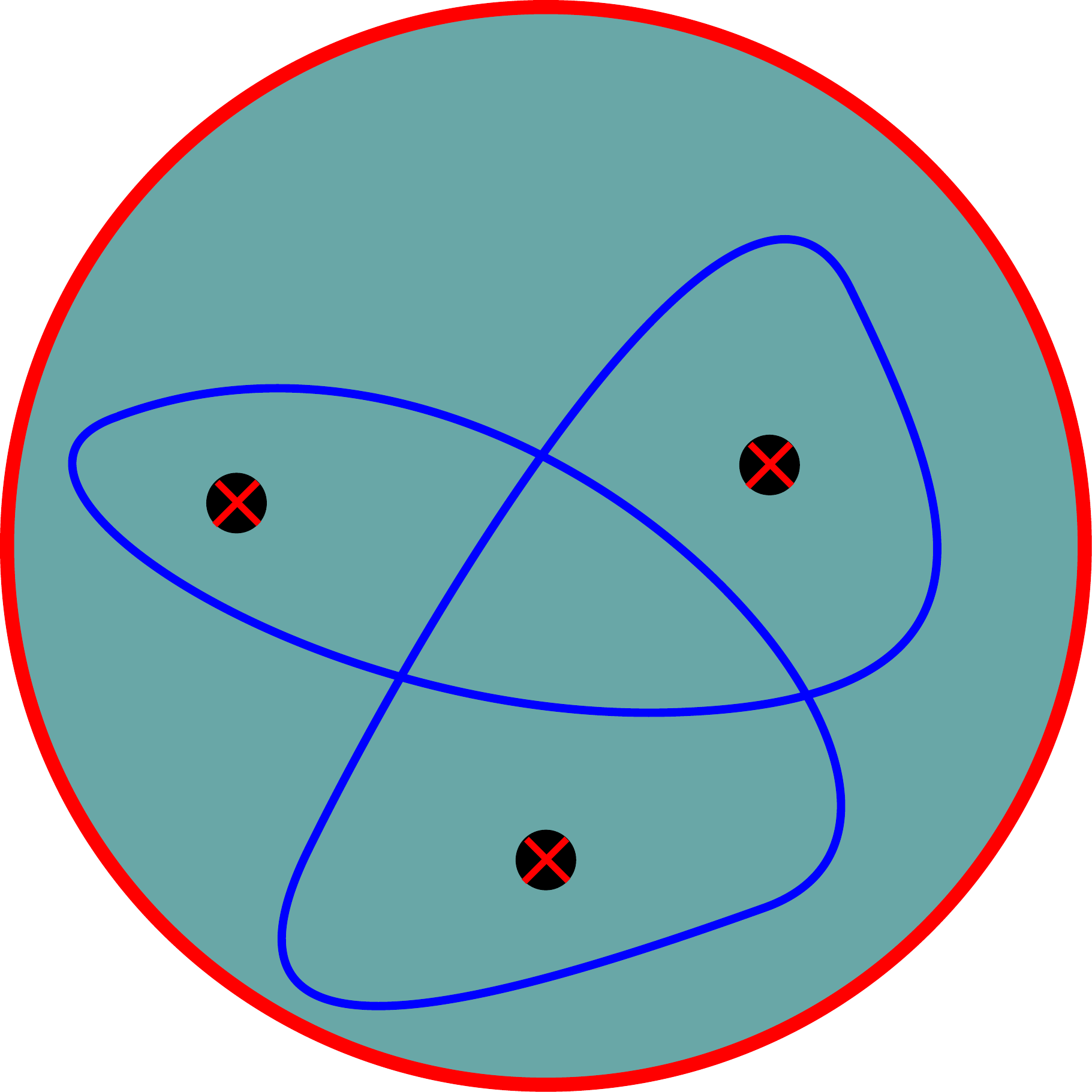}
\caption{Example of a bulk knot with three local $e^{i \beta \Phi(y_i)}$ punctures.}
\label{bulkknot}
\end{figure}

\begin{figure}[h]
\centering
\includegraphics[width=0.95\textwidth]{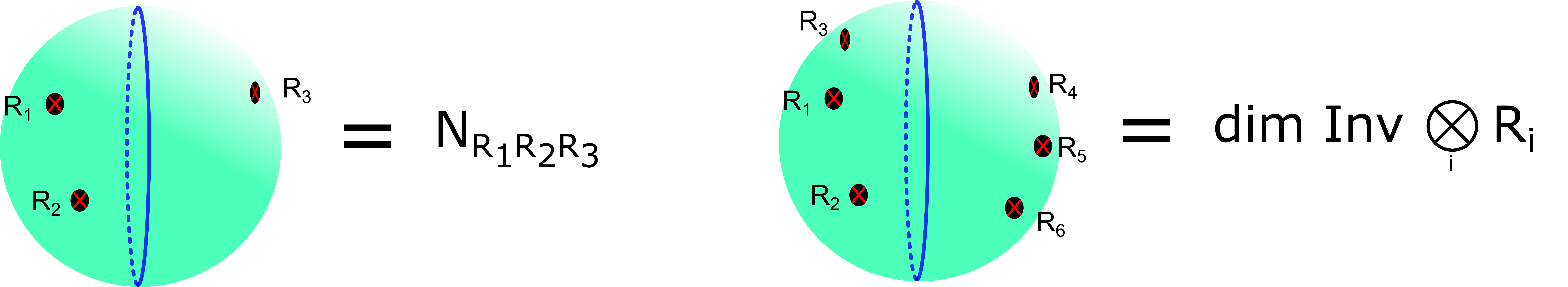}
\caption{Left: Fusion coefficient as three-holed sphere in BF. Right: Generic $n$-hole sphere amplitude gives the number of times the identity appears in the tensor product $\otimes_i \lambda_i$.}
\label{BFsphere}
\end{figure}

\begin{figure}[h]
\centering
\includegraphics[width=0.6\textwidth]{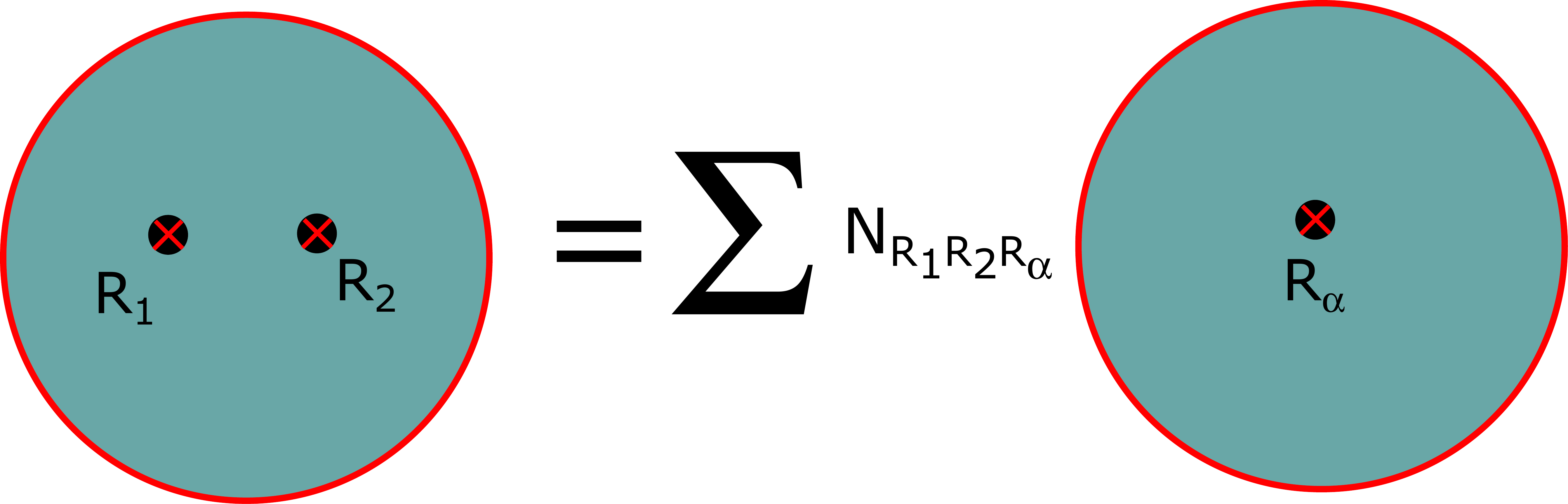}
\caption{Fusing two $\Phi$-punctures into a single one.}
\label{fusion}
\end{figure}

Specifying to 2d, the BF-path integral has two natural classes of observables:
\begin{equation}
\text{Tr}_R \mathcal{P} e^{i\oint A}, \quad \text{Tr}_R e^{i \Phi(y)},
\end{equation}
representing respectively Wilson loops and punctures on a 2d region of space. For concreteness, we continue with the Abelian case. The generalization to the non-Abelian case is quantitatively not straightforward (requiring knot-theoretic considerations \cite{Witten:1988hf}), but the canonical structure of the theory should be similar. The expectation values of these operators calculate their topological linking. Indeed, the path integral with insertions of both types:
\begin{equation}
\label{pibf}
\int \left[\mathcal{D}A\mathcal{D}\Phi\right] e^{i\alpha \oint A}  e^{i \beta\Phi(y)} e^{i \int \Phi F}
\end{equation}
leads to the constraint $F(x) = - \beta \delta(x-y)$, an Aharonov-Bohm vortex of magnetic flux at the location $y$. This means Wilson lines in $A$ can be freely deformed, as long as no $\Phi$-puncture is crossed. Correlators of this kind are determined by the Gauss intersection number $N$:
\begin{equation}
\left\langle e^{i\alpha \oint A}  e^{i \beta\Phi(y)}\right\rangle = e^{i\alpha\beta N}.
\end{equation}
It is instructive to analyze the canonical structure of this model, and reproduce the above topological result from this perspective. Canonical quantization of the 2d BF proceeds in analogy with that of 3d CS \cite{Witten:1988hf}, by treating $A_0$ as a Lagrange multiplier:
\begin{equation}
L = \text{Tr} \left(\Phi \partial_0 A_1 + \partial_1 \Phi A_0  + i \Phi \left[A_0,A_1\right]\right),
\end{equation}
leading to the equal-time canonical algebra:
\begin{equation}
\left[A_1^a(\mathbf{x}), \Phi^b(\mathbf{y})\right] = \delta^{ab} \delta(\mathbf{x}-\mathbf{y}),\label{canalgebra}
\end{equation}
with constraint $D_1 \Phi^a = 0$. Enforcing this constraint on physical wavefunctions, using $\Phi^a(\mathbf{x}) = -\frac{\delta}{\delta A_1^a(\mathbf{x})}$ leads immediately to what is called the Gauss' law constraint in EM, which in 2d leads to wavefunctionals of only the gauge-covariant Wilson lines. 
\\~\\
Choose now a disk and fix the coordinates as $A_0 = A_\rho$ and $A_1 = A_\phi$ in radial quantization of the theory for convenience (Figure \ref{ClosedChannel} left).
 Defining Wilson loops and punctures as:
\begin{equation}
\mathcal{W}_\alpha = e^{i \alpha \int_{0}^{2\pi}d\phi A_\phi(x)}, \quad \mathcal{W}_\beta = e^{i \beta \Phi(y)},
\end{equation}
the canonical algebra \eqref{canalgebra} results in the fundamental ``equal-time'' exchange algebra:
\begin{equation}
\mathcal{W}_\alpha \mathcal{W}_\beta = e^{i\alpha\beta} \mathcal{W}_\beta \mathcal{W}_\alpha,
\end{equation}
for a circular Wilson loop that intersects the local operator $\mathcal{W}_\beta$ (Figure \ref{ClosedChannel} middle). In a radially quantized path integral, this is to be read as the cost of moving one Wilson line through a puncture. This contour deforming situation is exactly as happens in 2d CFT.
This means that $\mathcal{W}_\beta$, as a local operator puncture, changes the holonomy eigenvalue of a state. Indeed, the above algebra implies
\begin{equation}
\mathcal{W}_\alpha \mathcal{W}_\beta \left|U= \mathbf{1}\right\rangle = e^{i\alpha\beta} \mathcal{W}_\beta \mathcal{W}_\alpha \left|U= \mathbf{1}\right\rangle = e^{i\alpha\beta} \mathcal{W}_\beta \left|U= \mathbf{1}\right\rangle,
\end{equation}
such that a holonomy eigenstate is obtained as $\mathcal{W}_\beta \left|U= \mathbf{1}\right\rangle = \left|U = e^{i \beta}\right\rangle$. We proved this result for the abelian case, but it seems evident that this will hold in the non-abelian case as well: $\text{Tr}_R e^{i \Phi(y)}$ creates a $U$-eigenstate using the well-known isomorphism between holonomies and irreps in group theory \cite{Elitzur:1989nr}:
\begin{equation}
U_\lambda = e^{-\frac{2\pi}{k} \lambda},\label{isomorph}
\end{equation}
for finite weight vector $\bm{\lambda}$ and vector of Cartan matrices $\bf{H}$, and $\lambda=\bm{\lambda}\cdot \bf{H}$. This is to be interpreted as a state-operator correspondence of a local irrep $\lambda$ puncture to a holonomy eigenstate $\left|U_\lambda\right\rangle$ that represents a tiny circle around the defect (Figure \ref{ClosedChannel} right). The partition function of BF on a disk with such a holonomy $U_\lambda$ defect in the interior results in the twisted partition function:
\begin{equation}
\sum_{R}\dim R\, \chi_R(U_\lambda) \, e^{-TH_R},\label{twisted}
\end{equation}
using concentric circle slicing of the disk. As discussed further on in appendix \ref{app:wzwcorr}, this is the $k\to\infty$ dimensional reduction of the torus path integral $\chi_{\hat{\lambda}}$ of 3d Chern-Simons with a bulk Wilson line in irrep $\lambda$ inserted along the dimension that is to be reduced, which is dual to a twisted chiral WZW on the torus boundary. This dimensionally reduces to a twisted particle-on-a-group whose partition function matches \eqref{twisted}. This makes explicit that the puncture in irrep $\lambda$ is the dimensional reduction of a 3d CS Wilson line in irrep $\lambda$.
\\~\\
A disk with $n$ punctures labeled by $\lambda_i$ and boundary holonomy $U$ results in 
\begin{equation}
\sum_R \chi_R(U) (\dim R)^{1-n}\prod_i \chi_R(U_{\lambda_i})e^{-T H_R}.
\end{equation}
As an additional feature, multiple punctures allow for non-trivial knots in the bulk of Wilson loops encircling the punctures (Figure \ref{bulkknot}).
\\~\\
As an example, a disk with two punctures $\lambda_{1,2}$ and holonomy $U_3$ on the outer boundary, is given by:
\begin{equation}
\sum_{R}\frac{\chi_R(U_{\lambda_1}) \chi_R(U_{\lambda_2}) \chi_R(U_3)}{\text{dim R}}\, e^{-TH_R}.
\end{equation}
The topological limit $H_R \to 0$ of this formula is the $k\to\infty$ limit of the Verlinde formula of the CFT fusion rules:
\begin{equation}
\label{verlinde}
\sum_{R}\frac{\chi_R(U_{\lambda_1}) \chi_R(U_{\lambda_2}) \chi_R(U_3)}{\text{dim R}} = \lim_{k\to\infty} \sum_R \frac{S_{\lambda_1 R}S_{\lambda_2 R}S_{\lambda_3 R}}{S_{0 R}}=  N_{\lambda_1 \lambda_2 \lambda_3}.
\end{equation}
This formula can be inverted, using orthonormality of the character basis, as 
\begin{equation}
\frac{\chi_R(U_{\lambda_1}) \chi_R(U_{\lambda_2})}{\text{dim R}} = \int d U_3 N_{\lambda_1\lambda_2 \lambda_3} \chi_R(U_3),\label{CGchar}
\end{equation}
which is the Clebsch-Gordan series for the characters. 
\\~\\
Equation \eqref{verlinde} can be seen as the amplitude of the three-holed sphere in BF, which has no physical boundary and hence no boundary Hamiltonian. Again, this is to be compared with the $k\to\infty$ limit of a configuration in 3d CS where three Wilson lines along $S^1$ in irreps $\lambda_i$ are inserted in the path integral on $S^2\cross S^1$. 
\\~\\
The disk state with two punctures, denoted as $\ket{U_{\lambda_1\otimes \lambda_2}}$, can be expanded in the complete basis of holonomy eigenstates, which by the state-operator correspondence \eqref{isomorph}, can be parametrized by the set of all single punctures on the disk:
\begin{equation}
\label{diskpunct}
\ket{U_{\lambda_1\otimes \lambda_2}} = \int d U_{\lambda_3} \bra{U_{\lambda_3}}\ket{U_{\lambda_1\otimes \lambda_2}} \ket{U_{\lambda_3}}
\end{equation}
The overlap amplitude $\bra{U_{\lambda_3}}\ket{U_{\lambda_1\otimes \lambda_2}}$ is computed by gluing the two disks together resulting in a three-holed sphere amplitude:
\begin{equation}
\bra{U_{\lambda_3}}\ket{U_{\lambda_1\otimes \lambda_2}} = Z_{S^2}(\lambda_1,\lambda_2,\lambda_3) = N_{\lambda_1 \lambda_2 \lambda_3},
\end{equation}
which by \eqref{verlinde} is just the fusion coefficient. This is depicted in Figure \ref{BFsphere} left.
In terms of correlators, applying $\bra{\mathbf{1}}e^{-T\hat{H}}$ to the left of \eqref{diskpunct}, allows us to expand a disk with two punctures into a sum over disks with one puncture (Figure \ref{fusion}):
\begin{equation}
Z_{B^1}(\lambda_1,\lambda_2)=\int dU_{\lambda_3}\, Z_{S^2}(\lambda_1,\lambda_2,\lambda_3)\, Z_{B^1}(\lambda_3)=\int dU_{\lambda_3}\, N_{\lambda_1 \lambda_2 \lambda_3}\,Z_{B^1}(\lambda_3),\label{CG}
\end{equation}
Explicitly, this can also be read off from the formula \eqref{CGchar}.
In terms of particle-on-a-group path integrals, using \eqref{CG} the twice-punctured disk amplitude decomposes into a sum of particle-on-a-group amplitudes with twisted boundary conditions:
\begin{equation}
Z_{B^1}(\lambda_1,\lambda_2)=\sum_{\lambda_3} N_{\lambda_1\lambda_2 \lambda_3} \int \left[\mathcal{D} g\right]_{g(\tau+\beta)=U_{\lambda_3} g(\tau)} e^{- S[g]}.
\end{equation}
Alternatively, from the BF path integral with two $\Phi$-insertions (as in \eqref{pibf}), integrating out $\Phi$ leads to a locally exact 1-form $A$. When going around the boundary of the disk, it picks up a non-trivial holonomy
\begin{equation}
\text{Tr}_{\lambda_3}\mathcal{P}e^{i\oint A} = U_{\lambda_3} = g(\tau+\beta)g(\tau)^{-1},
\end{equation}
the value of which is precisely determined by the above Clebsch-Gordan decomposition.
\\~\\
The generalization to a disk with $n$ punctures is straightforward: this can again be decomposed into a sum over disks with one puncture associated with a twisted particle on group, as:
\begin{equation}
Z_{B^1}(\lambda_1,\lambda_2,..,\lambda_n)=\int dU_{\lambda} \, Z_{S^2}(\lambda_1,\lambda_2,..,\lambda_n,\lambda)\, Z_{B^1}(\lambda).
\end{equation}
This is illustrated in Figure \ref{BFsphere} right. The $n$-point amplitude on the two sphere is just the number of times the identity appears in the tensor product decomposition $\otimes_i R_i$, or the dimension of the Chern-Simons Hilbert space on $S^2 \times \mathbb{R}$ \cite{Witten:1988hf}:
\begin{equation}
Z_{S^2}(\lambda_1,\lambda_2,..,\lambda_n) = \dim \mathcal{H}^{CS}_{S^2} = \dim \text{Inv}(\lambda_1 \otimes \lambda_2 \otimes \hdots \lambda_n).
\end{equation}

\subsection{Diagrammatic Rules}
The inclusion of the possibility of punctures in generic amplitudes adds an extra line to the diagrammatic rules of section \ref{s:pgdiag}:
\begin{itemize}
\item Within each region of the disk, labeled by irrep $R_i$, each puncture $\lambda_j$ contributes a weight $\frac{\chi_{R_i}(U_{\lambda_j})}{\dim R_i}$.
\end{itemize}
The denominator stems from the fact that each puncture augments the Euler characteristic of the disk by one. These rules allows for the calculation of a generic knot configuration in BF, possibly including boundary-anchored Wilson lines. 
\\~\\
Alternatively, one could calculate a given amplitude by stepwise simplifying it using the appropriate exchange algebra to pull non-Abelian Wilson lines through punctures, and eventually reducing the multiple punctures to a single puncture as explained above. One is left with a sum over twisted particle on a group diagrams with possibly boundary-anchored Wilson lines. The latter can then be calculated explicitly as explained in the main text. 

\subsection{Wilson Loop Crossings}
\label{app:A3}
\begin{figure}[h]
\centering
\includegraphics[width=0.95\textwidth]{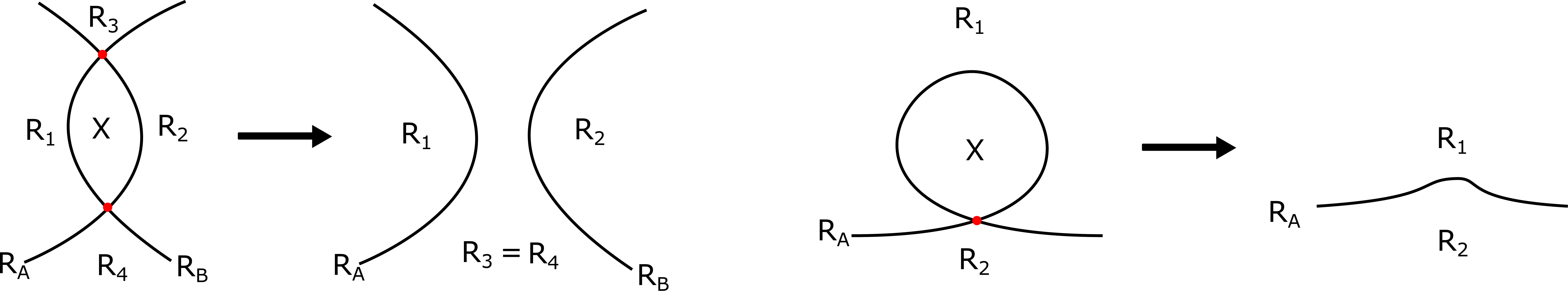}
\caption{Wilson line crossings that can be undone. Left: Double crossing of two lines. Right: Self-crossing of a single line.}
\label{fig:wilsonym}
\end{figure}

The canonical algebra of BF \eqref{canalgebra} shows that Wilson loops commute, meaning that they can be pulled through one another at will, and configurations including extensive crossings of Wilson lines can be greatly simplified before embarking on the calculation. Some examples of crossings that can be undone are drawn in Figure \ref{fig:wilsonym}.
\\~\\
\noindent It is amusing to see these moves explicitly in formulas as properties of $6j$-symbols. More explicitly, undoing one double crossing (Figure \ref{fig:wilsonym} left) is achieved by using the identity
\begin{equation}
\dim R \sum_{X}\dim X \sj{R_A}{R_1}{R_3}{R_B}{R_2}{X}\sj{R_A}{R_1}{R_4}{R_B}{R_2}{X}=\delta_{R_3 R_4}.
\end{equation}
Similarly, a self-crossing of a Wilson line (Figure \ref{fig:wilsonym} right) in the bulk can be undone by:
\begin{equation}
\sum_X \dim X \sj{R_A}{R_1}{R_2}{R_A}{R_1}{X}=1.
\end{equation}
Furthermore, a single bulk Wilson loop produces just an overall degeneracy factor $\dim R$. Indeed, the disk amplitude of a bulk Wilson line is obtained by taking the trace in \eqref{3R}:
\begin{equation}
\int dg \chi_{R_1}(g)\chi_{R_2}(g)\chi_{R_3}(g)=N_{R_1R_2R_3},
\end{equation}
such that 
\begin{equation}
\Big\langle\mathcal{W}_R\Big\rangle=\sum_{R_1,R_2}\dim R_2 \, N_{R_1R_2R}e^{-\beta H(R_1)} =\dim R \sum_{R_1} \dim R_1 e^{-\beta H(R_1)} =\dim R\, \left\langle \mathbf{1}\right\rangle,\label{bulkwilson}
\end{equation}
where we used the identity 
\begin{equation}
\sum_{R_2}\dim R_2 \, N_{R_1R_2R}=\dim R \dim R_1.
\end{equation}
Thus, Wilson lines in the bulk that do not enclose punctures produce just overall prefactors in amplitudes and are not interesting observables to consider. For the purpose of completeness, let us mention that such Wilson lines can be calculated from the particle-on-a-group perspective by taking the coincident limit of both endpoints of a single boundary-anchored Wilson line.

\section{Some Elementary Properties of Representation Matrices}
\label{app:groupth}
We collect some properties we need on integrals of representation matrices $R_{mn}(g) \equiv \bra{m} g \ket{n}$. Schur's orthogonality relation is given by:
\begin{equation}
\label{Schur}
\int dg R_{ab}(g)R'_{cd}(g^{-1}) = \frac{\delta_{RR'}}{\text{dim R}}\delta_{ad}\delta_{bc},
\end{equation}
also called the grand orthogonality theorem. The completeness relation
\begin{equation}
\label{compl}
\sum_{R,m,n}\text{dim R}\, R_{mn}(g_1)R_{nm}(g_2^{-1}) = \delta(g_1-g_2),
\end{equation}
with $\delta(g)$ defined in the Haar measure, is guaranteed by the Peter-Weyl theorem. These two properties combined allow us to write a complete normalized set of basis states $\left\{\left|R,mn\right\rangle\right\}$, whose wavefunctions are just the representation matrices:
\begin{equation}
\bra{g}\ket{R,mn} \equiv \sqrt{\text{dim R}}\, R_{mn}(g), \qquad \bra{R,mn}\ket{g} \equiv \sqrt{\text{dim R}}\, R_{nm}(g^{-1}),
\end{equation}
with orthonormality $\left\langle R_1,m_1,n_1\right|\left.R_2,m_2,n_2\right\rangle = \delta(R_1-R_2)\delta(m_1-m_2)\delta(n_1-n_2)$ and completeness $\sum_{R,m,n}\left|R,mn\right\rangle\left\langle R,mn\right| = \mathbf{1}$ following from \eqref{Schur} and \eqref{compl}.
Note that $\bra{R,mn}\ket{g} = \overline{\bra{g}\ket{R,mn}}$ implies $R_{nm}(g^{-1}) = \overline{R_{mn}(g)}$, implying only unitary representations are included.
\\~\\
The Kronecker tensor product of two representation matrices, can be expanded in a Clebsch-Gordan series as:
\begin{align}
R_{1m_1n_1}(g)R_{2m_2n_2}(g) &\equiv \bra{R_1,m_1} \otimes \bra{R_2, m_2} \, g \,  \ket{R_1,n_1} \otimes \ket{R_2,n_2} \nonumber \\
&= \sum_{R_3,\bar{R}_3,m_3,n_3}C_{R_1m_1,R_2m_2}^{R_3 m_3}C_{R_1n_1,R_2n_2}^{\bar{R}_3n_3} \bra{R_3, m_3} \, g \,  \ket{\bar{R}_3,n_3} \nonumber \\
&= \sum_{R_3,m_3,n_3}C_{R_1m_1,R_2m_2}^{R_3 m_3}C_{R_1n_1,R_2n_2}^{R_3n_3} R_{3,m_3n_3}(g),
\end{align}
a property well-known for the Wigner D-functions for SU$(2)$. Hence, we can write for the group integral of three representation matrices (using \eqref{Schur}):
\begin{align}
\label{3jiint}
\int d g R_{1,m_1n_1}(g) &R_{2,m_2n_2}(g) R_{3,n_3m_3}(g^{-1}) \nonumber \\
&= \sum_{R,m,n}C_{R_1m_1,R_2m_2}^{R m}C_{R_1n_1,R_2n_2}^{R n} \int d g R_{mn}(g) R_{3,n_3m_3}(g^{-1})\nonumber \\
&= \frac{C_{R_1m_1,R_2m_2}^{R_3 m_3}C_{R_1n_1,R_2n_2}^{R_3 n_3}}{\text{dim R$_3$}} \nonumber \\
&= (-)^{R_2-R_1+m_3}(-)^{R_2-R_1+n_3}\tj{R_1}{R_2}{R_3}{m_1}{m_2}{-m_3}\tj{R_1}{R_2}{R_3}{n_1}{n_2}{-n_3}.
\end{align}
We need this property in the main text. The $3j$-symbols satisfy the orthogonality property:
\begin{equation}
\label{orth3j}
\text{dim R}_3 \sum_{m_1,m_2}\tj{R_1}{R_2}{R_3}{m_1}{m_2}{m_3}\tj{R_1}{R_2}{\tilde{R}_3}{m_1}{m_2}{\tilde{m}_3} = \delta_{R_3\tilde{R}_3}\delta_{m_3\tilde{m}_3}.
\end{equation}
Throughout this work, we will absorb all minus signs in \eqref{3jiint} into the definition of the $3j$-symbols to streamline the notation.

\section{WZW and Particle-on-a-group}
\label{app:wzwcorr}
We provide some details on how the particle-on-a-group theory arises from 2d Wess-Zumino-Witten (WZW) CFT. The appearance of the particle-on-a-group theory from WZW is ubiquitous. We discuss the $k\to\infty$ dimensional reduction on a torus of both a chiral and a non-chiral WZW, resulting in the same theory.

\subsection{Non-Chiral WZW}
The non-chiral WZW model
\begin{equation}
S_{\text{WZW}}=\frac{k}{16\pi}\int d^2 z \text{Tr}(g^{-1}\partial g g^{-1}\bar{\partial} g) + k \Gamma_\text{WZ},\label{nonchiralS}
\end{equation}
with $\Gamma_{WZ}$ the Wess-Zumino term and $g(z,\bar{z})\in G$, defined on a torus worldsheet, leads to the thermal particle-on-a-group model after taking a double scaling limit where one of the cycles of the torus goes to zero, and the level $k\to +\infty$ keeping their product fixed. This procedure effectively sets $\partial, \bar{\partial} \to \partial_t$ and removes the Wess-Zumino term. The action \eqref{nonchiralS} almost trivially reduces to the action of quantum mechanics on the $G$-manifold. 
\\~\\
The WZW partition function is the diagonal modular invariant which contains a sum over all integrable representations $\hat{R}$ of $\hat{\mathfrak{g}}$. Taking the double-scaling limit directly in the growing channel, leads to:
\begin{equation}
Z(\tau) = \sum_{\hat{R}} \left|\chi_{\hat{R}}(\tau)\right|^2 \, \underset{\tau_2 \to +\infty,\, k\to \infty}{\longrightarrow}\, \sum_R \left|\text{dim R}\, e^{-\frac{\beta}{2}H_R}\right|^2, \qquad \beta = 2\pi \frac{\tau_2}{k+g^{\vee}}, \label{nonchiralZ}
\end{equation}
because the characters in this channel reduces to the zero-grade sector in the double-scaling limit \cite{Mertens:2017mtv}. States of WZW are labeled by $\ket{R,m \bar{m},n \bar{n}}$. The double-scaling limit projects on the zero-grade sector in the closed channel resulting in the particle-on-a-group spectrum $\left|R,m\bar{m}\right\rangle$. The partition function \eqref{nonchiralZ} should be read as constructed from this Hilbert space $\left|R,m\bar{m}\right\rangle$ with unity degeneracy.
\\~\\
Note that correlation functions of local WZW fields reduce in the limit to \emph{local} particle-on-a-group operators, and 3d Wilson lines in CS reduce to 2d Wilson lines in BF, both dual to bilocals. A priori without bulk knowledge, the resulting particle-on-a-group operators need not be associated with any bulk Wilson line. In the language of the particle-on-a-group calculation performed in Appendix \ref{app:pog}, this means that one could consider operators whose indices are not contracted. These operators are however not invariant under the global $G$ gauge redundancy and are hence not observables. Instead these operators are $G$-covariant. Such operators were studied in the Schwarzian model in \cite{Mertens:2017mtv}.

\subsection{Coadjoint Orbit}
The chiral WZW model (CWZW) associated with an integrable representation $\bm{\lambda}$ has action:
\begin{equation}
\label{CWZW}
S^\lambda_\text{CWZW} = \frac{k}{4\pi}\int dtd\phi \text{Tr}(g^{-1}\partial_\phi g g^{-1}(\partial_t-\partial_\phi) g) + k \Gamma_{\text{WZ}} + \text{Tr}(\lambda g^{-1}(\partial_t-\partial_\phi) g),
\end{equation}
where $\lambda = \bm{\lambda} \cdot \mathbf{H}$, with $\mathbf{H}$ the Cartan representation matrices. This is just the Kac-Moody coadjoint orbit action of the element $\bm{\lambda}$.\footnote{This is related to the more familiar Kirillov-Kostant formula for a single character of the group $G$, computed as a path-integral on the coadjoint orbit of $\lambda$ by restricting \eqref{CWZW} to the angular $\phi$-zero-mode:
\begin{equation}
S^\lambda_{\text{CWZW}} \,\to\, \int dt \text{Tr}(\lambda g^{-1}\partial_t g).
\end{equation}
Additionally a term $\text{Tr}\lambda^2$ could be added, giving the Casimir as Hamiltonian in the character.} It is found as the boundary action of Chern-Simons with a Wilson line in representation $\bm{\lambda}$ in the bulk, stretched in the time-direction. The path integral of this action yields the character $\chi_{\hat{\lambda}}$ in the time-direction. The untwisted particle-on-a-group action is obtained by dimensional reduction of the vacuum $\lambda=0$ coadjoint orbit along the $t$-direction.  
\\~\\
The partition function is the double scaling limit of the vacuum character $\chi_0$. This is most conveniently evaluated in the closed channel described above, for which in the double scaling limit the characters become traces over the zero grade sector only. The vacuum character decomposes into closed channel characters as:
\begin{align}
\chi_0(it) &= \text{Tr}q^{L_0} = \left\langle 0\right| e^{- \frac{\pi}{t} (L_0+\bar{L}_0)}\left|0\right\rangle = \sum_{\hat{R}} S_{0\hat{R}} \chi_{\hat{R}}(i/t), \qquad q=e^{-2\pi t} \nonumber \\ 
&\underset{t \to 0,\, k\to \infty}{\longrightarrow}\,  \sum_R \text{dim R} \left(\text{dim R}e^{-\beta H_R}\right), \qquad \beta t = \frac{2\pi}{k+g^{\vee}}.
\end{align}
The second equality of the first line illustrates that one should think of the Hilbert space again as a ``closed'' sector $\ket{R,m\bar{m}}$. Ordinarily, this makes no difference, as the brane boundary states project on the chiral sector anyway. But when inserting operators this becomes important. We have seen in the main text that the second $\bar{m}$ label in intermediate channels cannot be dropped for spinning operator insertions $\mathcal{O}_{R,M\bar{M}}$ with $M\neq \bar{M}$.
\\~\\
For a generic weight $\lambda$, dimensional reduction of \eqref{CWZW} to the $t$-zero-mode, with $k \int dt  = C$ held fixed, leads to:
\begin{equation}
\label{actred}
S^\lambda_\text{CWZW} \to -\frac{C}{4\pi}\int d\phi \text{Tr}(g^{-1}\partial_\phi g)^2 - \text{Tr}(\tilde{\lambda} g^{-1}\partial_\phi g).
\end{equation}
To obtain a finite second term, one is to consider weights scaling as $\lambda \sim k \to \infty$, resulting in a finite $\tilde{\lambda}=\lambda \int dt$. This particular scaling leads, up to some irrelevant multiplicative factors, to \cite{Witten:1991we,Cordes:1994fc}:
\begin{equation}
S_{\lambda R} \, \to \, \chi_R(U_\lambda),
\end{equation}
with $R\sim k^0$ and $\lambda\sim k^1$. The holonomy $U_\lambda$ is determined from the irrep $\lambda$ by \eqref{isomorph}. The scaling $\lambda \sim k$ results indeed in a finite holonomy in the double-scaling limit. The double-scaling limit of the $\hat{\lambda}$-coadjoint orbit hence becomes:
\begin{equation}
\chi_{\hat{\lambda}} = \sum_R S_{\hat{\lambda} R} \chi(-1/\tau) \, \to \,  \sum_R \chi_R(U_\lambda) \left(\text{dim R}e^{-\beta H_R}\right).\label{result}
\end{equation}
Alternatively, as shown in \cite{Barnich:2017jgw}, the second term in the action \eqref{actred} can be absorbed into the first using a new field $g$ that has twisted boundary conditions $g(\phi+\beta) = U_\lambda g(\phi)$, and the resulting twisted  particle-on-a-group computation indeed gives precisely this result \eqref{result}. 
\\~\\
A non-chiral WZW model can be decomposed into two chiral WZW models, see e.g. \cite{Falceto:1992bf}. The classical solution of the equations of motion of \eqref{nonchiralS} is given by:
\begin{equation}
g(x^+,x^-) = g_L(x^+) g_R^{-1}(x^-), \qquad g_{L/R}(x^\pm + 2\pi) = g_{L/R}(x^\pm)U,
\end{equation}
for holonomy $U \in G$. The phase space $\mathcal{M}$ consists of degrees of freedom contained in $g_{LR}$ and $U$, and using the association of a irrep $\lambda$ with $U$ \eqref{isomorph}, it can be written as
\begin{equation}
\mathcal{M} = \sum_{\hat{\lambda}} \mathcal{M}_\lambda = \sum_{\hat{\lambda}}\frac{LG_{\lambda} \times LG_{\lambda}}{G}.
\end{equation}
The non-chiral WZW path integral, is written in phase space as
\begin{equation}
\int_{\mathcal{M}} \left[\mathcal{D}g \mathcal{D}\pi_g\right]e^{-\frac{k}{16\pi}\int d^2 z \text{Tr}\left[\pi_g g^{-1}\dot{g} - \frac{1}{2}(\pi_g^2 + (g^{-1}g')^2)\right] + k \Gamma_\text{WZ}}.
\end{equation}
Changing field variables from $(g,\pi_g)$ into $(g_L,g_R)$ as
\begin{equation}
\label{pstf}
g = g_L g_R^{-1}, \quad \pi_g = g_R g_L^{-1}\, \partial_\phi g_L \, g_R^{-1} + \partial_\phi g_R \, g_R^{-1},
\end{equation}
including an integral over the twist $U$, one finds the action reduces to a sum of two chiral WZW models:
\begin{equation}
\sum_{\hat{\lambda}}\int_{\mathcal{M}_\lambda} \left[\mathcal{D}g_L\right] \left[\mathcal{D}g_R\right]e^{-S^{\lambda}_{\text{CWZW}}[g_L] - S^{\lambda}_{\text{CWZW}}[g_R]},
\end{equation}
with $S^{\lambda}_{\text{CWZW}}[g]$ given by \eqref{CWZW}. Note that when considering the non-chiral theory between branes (take e.g. one vacuum brane and one arbitrary brane), the additional boundary conditions enforce $g_L= g_R$ and fix $U_\lambda$ in terms of the specific boundary state $\ket{\lambda}$. In this case, the transformation \eqref{pstf} then leads to a single chiral WZW model, with phase space $LG_\lambda/G$, a perspective made explicit for the Schwarzian theory in \cite{Mertens:2018fds}.

\subsection{Branes}
\begin{figure}[h]
\centering
\includegraphics[width=0.95\textwidth]{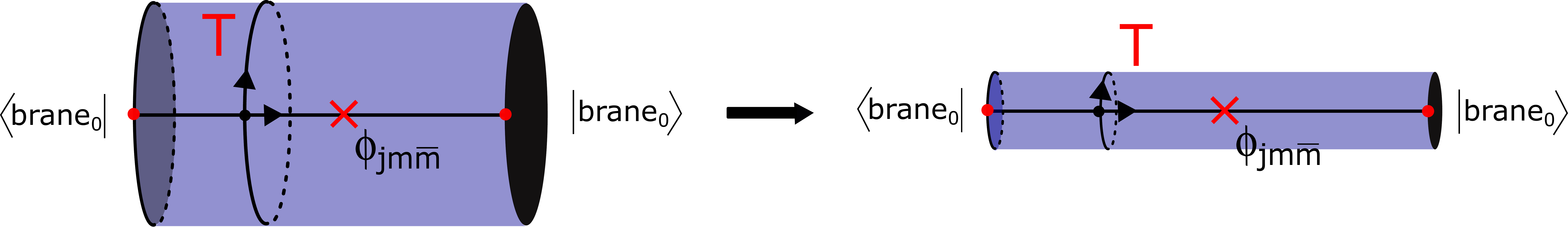}
\caption{Left: Cylinder amplitude in WZW between vacuum branes $\left|\text{brane}_0\right\rangle$. Right: Double scaling limit where the circumference of the cylinder $T\to 0$, as $k\to\infty$ keeping the product fixed $kT \sim C$. A local WZW puncture gives a bilocal operator in the dimensionally reduced particle-on-a-group model.}
\label{WZWcylinder}
\end{figure}

The calculations in \cite{Mertens:2018fds} relating WZW to particle-on-a-group considered non-chiral WZW between vacuum branes as ancestor of particle-on-a-group (Figure \ref{WZWcylinder}). This is by definition the chiral WZW model described above.
In the closed channel, a vacuum brane is characterized by an initial (or final) state $\ket{0}=\sum_\lambda \sqrt{S_{0 \lambda}}| \hat{\lambda} \rangle\hspace{-0.2em}\rangle$ with $| \hat{\lambda} \rangle\hspace{-0.2em}\rangle$ the Ishibashi state. In the double-scaling limit, the Ishibashi state reduces to a sum over diagonal states $| \hat{\lambda} \rangle\hspace{-0.2em}\rangle=\sum_{m}\ket{\lambda,m m}$. Crucially, this construction in no way alters the Hilbert space of the theory which is a local construction independent of any branes.
\\~\\
Changing a single brane boundary state from $\left|0\right\rangle$ into $\left|\lambda\right\rangle$, leads to the character $\chi_{\hat{\lambda}}$. The fact that the dimensional reduction of this configuration does not yield a nice thermal partition function \eqref{result} resonates with the fact that in the Schwarzian theory, other branes do not give a well-defined 1d thermal system \cite{Mertens:2017mtv,Mertens:2018fds}. 
\\~\\
Local operators in non-chiral WZW between branes are in the double scaling limit taken to bilocal operator insertions in the particle-on-a-group model \cite{Mertens:2018fds}:\footnote{Actually one should think of the higher dimensional local operators as local in $\phi$ but smeared in $t$, because only the zero-grade operators survive the double scaling.}
\begin{equation}
\mo_{j,m \bar{m}}(\tau_1,\tau_2)=R_{j,m\bar{m}}(g(\tau_2)g^{-1}(\tau_1)).\label{bilocal}
\end{equation}
Correlators of $n$ of these bilocals were calculated as the limit of $n$-point functions of primary operators in WZW. Upon expanding the vacuum brane into Ishibashi states, and introducing complete sets of states between each local WZW insertion, such a computation is reduced to a number of matrix elements of primary operators. Using group theory, these three-point functions can then be directly computed in coordinate space as:
\begin{equation}
\bra{j_1,m_1, \bar{m}_1}\mo_{j,m\bar{m}}\ket{j_2,m_2,\bar{m}_2}= \int d g \bra{j_1,m_1, \bar{m}_1}\ket{g}R_{j,m\bar{m}}(g) \bra{g}\ket{j,m_2,\bar{m}_2},
\end{equation}
where we introduced a completeness relation. Using expressions \eqref{wavefctrational} and \eqref{3R}, the wavefunctions can be written out and the integral performed:\footnote{In \cite{Mertens:2018fds}, only spinless operators were considered, effectively restricting all sums to the diagonal sector $m_i=\bar{m}_i$. We present the general situation here.}
\begin{equation}
\bra{j_1,m_1, \bar{m}_1}\mo_{j,m\bar{m}}\ket{j_2,m_2,\bar{m}_2}=\sqrt{\dim j_1}\sqrt{\dim j_2}\tj{j_1}{j}{j_2}{m_1}{m}{m_2}\tj{j_1}{j}{j_2}{\bar{m}_1}{\bar{m}}{\bar{m}_2}.\label{3ptminisuper}
\end{equation}
The derivation of this relation provides a faster route towards the $3j$-symbol decomposition \eqref{pg3j} of particle-on-a-group correlators than the one presented in \cite{Mertens:2017mtv}, where a 2d argument in terms of the Wigner-Eckart theorem was given. 

\section{Correlation Functions in Particle-on-a-group}
\label{app:pog}
In the bulk text we have calculated particle-on-a-group correlators using holography. In \cite{Mertens:2018fds}, these were calculated using the embedding of particle-on-a-group in WZW. In this appendix we provide details of the third way of calculating these correlation functions: by means of a direct calculation in particle-on-a-group.

\subsection*{Four-Point Functions}

\begin{figure}[h]
\centering
\includegraphics[width=0.55\textwidth]{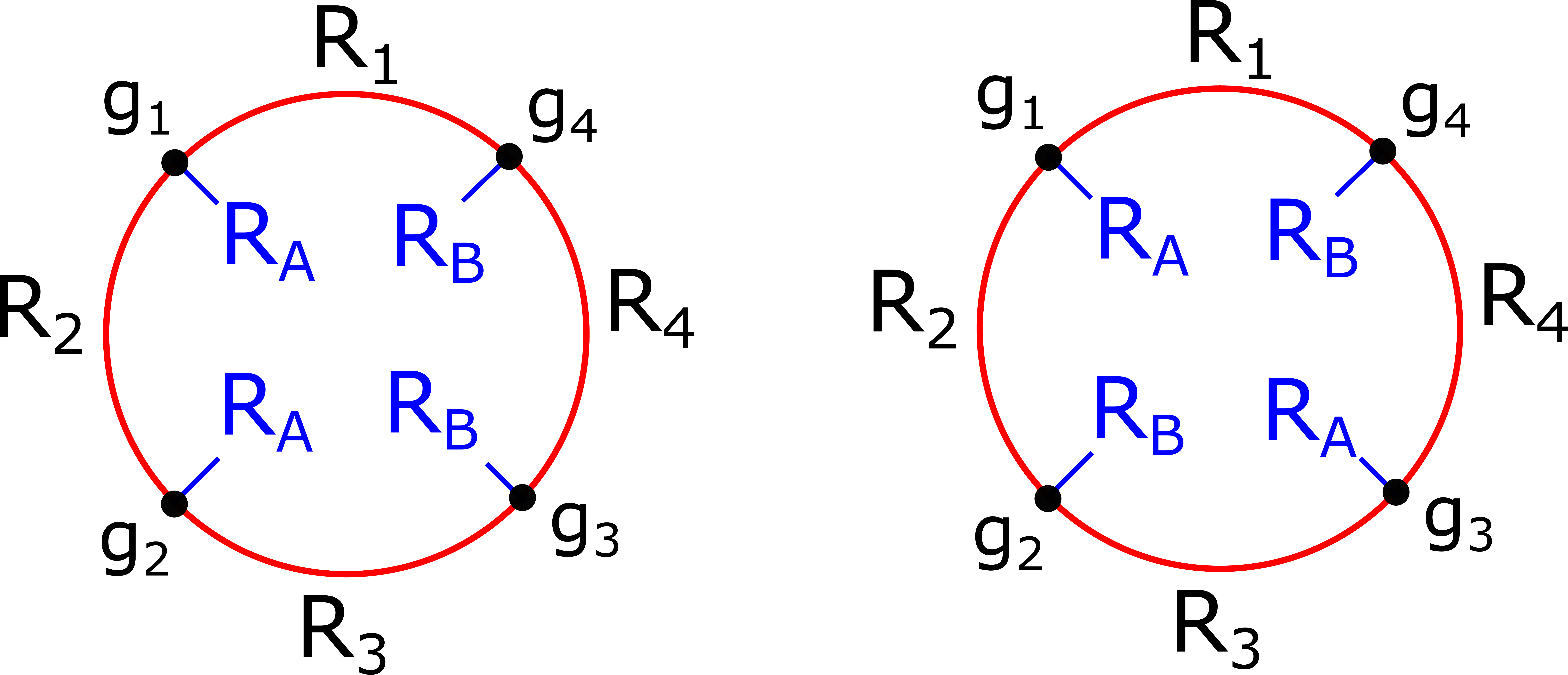}
\caption{Left: Four-point correlator for the thermal particle-on-a-group model. Right: Four-point correlator with crossed bilocal operators.}
\label{fig:pog}
\end{figure}

Using that the bilocal operators are of the form:
\begin{align}
\mathcal{W}_{R_A,m_A\bar{m}_A} &= R_A(g_2)_{m_A \alpha} R_A(g_1^{-1})_{\alpha \bar{m}_A}, \nonumber \\
\mathcal{W}_{R_B,m_B\bar{m}_B} &= R_B(g_4)_{m_B \beta} R_B(g_3^{-1})_{\beta \bar{m}_B},
\end{align}
the time-ordered finite-temperature correlator (Figure \ref{fig:pog} left) is computed as
\begin{equation}
\text{Tr}\left[e^{-\beta H} R_B(\hat{g}_4)_{m_B \beta } R_B(\hat{g}_3^{-1})_{\beta \bar{m}_B} R_A(\hat{g}_2)_{m_A \alpha} R_A(\hat{g}_1^{-1})_{\alpha \bar{m}_A} \right].
\end{equation}
Expanding in coordinate space, one obtains
\begin{align}
\sum_{\alpha,\beta} \int dg_i \bra{g_1} e^{-T_4 H} \ket{g_4} R_B(g_4)_{m_B \beta } \bra{g_4} e^{-T_3 H} \ket{g_3} \nonumber \\
R_B(g_3^{-1})_{\beta \bar{m}_B} \bra{g_3} e^{-T_2 H} \ket{g_2}  R_A(g_2)_{m_A \alpha} \bra{g_2} e^{-T_1 H}\ket{g_1} R_A(g_1^{-1})_{\alpha \bar{m}_A}.
\end{align}
Introducing four complete sets of states to compute each of the four matrix elements, from right to left denoted as $\left|R_i, m_i \bar{m}_i \right\rangle, i=1,2,3,4$, one can write
\begin{align}
\sum_{\alpha,\beta,R_i,m_i \bar{m}_i} &\int dg_i \text{dim R}_i  e^{-T_4 \mathcal{C}_4}e^{-T_3 \mathcal{C}_3} e^{-T_2 \mathcal{C}_2} e^{-T_1 \mathcal{C}_1} \nonumber \\
&\times R_4(g_1)_{m_4\bar{m}_4} R_4(g_4^{-1})_{\bar{m}_4m_4} R_B(g_4)_{m_B \beta } R_3(g_4)_{m_3\bar{m}_3}  R_3(g_3^{-1})_{\bar{m}_3m_3} R_B(g_3^{-1})_{\beta \bar{m}_B} \nonumber \\
&\times R_2(g_3)_{m_2\bar{m}_2}  R_2(g_2^{-1})_{\bar{m}_2m_2} R_A(g_2)_{m_A \alpha} \bra{g_2} R_1(g_2)_{m_1\bar{m}_1}  R_1(g_1^{-1})_{\bar{m}_1m_1} R_A(g_1^{-1})_{\alpha \bar{m}_A},
\end{align}
which gives
\begin{align}
\sum_{\alpha,\beta,R_i,m_i \bar{m}_i} &\text{dim R}_i  e^{-T_4 \mathcal{C}_4}e^{-T_3 \mathcal{C}_3} e^{-T_2 \mathcal{C}_2} e^{-T_1 \mathcal{C}_1} \nonumber \\
&\times \tj{R_4}{R_1}{R_A}{m_4}{m_1}{m_A} \tj{R_4}{R_1}{R_A}{\bar{m}_4}{\bar{m}_1}{\alpha}  \tj{R_3}{R_4}{R_B}{m_3}{m_4}{\bar{m}_B} \tj{R_3}{R_4}{R_B}{\bar{m}_3}{\bar{m}_4}{\beta} \nonumber \\
&\times \tj{R_2}{R_3}{R_B}{m_2}{m_3}{m_B} \tj{R_2}{R_3}{R_B}{\bar{m}_2}{\bar{m}_3}{\beta} \tj{R_1}{R_2}{R_A}{m_1}{m_2}{\bar{m}_A} \tj{R_1}{R_2}{R_A}{\bar{m}_1}{\bar{m}_2}{\alpha} \nonumber \\
&= \sum_{R_1,R_2,R_3} \text{dim R}_i  e^{-T_4 \mathcal{C}_4}e^{-T_3 \mathcal{C}_3} e^{-T_2 \mathcal{C}_2} e^{-T_1 \mathcal{C}_1} \nonumber \\
&\times \sum_{m_i}\tj{R_2}{R_1}{R_A}{m_4}{m_1}{m_A} \tj{R_3}{R_2}{R_B}{m_3}{m_4}{\bar{m}_B} \tj{R_2}{R_3}{R_B}{m_2}{m_3}{m_B}  \tj{R_1}{R_2}{R_A}{m_1}{m_2}{\bar{m}_A},
\end{align}
where we used \eqref{orth3j} to evaluate several summations, reproducing precisely the time-ordered four-point correlator.
\\~\\
This procedure can be extended to the OTO (or crossed) computation (Figure \ref{fig:pog} right) as we now demonstrate. The crossed correlator is computed in the particle-on-a-group model as
\begin{equation}
\text{Tr}\left[e^{-\beta H} R_B(\hat{g}_4)_{m_B \beta } R_A(\hat{g}_3)_{m_A \alpha} R_B(\hat{g}_2^{-1})_{\beta \bar{m}_B} R_A(\hat{g}_1^{-1})_{\alpha \bar{m}_A} \right].
\end{equation}
The index contractions contain all the information on the topological crossing that happens in the bulk. Expanding in coordinate space, one obtains
\begin{align}
\sum_{\alpha,\beta} \int dg_i \bra{g_1} e^{-T_4 H} \ket{g_4} R_B(g_4)_{m_B \beta} \bra{g_4} e^{-T_3 H} \ket{g_3} \nonumber \\
R_A(g_3)_{m_A \alpha} \bra{g_3} e^{-T_2 H} \ket{g_2} R_B(g_2^{-1})_{\beta \bar{m}_B} \bra{g_2} e^{-T_1 H}\ket{g_1} R_A(g_1^{-1})_{\alpha \bar{m}_A}.
\end{align}
Introducing again four complete sets of states to compute each of the four matrix elements, from right to left denoted as $\left|R_i, m_i \bar{m}_i \right\rangle, i=1,2,3,4$, one can write
\begin{align}
\sum_{\alpha,\beta,R_i,m_i \bar{m}_i} &\int dg_i \text{dim R}_i  e^{-T_4 \mathcal{C}_4}e^{-T_3 \mathcal{C}_3} e^{-T_2 \mathcal{C}_2} e^{-T_1 \mathcal{C}_1} \nonumber \\
&\times R_4(g_1)_{m_4\bar{m}_4} R_4(g_4^{-1})_{\bar{m}_4m_4} R_B(g_4)_{m_B \beta} R_3(g_4)_{m_3\bar{m}_3}  R_3(g_3^{-1})_{\bar{m}_3m_3} R_A(g_3)_{m_A \alpha} \nonumber \\
&\times R_2(g_3)_{m_2\bar{m}_2}  R_2(g_2^{-1})_{\bar{m}_2m_2} R_B(g_2^{-1})_{\beta \bar{m}_B} R_1(g_2)_{m_1\bar{m}_1}  R_1(g_1^{-1})_{\bar{m}_1m_1} R_A(g_1^{-1})_{\alpha \bar{m}_A},
\end{align}
which can be computed as
\begin{align}
\sum_{\alpha,\beta,R_i,m_i \bar{m}_i} &\text{dim R}_i  e^{-T_4 \mathcal{C}_4}e^{-T_3 \mathcal{C}_3} e^{-T_2 \mathcal{C}_2} e^{-T_1 \mathcal{C}_1} \nonumber \\
&\times \tj{R_4}{R_1}{R_A}{m_4}{m_1}{\bar{m}_A} \tj{R_4}{R_1}{R_A}{\bar{m}_4}{\bar{m}_1}{\alpha}  \tj{R_3}{R_4}{R_B}{m_3}{m_4}{m_B} \tj{R_3}{R_4}{R_B}{\bar{m}_3}{\bar{m}_4}{\beta} \nonumber \\
&\times \tj{R_2}{R_3}{R_A}{m_2}{m_3}{m_A} \tj{R_2}{R_3}{R_A}{\bar{m}_2}{\bar{m}_3}{\alpha} \tj{R_1}{R_2}{R_B}{m_1}{m_2}{\bar{m}_B} \tj{R_1}{R_2}{R_B}{\bar{m}_1}{\bar{m}_2}{\beta} \nonumber \\
&= \sum_{R_i,m_i} \text{dim R}_i  e^{-T_4 \mathcal{C}_4}e^{-T_3 \mathcal{C}_3} e^{-T_2 \mathcal{C}_2} e^{-T_1 \mathcal{C}_1} \sj{R_B}{R_1}{R_4}{R_A}{R_3}{R_2} \nonumber \\
&\times \tj{R_4}{R_1}{R_A}{m_4}{m_1}{\bar{m}_A} \tj{R_3}{R_4}{R_B}{m_3}{m_4}{m_B} \tj{R_2}{R_3}{R_A}{m_2}{m_3}{m_A}  \tj{R_1}{R_2}{R_B}{m_1}{m_2}{\bar{m}_B}, 
\end{align}
reproducing precisely the crossed four-point correlator.

\subsection*{Higher-Point Functions}
\begin{figure}[h]
\centering
\includegraphics[width=0.98\textwidth]{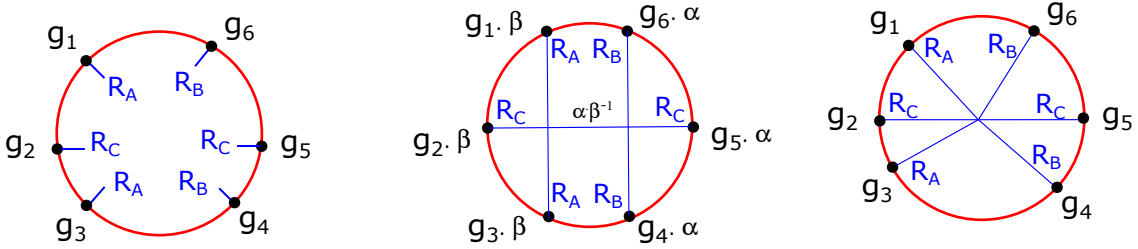}
\caption{Left: Particle-on-a-group evaluation. Middle: Choice of group elements that agrees with the bulk BF computation. Right: Bulk interpretation of the particle-on-a-group computation.}
\label{PoG6pt2}
\end{figure}

\begin{figure}[h]
\centering
\includegraphics[width=0.7\textwidth]{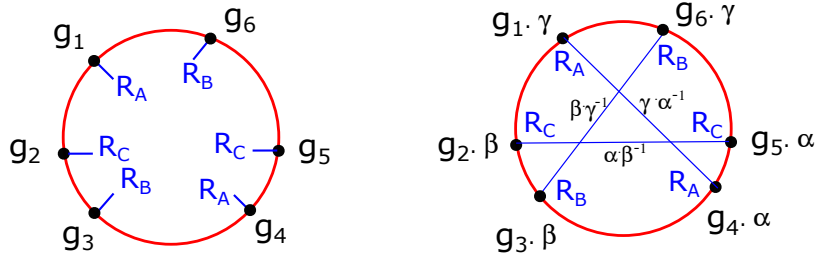}
\caption{Left: Particle-on-a-group evaluation. Right: Choice of group elements that agrees with the bulk BF computation.}
\label{PoG6pt}
\end{figure}

For higher-point functions, the above procedure does not manifestly match with the BF calculation of the main text. But, manipulating the group expressions, we show here that they do match.
\\
First note that for any group, we can parametrize a group element in a redundant way as $g\cdot \alpha$, leading to the equality:
\begin{equation}
\int d g R_1(g\cdot \alpha)R_2(g\cdot \alpha) R_3(g\cdot \alpha)^*=\int d g R_1(g)R_2(g) R_3(g)^*.
\end{equation}
Hence integrating the l.h.s. of this equation over $\alpha$, just produces an overall factor $\text{Vol}\, G=1$. The right hand side are the basic building blocks of the particle-on-a-group amplitudes, and hence this shows that particle-on-a-group amplitudes are unaffected by replacing the group integrals over $g_i$, inserted at each operator insertion $i$, by $g_i\cdot \alpha_i$ with a double integral over $g_i$ and $\alpha_i$. In fact we can go even more exotic, and choose several of the $\alpha_i$ identical. All fundamental building blocks are $\alpha_i$-independent, and all such choices yield the same answer. Choosing these variables wisely around the thermal circle allows us to obtain formulas that manifestly match with a BF computation.
\\
One can use this trick to rewrite all higher point functions to manifestly match a BF computation. We choose to demonstrate this on two examples of six-point functions which highlight all essential features (Figures \ref{PoG6pt2} and \ref{PoG6pt}).
\\~\\
A few useful facts are:
\begin{align}
\left\langle g_i \cdot \alpha \right| e^{-T H}\left|g_j \cdot \alpha\right\rangle &= \left\langle g_i \right| e^{-T H}\left|g_j \right\rangle, \\
\mathcal{W}_A = R_A\left(g_f g_i^{-1}\right)&= R_A\left((g_f \alpha) (g_i \alpha)^{-1}\right),
\end{align}
both of which simply represent overall $G$-invariance of the particle-on-a-group model. Using these two properties, the expressions can be manipulated such that only non-triangular regions reaching the boundary and interior pieces of Wilson lines feel the presence of these $\alpha_i$'s.
\\
Explicitly, for the case of Figure \ref{PoG6pt2}, one rewrites the $R_C$ Wilson line as, with $h\equiv \alpha \beta^{-1}$:
\begin{equation}
\mathcal{W}_{C} = R_C(g_5)\cdot R_C(h)\cdot R_C(g_2^{-1}),
\end{equation}
Similarly, the upper square region is manipulated as:
\begin{equation}
\left\langle g_1 \cdot \beta \right| e^{-T H}\left|g_6 \cdot \alpha\right\rangle = \left\langle g_1 \right| e^{-T H}\left|g_6 \cdot h\right\rangle = \sum_{R,a,b,c}\text{dim }R\,  R(g_1)_{ab} R (h^{-1})_{bc} R\left(g_6^{-1}\right)_{ca}.
\end{equation}
The same happens for the bottom square region. These building blocks manifestly match the BF ones, with a group element $h= \alpha\beta^{-1}$ associated with the interior line, to be integrated over in the end. This matches with the BF calculation.
\\
The particle-on-a-group perspective we started with, leads on the other hand to a decomposition of this calculation using wedge-shaped regions (Figure \ref{PoG6pt2} right). The equivalence of the computations we proved above, then leads to an important conclusion: in BF theory, any $n$-point function diagram is equivalent to a diagram where all lines intersect and converge in the bulk at a single point. This is related to a property of the bulk calculation of Wilson line networks, where all bulk Wilson line vertices can be contracted to a single point (see e.g. \cite{Besken:2016ooo}).
\\~\\
As a second illustrative example consider the diagram in Figure \ref{PoG6pt} which contains an interior region disconnected from the boundary.
Following the above logic, three group elements appear on the three interior Wilson line sections. Note that these interior group elements $h_1=\alpha\beta^{-1}$, $h_2=\beta \gamma^{-1}$ and $h_3=\gamma \alpha^{-1}$ multiply to $\mathbf{1}$, since there is no defect in the bulk region that could yield a non-trivial holonomy. Integrating over the $h_i$ hence requires a delta-function constraint:
\begin{equation}
\delta (h_1\cdot h_2 \cdot h_3 - \mathbf{1}) = \sum_R \text{dim }R \, \chi_R(h_1 \cdot h_2 \cdot h_3) = \sum_{R,a,b,c} \text{dim }R \, R(h_1)_{ab}R(h_2)_{bc}R(h_3)_{ca}.\label{delta}
\end{equation}
The building blocks now again manifestly match with the BF calculations. In particular \eqref{delta} shows how the $\dim R$ factors and the necessary group integrals arise for interior regions. This procedure is readily generalizable to other diagrams.
\section{Representation Theory of $\slrp$}
\label{app:rep}
We present the representation theory of $\slrp$. It is very closely related to that of $\slr$ itself, and large parts of it can be found in the available literature \cite{V, VK, Dijkgraaf:1991ba}.

\subsection{Subsemigroup $\slrp$}
The semigroup $\slrp$ is defined as the set of positive $\slr$ matrices with the usual matrix operations:
\begin{equation}
\left(\begin{array}{cc}
a & b \\
c & d \\
\end{array}\right), \qquad ad-bc = 1, \quad a,b,c,d > 0.
\end{equation}
In spite of the lack of an inverse, hence the name \emph{semi}group, it is possible to set up a meaningful representation theory in the sense that
\begin{equation}
R(g_1\cdot g_2) = R(g_1)\cdot R(g_2).
\end{equation}
It has an action on $L^2(\mathbb{R}^+)$ as:\footnote{That this defines a representation is seen by checking the composition rule:\begin{align}
\label{compogroup}
\left\langle f_1\right|g_1g_2\left|f_2\right\rangle &\equiv \int dx f_1(x) (g_1g_2\cdot f_2)(x), \qquad\qquad g_i = \left(\begin{array}{cc}
a_i & b_i \\
c_i & d_i \\
\end{array}\right),  \nonumber \\
&= \int dx f_1(x) g_1 \cdot \left(\left|b_2x+d_2\right|^{2j}f_2\frac{a_2x+c_2}{b_2x+d_2}\right) \nonumber \\
&= \int dx f_1(x) \left|bx+d\right|^{2j} f_2\left(\frac{ax+c}{bx+d}\right)
,\qquad g_1 g_2 = \left(\begin{array}{cc}
a & b \\
c & d \\
\end{array}\right).
\end{align}}
\begin{equation}
\label{repgroup}
f_j(x) \to (g \cdot f_j)(x) = \left|bx+d\right|^{2j}f_j\left(\frac{ax+c}{bx+d}\right),
\end{equation}
Due to the positivity of all matrix entries, this operation is internal in $\mathbb{R}^+$ and is well-defined. 
\\~\\
The $\mathfrak{sl}(2,\mathbb{R})$ algebra generators are defined as the traceless matrices:
\begin{eqnarray}
\label{alggen}
J^+=\left(\begin{array}{cc}
0 & -i \\
0 & 0 \\
\end{array}\right) , \quad J^- = \left(\begin{array}{cc}
0 & 0 \\
-i & 0 \\
\end{array}\right), \quad J^0 = \frac{1}{2}\left(\begin{array}{cc}
i & 0 \\
0 & -i \\
\end{array}\right),
\end{eqnarray}
satisfying the algebra:
\begin{equation}
\label{alge}
\left[J^0,J^\pm\right] = \pm i J^\pm, \quad \left[J^+,J^-\right] = 2iJ^0.
\end{equation}
Studying the group transformation \eqref{repgroup} infinitesimally, using $g = \mathbf{1} + i \epsilon_a J^a$, leads to the Borel-Weil realization of the $\mathfrak{sl}(2,\mathbb{R})$ algebra:
\begin{align}
\label{BW}
i\hat{J}^- &= \partial_x, \nonumber \\
i\hat{J}^0 &= -x\partial_x + j, \nonumber \\
i\hat{J}^+ &= -x^2\partial_x + 2j x,
\end{align}
satisfying the same relations \eqref{alge}. The Casimir associated to this representation $j$ computed in this realization equals $\mathcal{C} = \left(\hat{J}^0\right)^2 + \frac{1}{2}\left(\hat{J}^+\hat{J}^- + \hat{J}^-\hat{J}^+\right) = -j(j+1)$, and demonstrates that \eqref{repgroup} indeed constructs a spin-$j$ representation.
\\~\\
A matrix element in a representation $j$ is defined as the overlap:
\begin{equation}
R_{ab}(g) \equiv \left\langle j a\right| g \left|j b\right\rangle = \int_{0}^{+\infty} dx f_{ja}^*(x) (g \cdot f_{jb}(x)) .\label{rab}
\end{equation}
The continuous representations form a complete set in the sense that:\footnote{This is the classical $b\to 0$ limit of the Plancherel decomposition of $L^2 ($SL${}_q^+(2,\mathbb{R}))$. The latter was conjectured to hold by Ponsot and Teschner \cite{Ponsot:1999uf,Ponsot:2000mt}, but proven later in the Mathematics literature \cite{Ip}.}
\begin{equation}
\label{plusplanch}
\boxed{L^2 (\slrp) \simeq \int_{0}^{\infty}dk \, k \sinh(2\pi k) P_k \otimes P_k},
\end{equation}
in terms of the (unitary) principal continuous irreps $P_k$, which are the same as those of $\slr$. This can be compared to the analogous decomposition \eqref{plslr} of the full $\slr$ manifold.
\\~\\
The above Plancherel decomposition \eqref{plusplanch} implies that wavefunctions of the particle on $\slrp$ model, can be decomposed into these irreps using the complete orthogonal set of functions:
\begin{equation}
\psi^k_{ab}(g) = \sqrt{\text{dim }k} R_{k,ab}(g), \qquad \text{dim }k = k \sinh(2\pi k),\label{orthoset}
\end{equation}
in terms of the matrix elements $R_{k,ab}(g)$.

\subsection{Hyperbolic Basis}
The matrix elements of the subsemigroup $\slrp$ can be found as a subset of the $\slr$ matrix elements when diagonalizing the $J^0$ generator. This basis is called the \emph{hyperbolic} basis. First let's discuss $\slr$. The eigenfunctions ($x>0$)
\begin{align}
\left\langle x\right|\left.s\right\rangle = \frac{1}{\sqrt{2\pi}}x^{is-1/2}, \qquad \left\langle s\right|\left.x\right\rangle = \frac{1}{\sqrt{2\pi}}x^{-is-1/2},\label{hypbas}
\end{align}
are a basis on $\mathbb{R}^+$:
\begin{align}
\label{basis}
\int_{0}^{+\infty} \frac{dx}{x}\, x^{is}x^{-is'} &= 2\pi \delta(s-s'), \\
\int_{-\infty}^{+\infty} ds \, x^{is-1/2} x'^{-is-1/2} &= 
2\pi \delta(x-x'),
\end{align}
with parameter $s$ related to the $J^0$-eigenvalue by \eqref{BW}. An analogous basis is constructed on $\mathbb{R}^-$. 
Defining the four matrix elements
\begin{equation}
K_{\pm \pm}(g; s_1,s_2) \equiv \left\langle s_1,\pm\right| g \left|s_2,\pm\right\rangle,\label{kdef}
\end{equation}
linking basis functions in the $x<0$ $(-)$ or $x>0$ $(+)$ sector with one another, we can write the matrix elements of  $\slr$ on $L^2(\mathbb{R})$ in the hyperbolic basis as a $2\times 2$ matrix of matrix elements:
\begin{equation}
\mathbf{K}(g) = \left(\begin{array}{cc}
K_{++} & K_{+-} \\
K_{-+} & K_{--} \\
\end{array}\right).
\end{equation}
This matrix composes under group transformations using matrix multiplication: $\mathbf{K}(g_1 \cdot g_2) = \mathbf{K}(g_1) \cdot \mathbf{K}(g_2)$. 
Specifying now to $\slrp$, it is obvious from \eqref{kdef} that the matrix elements in the hyperbolic basis of $\slrp$ are just $K_{++}$.\footnote{Also their $q$-deformed variants are known, which reduce to these in the classical limit again \cite{Ip}.} A consistency check is that for the subsemigroup elements $g_1$, $g_2$ the composition law of $\slr$ irrep matrices implies the composition law of $\slrp$ irrep matrices: $K_{++}(g_1 \cdot g_2) = K_{++}(g_1) \cdot K_{++}(g_2)$. 
This matrix element can be explicitly computed by evaluating the defining integral:
\begin{equation}
K_{++}(g; s_1,s_2) = \left\langle s_1\right| g \left|s_2\right\rangle = \int_{0}^{+\infty} dx x^{-is_1-1/2}(g \cdot x^{is_2-1/2}).
\end{equation}
The Gauss decomposition of a generic $\slrp$ matrix is given by:
\begin{eqnarray}
\label{Gauss}
g=e^{i\gamma_L J^-}e^{2i\phi J^0}e^{i \gamma_R J^+} = \left(\begin{array}{cc}
1 & 0 \\
\gamma_L & 1 \\
\end{array}\right)\left(\begin{array}{cc}
e^{-\phi} & 0 \\
0 & e^{\phi} \\
\end{array}\right)\left(\begin{array}{cc}
1 & \gamma_R \\
0 & 1 \\
\end{array}\right), \quad \gamma_L,\gamma_R >0,
\end{eqnarray}
and provides a complete covering of the $\slrp$ manifold. It has corresponding metric
\begin{equation}
ds^2 = \frac{1}{2} \text{Tr}\left[(g^{-1}dg)^2\right] = d\phi^2 + e^{-2\phi}d\gamma_Ld\gamma_R , \quad \gamma_L,\gamma_R >0.
\end{equation}
Note here that for $g\in\slrp$, though no inverse exists in the semigroup, $g^{-1}$ is well-defined because $\slrp$ is a subregion of $\slr$. This is crucial \cite{toap}, as otherwise the construction of a $\slrp$ BF theory would not be valid. 
\\
For each of the three constituents of \eqref{Gauss}, one obtains the matrix elements $(j=-\frac{1}{2}+ik)$:
\begin{align}
K_{++}(s_1,s_2;\phi) &= e^{2i(k-s_2)\phi}\delta(s_1-s_2), \\
K_{++}(s_1,s_2;\gamma_L) &= \frac{1}{2\pi}\frac{\Gamma(-is_1+1/2)\Gamma(is_1-is_2)}{\Gamma(-is_2+1/2)}\gamma_L^{is_2-is_1}, \\
K_{++}(s_1,s_2;\gamma_R) &= \frac{1}{2\pi}\frac{\Gamma(is_2-is_1)\Gamma(is_1+1/2-2ik)}{\Gamma(is_2+1/2-2ik)}\gamma_R^{is_1-is_2}.
\end{align}
The generic matrix element can then be readily computed as
\begin{equation}
K_{++}(s_1,s_2;g) = \int_{-\infty}^{+\infty} d\alpha d\beta \, K_{++}(s_1,\alpha;\gamma_L) K_{++}(\alpha,\beta;\phi) K_{++}(\beta,s_2;\gamma_R).
\end{equation} 
The orthogonal wavefunctions are, following \eqref{orthoset} obtained as
\begin{equation}
\psi^k_{s_1 s_2}(g)=\sqrt{k\sinh 2\pi k}\, K_{++}(s_1,s_2;g).
\end{equation}

\subsection{Parabolic Matrix Elements and Whittaker Functions}
To find parabolic eigenfunctions, we diagonalize the Borel subsemigroup:
\begin{eqnarray}
\mathbf{h_-}=\left(\begin{array}{cc}
1 & 0 \\
t & 1 \\
\end{array}\right) = \exp i t J ^-, \quad t \in \mathbb{R}^+,
\end{eqnarray}
in terms of damped exponentials $f_{j\nu}(x) = e^{-\nu x}$, which indeed diagonalize $\mathbf{h_-}$ as
\begin{equation}
\mathbf{h_-} \cdot f_{j\nu}(x) = f_{j\nu}(x+t) = e^{- \nu t} f_{j\nu}(x),
\end{equation}
with $J^- = i\nu$. 
We denote this set $\left\{\left| \nu_L\right\rangle, \,\, \nu \in \mathbb{R}\right\}$:
\begin{equation}
\label{whittakvect}
\left\langle x\right|\left.\nu_L\right\rangle = \frac{1}{\sqrt{2\pi}} e^{-\nu x}, \quad \left\langle \nu_L\right|\left.x\right\rangle = \frac{1}{\sqrt{2\pi}} e^{-\nu x}.
\end{equation}
Likewise, the subsemigroup
\begin{eqnarray}
\mathbf{h_+} =\left(\begin{array}{cc}
1 & t \\
0 & 1 \\
\end{array}\right) = \exp i t J ^+, \quad t \in \mathbb{R}^+,
\end{eqnarray}
is diagonalized by $f_{j\nu}(x) = \left|x\right|^{2j} e^{-\frac{\nu}{x}}$ with eigenvalue $e^{-\nu t}$ and hence $J^+ = i\nu$. We denote them as $\left\{\left|\nu_R\right\rangle, \,\, \nu \in \mathbb{R}\right\}$:
\begin{align}
\label{Lbasis}
\left\langle x\right|\left.\nu_R\right\rangle = x^{2j} e^{- \frac{\nu}{x}}, \qquad \left\langle \nu_R\right|\left.x\right\rangle = x^{-2j-2} e^{-\frac{\nu}{x}}.
\end{align} 
These state vectors are \emph{not} basis vectors. E.g. \eqref{whittakvect} diagonalizes the operator $i\frac{\partial}{\partial x}$, which is self-adjoint on $\mathbb{R}$, but not on $\mathbb{R}^+$. This is a big difference between $\slr$ and $\slrp$, where in the former case the parabolic eigenfunctions do form a basis. In the case of $\slrp$, they do not. 
$J^0$ on the other hand \emph{is} self-adjoint on $\mathbb{R}^+$, and leads to the hyperbolic basis \eqref{hypbas} we constructed above.
The vectors \eqref{whittakvect} and \eqref{Lbasis} can then be decomposed in the hyperbolic basis using the Cahen-Mellin integral:
\begin{equation}
e^{-y} = \frac{1}{2\pi i}\int_{c-i\infty}^{c+i\infty}ds \, \Gamma(s) y^{-s}, \qquad y>0.\label{mellin}
\end{equation}
One finds for the overlap:
\begin{alignat}{4}
&\left\langle \nu_L\right|\left.s\right\rangle = &&\frac{1}{2\pi} \Gamma\Big(is+\frac{1}{2}\Big) \nu^{-is-\frac{1}{2}}, \qquad 
&&\left\langle s\right|\left.\nu_L\right\rangle = &&\frac{1}{2\pi} \Gamma\Big(-is+\frac{1}{2}\Big) \nu^{+is-\frac{1}{2}}, \\
&\left\langle \nu_R\right|\left.s\right\rangle = &&\frac{1}{2\pi} \Gamma\Big(-is+\frac{1}{2}+2ik\Big) \nu^{is+\frac{1}{2}-2ik}, \qquad 
&&\left\langle s\right|\left.\nu_R\right\rangle = &&\frac{1}{2\pi} \Gamma\Big(is+\frac{1}{2}-2ik\Big) \nu^{-is+\frac{1}{2}+2ik}.
\end{alignat}
This transition is the same as that linking Minkowski eigenmodes to Rindler modes.\footnote{Mellin transform and inverse:
\begin{align}
f(x) &= \frac{1}{\sqrt{2\pi}}\int_{-\infty}^{+\infty}ds F(s) x^{-is-1/2}, \\
F(s) &= \frac{1}{\sqrt{2\pi}} \int_{0}^{+\infty}dx f(x) x^{is-1/2},
\end{align}
which can be checked from \eqref{basis}. The Mellin transform maps a function on $\mathbb{R}^+$ to a function on the entire real axis.
}
\\~\\
Imposing the Schwarzian constraints $\bar{J^-} = -i\sqrt{\mu}, \, J^+ = i\sqrt{\mu}$ on a matrix element $R_{ab}(g)$ means that the $a$-index needs to be an eigenvalue of $J^-$ and the $b$-index needs to be an eigenvalue of $J^+$. The more general representation matrix elements to consider in the main text are hence matrix elements of a \emph{mixed parabolic} type, where the ket is taken from \eqref{Lbasis} and the bra from \eqref{whittakvect}:
\begin{align}
\label{Lcons1}
J^+\left| \lambda_R \right\rangle &= i\lambda \left|\lambda_R \right\rangle, \\
\label{Lcons2}
\left\langle \nu_L\right| J^- &= \left\langle  \nu_L\right| i\nu.
\end{align}
Taking the Gauss parametrization \eqref{Gauss}, a matrix element of this type diagonalizes the upper and lower triangular matrices. The matrix element of the remaining middle Cartan element $e^{2i\phi J^0}$ is called the Whittaker function (or coefficient). The elementary basis functions $f_\nu $ and $f_\lambda$ are called the Whittaker vectors in the Mathematics literature. The study of such functions on group manifolds goes back to the '60-'70s \cite{Jacquet,Schiffmann,Hashizume1,Hashizume2}. 
\\
The overlap between these states is easily found as
\begin{equation}
\bra{\nu_L}\ket{\lambda_R}=\frac{1}{2\pi}\int_0^\infty \frac{dx}{x} x^{ik} e^{-\nu x}e^{-\frac{\lambda}{x}}=\frac{1}{\pi} \left(\frac{\lambda}{\nu}\right)^{ik} K_{2ik}(2\sqrt{\nu\lambda}).\label{nulambda}
\end{equation}
The full mixed parabolic matrix element is:\footnote{This matches, including all prefactors, with the classical limit $b\to0$ of the Whittaker function of $U_q(\mathfrak{sl}(2,\mathbb{R})) \otimes U_{\tilde{q}}(\mathfrak{sl}(2,\mathbb{R}))$ \cite{Kharchev:2001rs}.}
\begin{align}
R_{j,\nu\lambda}(g) \equiv \left\langle \nu_L\right| g(\phi,\gamma_L,\gamma_R) \left|\lambda_R\right\rangle &=  \int ds_1 ds_2 \left\langle \nu_L\right|\left.s\right\rangle \left\langle s_1\right| g \left|s_2\right\rangle \left\langle s_2\right|\left.\lambda_R\right\rangle \nonumber \\
&= \frac{1}{\pi} \left(\frac{\lambda}{\nu}\right)^{ik} e^{-\phi}K_{2ik}(2\sqrt{\nu\lambda}e^{\phi})e^{-\nu\gamma_L-\lambda\gamma_R}.
\end{align}
Applying the Schwarzian constraints, this becomes:
\begin{align}
R_{j,\sqrt{\mu}\sqrt{\mu}}(g) = \left\langle \sqrt{\mu}_L\right|g(\phi,\gamma_L,\gamma_R) \left|\sqrt{\mu}_R\right\rangle = \frac{1}{\pi} e^{\phi} K_{2ik}\left(2\sqrt{\mu}e^{\phi}\right)e^{-\sqrt{\mu}(\gamma_L + \gamma_R)}.
\end{align}
An addition theorem can be found by inserting a complete set of intermediate states in the \emph{hyperbolic} basis:
\begin{equation}
\label{addtheo}
\left\langle \nu_L\right| g_1\cdot g_2 \left|\lambda_R\right\rangle = \int_{-\infty}^{+\infty} ds \left\langle \nu_L\right| g_1 \left|s\right\rangle\left\langle s\right| g_2 \left|\lambda_R\right\rangle.
\end{equation}
Doing this to split up internal lines in the diagram, we are led to computing integrals of the type:\footnote{To deal with splitting up discrete lines in the hyperbolic basis, one can use formulas known from $\slr$: upon setting the labels $m=n=0$ (as we ought to do for the parabolic endpoints of the Wilson line), the treatment is exactly the same as the $\slr$ case discussed in section 7.7.7 in \cite{VK}. Transferring to the hyperbolic basis is done by again performing a Mellin transform on the amplitudes.}
\begin{equation}
\int dg \, \left\langle \nu_L\right| g \left|s_1\right\rangle \left\langle \nu'_L\right| g \left|s_2\right\rangle \left\langle m_L\right| g \left|s_2\right\rangle.\label{para3}
\end{equation}
This is the triple Mellin transform of 
\begin{equation}
\int d g \bra{s_a}g\ket{s_1}\bra{s_b}g\ket{s_2}\bra{s_c}g\ket{s_3},
\end{equation}
following \eqref{mellin}. Doing this integral results in the product of hyperbolic $3j$-symbols of $\slrp$. The objects that appear at Wilson line endpoints on the boundary are then parabolic $3j$-symbols, defined as the triple Mellin transform of a hyperbolic $3j$-symbol:
\begin{equation}
\tj{k_1}{k_2}{\ell}{\nu}{\nu'}{m}_{\text{par}} \equiv  \int ds_a ds_b ds_c \left\langle \nu_L\right|\left.s_a\right\rangle \left\langle \nu'_L\right|\left.s_b\right\rangle \left\langle m_L\right|\left.s_c\right\rangle \tj{k_1}{k_2}{\ell}{s_a}{s_b}{s_c}_{\text{hyp}}.
\end{equation}
This is the precise meaning of \eqref{cgsch} in the main text.
\\~\\
When considering diagrams with intersections in the bulk, we can use the addition theorem \eqref{addtheo} to split the lines, where intermediate labels are always in the hyperbolic basis. The intermediate hyperbolic channel $s$ is integrated over, and hence these do \emph{not} satisfy the Liouville constraints \eqref{Lcons1}, \eqref{Lcons2}. This means that this step is only possible due to our knowledge on how to embed Liouville / Schwarzian within an $\slrp$ system. This procedure of splitting up internal lines is only required when calculating crossed Wilson lines (OTO-correlators) and resonates with the fact that also in \cite{Mertens:2017mtv} we need to go outside the strict 1d theory in order to have a natural understanding on how folded time contours work. 
\\~\\
As an example, let's consider how the $6j$-symbol appears. Drawing parabolic labels in red, and hyperbolic labels in black, the $6j$-symbol computation corresponds to the following picture:
\begin{align}
\label{6jpict2}
\begin{tikzpicture}[scale=1, baseline={([yshift=0cm]current bounding box.center)}]
\draw[thick,red] (-1.0,1.15) -- (-0.5,0.65);
\draw[thick] (-0.5,0.65) -- (0,0.15);
\draw[thick] (0.15,0) -- (0.65,-0.5);
\draw[thick,red] (0.65,-0.5) -- (1.15,-1.0);
\draw[thick,red] (-1.05,1.05) -- (-0.55,0.55);
\draw[thick] (-0.55,0.55) -- (0.55,-0.55);
\draw[thick,red] (0.55,-0.55) -- (1.05,-1.05);
\draw[thick,red] (-1.15,1.00) -- (-0.65,0.5);
\draw[thick] (-0.65,0.5) -- (-0.15,0);
\draw[thick] (0,-0.15) -- (0.5,-0.65);
\draw[thick,red] (0.5,-0.65) -- (1.00,-1.15);
\draw[thick,red] (-1.05,-1.05) -- (-0.55,-0.55);
\draw[thick] (-0.55,-0.55) -- (0.55,0.55);
\draw[thick,red] (0.55,0.55) -- (1.05,1.05);
\draw[thick,red] (-1.15,-1.00) -- (-0.65,-0.5);
\draw[thick] (-0.65,-0.5) -- (-0.15,0);
\draw[thick] (0,0.15) -- (0.5,0.65);
\draw[thick,red] (0.5,0.65) -- (1.00,1.15);
\draw[thick,red] (-1.00,-1.15) -- (-0.5,-0.65);
\draw[thick] (-0.5,-0.65) -- (0,-0.15);
\draw[thick] (0.15,0) -- (0.65,0.5);
\draw[thick,red] (0.65,0.5) -- (1.15,1.00);
\draw[thick] (0,0) circle (1.5);
\draw[fill,red] (-1.05,-1.05) circle (0.1);
\draw[fill,red] (1.05,-1.05) circle (0.1);
\draw[fill,red] (-1.05,1.05) circle (0.1);
\draw[fill,red] (1.05,1.05) circle (0.1);
\draw (-0.9,-0.4) node {\footnotesize  $g_1$};
\draw (0.9,-0.4) node {\footnotesize  $g_2$};
\draw (0.9,0.4) node {\footnotesize  $g_3$};
\draw (-0.9,0.4) node {\footnotesize  $g_4$};
\end{tikzpicture}
\end{align}
Each of the four $g$-integrals (each with three matrix elements) leads to a product of one hyperbolic $3j$-symbol (the interior black lines), and one parabolic $3j$-symbol (the exterior red lines). The $6j$-symbol computation then only contains the four interior hyperbolic $3j$-symbols and proceeds along the same lines as in the compact case, with the only difference the continuous hyperbolic labels $s$ that are integrated over (with flat measure):
\begin{align}
\sj{R_B}{R_1}{R_4}{R_A}{R_3}{R_2} = \int ds_i \tj{R_1}{R_2}{R_A}{s_1}{s_2}{s_A} \tj{R_2}{R_3}{R_B}{s_2}{s_3}{s_B} \tj{R_3}{R_4}{R_A}{s_3}{s_4}{s_A} \tj{R_4}{R_1}{R_B}{s_1}{s_2}{s_B}.
\end{align}
Indeed, within the $q$-deformed theory, Ponsot and Teschner determined the $q$-deformed $6j$-coefficients for the Virasoro case using the integral of four hyperbolic $q$-deformed $3j$-coefficients \cite{Ponsot:1999uf,Ponsot:2000mt}. The classical limit of their expressions results in the $\slrp$ $6j$-symbol recovered here.

Finally, note that matrix elements of the mixed parabolic type do not give unity when $\phi=\gamma_L=\gamma_R=0$, as made explicit in \eqref{nulambda}. In other bases (e.g. as discussed recently in \cite{Kitaev:2017hnr}), this property $R_{mn}(\mathbf{1})=\delta_{m,n}$ fixes the normalization of the matrix element. The analogue of the unit group element is played by setting 
\begin{eqnarray}
g = \left(\begin{array}{cc}
0 & -1 \\
1 & 0 \\
\end{array}\right).
\end{eqnarray}
Indeed, $g$ transforms an $L$-eigenstate \eqref{whittakvect} into an $R$-eigenstate \eqref{Lbasis} and vice versa, through \eqref{repgroup}, such that mixed parabolic matrix elements of $g$ \emph{are} diagonal. In Gauss coordinates \eqref{Gauss}, this element can be written as the locus:
\begin{equation}
\label{locus}
\phi \to \infty, \quad \gamma_L = e^{\phi}, \quad \gamma_R = -e^{\phi}.
\end{equation}
This fact is important in the main text for pinpointing precisely how to implement the vacuum boundary states in the JT amplitudes, and is crucial in obtaining the correct density of states in regions connected to the outer disk boundary.

\section{From the $\mathfrak{sl}(2,\mathbb{R})$ Casimir to the minisuperspace Liouville Hamiltonian}
\label{app:eigenvalue}
\begin{figure}[h]
\centering
\includegraphics[width=0.35\textwidth]{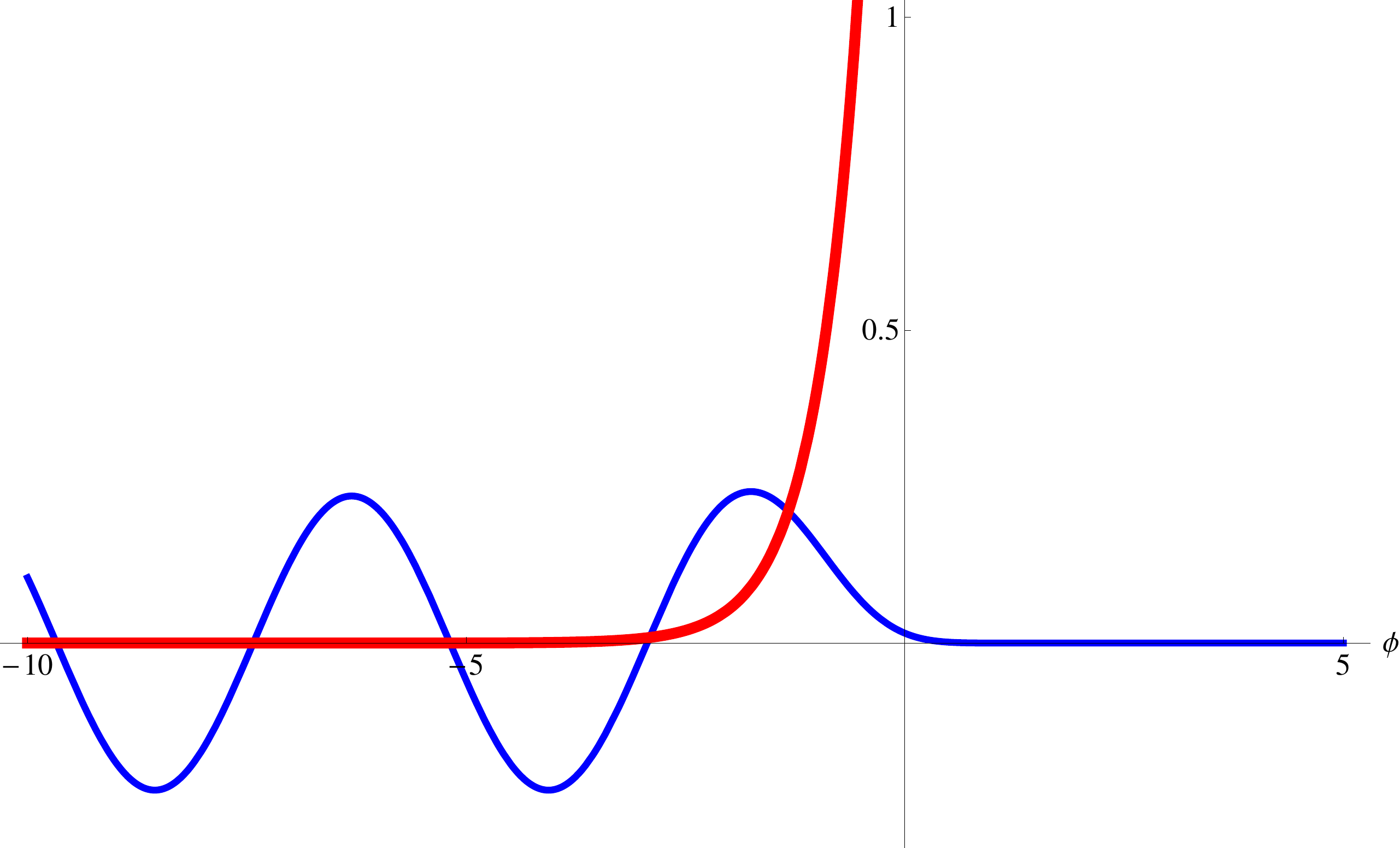}
\caption{1d Liouville potential (red), with normalizable solution $\psi_s(\phi)$ superimposed (blue).}
\label{fig:ExpPotential}
\end{figure}

To the Gauss parametrization \eqref{Gauss}, one can associate the left- and right-regular realization of the algebra:
\begin{alignat}{4}
&i\hat{\mathcal{D}}^-_L &&= \partial_{\gamma_L}, \qquad && i\hat{\mathcal{D}}^-_R &&= -\gamma_R^2 \partial_{\gamma_R} - \gamma_R \partial_\phi + e^{2\phi} \partial_{\gamma_L}, \nonumber \\
&i\hat{\mathcal{D}}^0_L &&= -\gamma_L \partial_{\gamma_L} - \frac{1}{2}\partial_\phi,  \qquad && i\hat{\mathcal{D}}^0_R &&= -\gamma_R \partial_{\gamma_R} - \frac{1}{2}\partial_\phi, \\
&i\hat{\mathcal{D}}^+_L &&= -\gamma_L^2 \partial_{\gamma_L} - \gamma_L \partial_\phi + e^{2\phi} \partial_{\gamma_R}, \qquad && i\hat{\mathcal{D}}^+_R &&= \partial_{\gamma_R} \nonumber ,
\end{alignat}
with Casimir operator, computed in either realization:
\begin{equation}
\hat{\mathcal{C}} = \left(\hat{\mathcal{D}}^0\right)^2 + \frac{1}{2}\left(\hat{\mathcal{D}}^+\hat{\mathcal{D}}^- + \hat{\mathcal{D}}^-\hat{\mathcal{D}}^+\right) = - \frac{1}{4}\partial_\phi^2 + \frac{1}{2}\partial_\phi - e^{2\phi}\partial_{\gamma_L}\partial_{\gamma_R}.
\end{equation}
The $\mathfrak{sl}(2,\mathbb{R})$ Casimir eigenvalue problem 
\begin{equation}
\hat{\mathcal{C}} \, \chi(\phi,\gamma_L,\gamma_R) = \left(\frac{1}{4} + s^2\right) \chi(\phi,\gamma_L,\gamma_R),
\end{equation}
complemented with the Liouville constraints $i\partial_{\gamma_{L}} = -i\partial_{\gamma_{R}} = i\sqrt{\mu}$, and upon extracting a factor $e^\phi$ from the eigenfunction as $\chi = e^{\phi}\psi$, simplifies to
\begin{equation}
\left(- \frac{1}{4}\partial_\phi^2 + \mu e^{2\phi}\right) \psi_s(\phi) =  s^2 \psi_s(\phi),
\end{equation}
which is a 1d Liouville equation, with normalizable solution:
\begin{equation}
\psi_s(\phi) = K_{2is}(2\sqrt{\mu}e^{\phi}).
\end{equation}
The potential and its solution are sketched below (Figure \ref{fig:ExpPotential}).
This reduces the $\mathfrak{sl}(2,\mathbb{R})$ system to the 1d Liouville model, where only the continuous sector survives as normalizable solutions. This construction can be compared to that given in section \ref{sect:Liou} in 2d.

\section{Representation Theory of SL$^+(n,\mathbb{R})$}
\label{app:higher}
The generalization to the Hamiltonian reduction of SL$(n,\mathbb{R})$ is known, and it is useful to present some of the details. We follow \cite{Bershadsky:1989mf,Gerasimov:1996zk} and focus on SL$(3,\mathbb{R})$. We will make some comments of the general case as we go along. The eight traceless SL$(3,\mathbb{R})$ generators can be written:
\begin{align}
H_1 &= \left(\begin{array}{ccc}
i & 0 & 0\\
0 & -i & 0\\
0 & 0 & 0
\end{array}\right), \quad
H_2 = \left(\begin{array}{ccc}
0 & 0 & 0\\
0 & i & 0\\
0 & 0 & -i
\end{array}\right), \nonumber \\
J_1^+ &= \left(\begin{array}{ccc}
0 & -i & 0\\
0 & 0 & 0\\
0 & 0 & 0
\end{array}\right), \quad
J_2^+ = \left(\begin{array}{ccc}
0 & 0 & 0\\
0 & 0 & -i\\
0 & 0 & 0
\end{array}\right), \quad
J_{12}^+ = \left(\begin{array}{ccc}
0 & 0 & -i\\
0 & 0 & 0\\
0 & 0 & 0
\end{array}\right), \nonumber \\
J_1^- &= \left(\begin{array}{ccc}
0 & 0 & 0\\
i & 0 & 0\\
0 & 0 & 0
\end{array}\right), \quad
J_2^- = \left(\begin{array}{ccc}
0 & 0 & 0\\
0 & 0 & 0\\
0 & i & 0
\end{array}\right), \quad
J_{12}^- = \left(\begin{array}{ccc}
0 & 0 & 0\\
0 & 0 & 0\\
i & 0 & 0
\end{array}\right).
\end{align}
The generators with multiple indices are associated to non-simple roots, and the indexation refers to how they are created from commutators. All off-diagonal matrices generate parabolic subgroups of SL$(n,\mathbb{R})$ as 
\begin{equation}
\exp{i t J_1^+} = \left(\begin{array}{ccc}
1 & t & 0\\
0 & 1 & 0\\
0 & 0 & 1
\end{array}\right), \qquad \text{etc.}
\end{equation}
The Hamiltonian reduction of the positive semisubgroup SL${}^+(n,\mathbb{R})$ is obtained by fixing all generators of simple roots $J_i^+ = i\sqrt{\mu_i}$, and all remaining ones (in this case only $J_{12}$) to zero:
\begin{align}
J_i^+ = i\sqrt{\mu_i}, \qquad J_{ij \hdots} = 0, \qquad \bar{J}_i^- = - i\sqrt{\mu_i}.
\end{align}
The Gauss decomposition can be generalized to this case, as
\begin{equation}
g = e^{i \gamma_L^{ij \hdots} J^-_{ij \hdots}} \,\, \mathbf{h}\,\, e^{i \gamma_R^{ij \hdots} J^+_{ij \hdots}}
\end{equation}
with $\mathbf{h} = \prod_i \exp{i \phi_i H_i}$ the Cartan group element, in this case $\mathbf{h} = \exp{i \phi_1 H_1} \exp{i \phi_2 H_2}$, for $N-1$ fields $\phi_i$, combined in the vector $\bm{\phi}$. These will become the Toda CFT fields.
\\~\\
States in a representation are labeled with the eigenvalue of the Cartan subgroup (in this case $H_1$ and $H_2$) on the highest weight state. We denote the labels as $j_i$, combined in the vector $\mathbf{j}$. 
\\~\\
As for the $n=2$ case, two different bases are constructed. The $R$-basis diagonalizes the $J^+_i$ matrices, the $L$-basis diagonalizes the $J^-_i$ matrices. Matrix elements is the resulting mixed basis can be computed, and represent a solution to the $\mathfrak{sl}(n,\mathbb{R})$ Toda minisuperspace eigenvalue problem, just like the mixed parabolic matrix elements of $\slrp$ satisfy the 1d Liouville equation. The simplest way to obtain these wavefunctions is to diagonalize the operators $J^+_i$ or $J^-_i$ in the Borel-Weil realization \cite{Gerasimov:1996zk,Chervov}, analogous to \eqref{BW}. For $n=3$, the coordinate space wavefunctions are given by:
\begin{align}
\label{Rbasistoda}
\left\langle x_1,x_2,x_{12}\right|\left.\sqrt{\mu_1}_L\sqrt{\mu_2}_L\right\rangle &= e^{- \sqrt{\mu_1} x_1 - \sqrt{\mu_2} x_2}, \\
\label{Lbasistoda}
\left\langle \sqrt{\mu_1}_R\sqrt{\mu_2}_R \right|\left.x_1,x_2,x_{12}\right\rangle &= (x_{12}-x_1x_2)^{-2j_2-2}x_{12}^{-2j_1-2}e^{- \sqrt{\mu_2} \frac{x_1}{x_{12}}}e^{-\sqrt{\mu_1} \frac{x_2}{x_1x_2-x_{12}}},
\end{align}
and the overlap $\left\langle \sqrt{\mu_1}_L\sqrt{\mu_2}_L \right|\left.\sqrt{\mu_1}_R\sqrt{\mu_2}_R\right\rangle$ can be computed in principle. In practice, only the $n=2$ case allows a direct evaluation of the integrals. The general matrix element of interest is then given by
\begin{equation}
R_{\mathbf{j},\sqrt{\mu_i}\sqrt{\mu_i}}(\bm{\phi}) \equiv \left\langle \sqrt{\mu_i}_L \right| \mathbf{h} \left|\sqrt{\mu_i}_R\right\rangle,
\end{equation}
and is called the Whittaker function. These representation matrices are to be inserted in the computation in the main text to produce $3j$- and eventually $6j$-symbols. As earlier, the group integrals over the $\gamma_L^{ij\hdots}$ and $\gamma_R^{ij\hdots}$ variables reduce to delta-functions, and the computation boils down in essence to a Toda calculation:
\begin{equation}
\int d\bm{\phi} R_{\mathbf{j}_1,\sqrt{\mu_i}\sqrt{\mu_i}}(\bm{\phi}) V_\mathbf{l}(\bm{\phi}) R_{\mathbf{j}_2,\sqrt{\mu_i}\sqrt{\mu_i}}(\bm{\phi}) = \tj{\mathbf{j}_1}{\mathbf{l}}{\mathbf{j}_2}{\sqrt{\mu_i}}{\mathbf{0}}{\sqrt{\mu_i}}^2,\label{3Rtoda}
\end{equation}
where $V_\mathbf{l} = e^{\left\langle \mathbf{l},\bm{\phi}\right\rangle}$ is the Toda vertex operator. This in principle solves the problem. It remains to be seen whether these expressions can be made more explicit.

\end{document}